\documentclass[aps,prb,twocolumn,superscriptaddress,nofootinbib,showpacs]{revtex4}
\usepackage{amssymb}
\usepackage{graphicx}
\usepackage{epsf,epsfig}
\usepackage{bm}
\usepackage{bbm}
\usepackage{amsfonts}
\usepackage{amsmath}
\usepackage{graphics}
\usepackage[normalem]{ulem}
\usepackage{simplewick}
\usepackage{enumerate}

\newcommand{\be}{\begin{eqnarray}}
\newcommand{\ee}{\end{eqnarray}}
\newcommand{\ba}{\begin{array}}
\newcommand{\ea}{\end{array}}
\newcommand{\no}{\nonumber}
\newcommand{\tr}{\mbox{tr}}
\newcommand{\Tr}{\mbox{Tr}}

\newcommand{\eps}{\varepsilon}

\newcommand{\bfr}{{\bf r}}

\newcommand{\bfp}{{\bf p}}

\newcommand{\bfq}{{\bf q}}
\newcommand{\bfs}{{\bf s}}

\newcommand{\bfsigma}{{\bm \sigma}}
\newcommand{\bfvarphi}{{\bm \varphi}}

\newcommand{\bftheta}{{\bm \theta}}

\begin{document}

\title{Keldysh approach to the renormalization group analysis of the disordered electron liquid}
\author{G. Schwiete}
\email{schwiete@zedat.fu-berlin.de} \affiliation{Dahlem Center for Complex Quantum Systems and Institut f\"ur Theoretische
Physik, Freie Universit\"at Berlin, 14195 Berlin, Germany
}
\author{A. M. Finkel'stein}
\affiliation{Department of Physics and Astronomy, Texas A\&M University, College Station, TX 77843-4242, USA}
\affiliation{Department of
Condensed Matter Physics, The Weizmann Institute of Science, 76100
Rehovot, Israel}
\date{\today}

\begin{abstract}
We present a Keldysh nonlinear sigma-model approach to the renormalization group analysis of the disordered electron liquid. We include both the Coulomb interaction and Fermi-liquid type interactions in the singlet and triplet channels into the formalism. Based on this model, we reproduce the coupled renormalization group equations for the diffusion coefficient, the frequency and interaction constants previously derived with the replica model in the imaginary time technique. With the help of source fields coupling to the particle-number and spin densities we study the density-density and spin density-spin density correlation functions in the diffusive regime. This allows us to obtain results for the electric conductivity and the spin susceptibility and thereby to re-derive the main results of the one-loop renormalization group analysis of the disordered electron liquid in the Keldysh formalism.
\end{abstract}

\pacs{71.10.Ay, 72.10.-d, 72.15.Eb, 73.23.-b} \maketitle

\section{Introduction}

In disordered conductors, perturbations of charge and spin relax diffusively at low frequencies and large distances. In a system obeying time-reversal symmetry, the low-energy modes in the Cooper channel also have a diffusive form. These modes, Diffusons and Cooperons, describe the low-energy dynamics of disordered electrons. The electron-electron (e-e) interaction causes a scattering of the diffusion modes. As a result, the diffusion constant, frequency, and interaction constants acquire corrections, which in two dimensions are logarithmically divergent at low temperatures.\cite{Altshuler85,Lee85,Finkelstein90,Belitz94RMP,DiCastro04,Finkelstein10} The procedure that handles these mutually coupled corrections corresponds to a renormalization group (RG) analysis.\cite{Finkelstein83,Castellani84,Baranov99} The derivation of the coupled RG equations is conveniently based on a generalized nonlinear sigma model (NL$\sigma$M) that includes the effects of electron-electron interactions.\cite{Finkelstein83} The structure of the theory remains intact during the course of renormalization, albeit with effective temperature-dependent parameters. Among other things, the RG analysis reveals the importance of spin\cite{Finkelstein90} (as well as valley\cite{Punnoose01}) fluctuations for establishing the strange metallic phase at low temperatures, which does not exist in two dimensions in the absence of e-e interactions.\cite{Abrahams79,Abrahams01,Spivak10} Based on this theory, both quantitative and qualitative statements about transport and thermodynamic quantities close to the metal-insulator transition in two-dimensional electron systems can be obtained for the case when it is driven by disorder and interactions.\cite{Punnoose05,Anissimova07,Finkelstein10,Punnoose10}

By its essence, the NL$\sigma$M is a minimal microscopic theory, which incorporates all symmetry constraints and conservation laws relevant for the low-energy dynamics of electrons in disordered conductors. Phenomenologically, such a theory may be considered as an analog of the Fermi liquid theory for the diffusion modes. As such, the range of applicability of the NL$\sigma$M can be broader than the conditions of its derivation. 

The original formulations of the NL$\sigma$M for non-interacting\cite{Wegner79,Efetov80,Pruisken84} as well as for interacting systems\cite{Finkelstein83} were based on the replica method,\cite{Edwards75} in combination with the imaginary time technique.\cite{AGD63} In this scheme, the partition function is replicated $n$ times before the averaging over disorder-configurations is performed; at the end of the calculation, the limit $n\rightarrow 0$ needs to be taken in order to remove certain unphysical terms that are present in the theory for finite $n$. As the main object of study is the equilibrium partition function, the theory can serve as a platform for studying thermodynamic quantities as well as the response to weak perturbations through the calculation of equilibrium correlation functions. The replica sigma model is very convenient for perturbative RG calculations, which are at the heart of the mentioned successes of this approach.

Despite these successes, the theory in its original formulation has certain limitations. The study of equilibrium correlation functions may be obscured by the required analytical continuation from imaginary frequencies to real ones, which can be very involved. Most notably, however, true non-equilibrium phenomena are beyond the scope of this theory as it is constructed with the help of the equilibrium imaginary time technique. An alternative approach to interacting many-body systems, which is free of these limitations, is the so-called Keldysh technique.\cite{Schwinger61,Kadanoff62,Keldish65,Kamenev11} It is closely related to real-time techniques developed for classical systems.\cite{Martin73,DeDominicis78,Cugliandolo02,Kamenev11,Schwiete13}
In these approaches, correlation functions are calculated directly in real time, thereby rendering the analytical continuation unnecessary. The range of applicability of the Keldysh approach includes systems in thermodynamic equilibrium as well as non-equilibrium problems. In this context, the intimate connection to quantum kinetics is of particular advantage. An additional property is very convenient when treating quenched disordered systems: the normalization of the Keldysh partition function is independent of the disorder potential. The disorder averaging can therefore be performed straightforwardly without introducing replicated fields as was already noted in Refs.~\onlinecite{DeDominicis78} and \onlinecite{Aronov85}.

In this work, we analyze a Keldysh NL$\sigma$M for e-e interactions in disordered electron systems. The Keldysh NL$\sigma$M was first employed for non-interacting electrons in Ref.~\onlinecite{Horbach93}. [A combination of replicas and the Keldysh approach was already used in Ref.~\onlinecite{Dugaev86}.] For disordered fermions with short-range interactions a Keldysh sigma model was constructed in Ref.~\onlinecite{Chamon99}, and the RG equations\cite{Finkelstein83,Castellani84} were re-derived for this case. A sigma model for electrons with long-range Coulomb interaction was introduced in Ref.~\onlinecite{Kamenev99}, and generalized to include the interaction in the Cooper channel in a subsequent work. \cite{Feigelman00} 
Our study differs from previous related works \cite{Kamenev99,Chamon99,Feigelman00} in several aspects. In contrast to Ref.~\onlinecite{Kamenev99}, we account for both the Coulomb interaction and Fermi liquid-type interactions in the singlet and the triplet channels in order to find the Keldysh analog of the original model of Ref.~\onlinecite{Finkelstein83}. The obtained model allows us to perform the full RG analysis in the presence of a perturbation that violates the time-reversal symmetry, i.e., when the Cooperons can be neglected. Unlike Ref.~\onlinecite{Chamon99}, we implement the procedure directly in the Larkin-Ovchinnikov representation, (for a review, see Ref.~\onlinecite{Kamenev11}), which is very convenient for the calculation of retarded correlation functions. We also introduce source fields coupled to the particle and spin densities. They allow us to derive the density-density and spin-density spin-density correlation functions. This requires an analysis of the static and dynamic parts of the correlation functions, including vertex corrections, and enables us, in particular, to obtain the low-temperature behavior of the electric conductivity and the spin susceptibility. In this way, we re-derive the main results of the RG theory of the disordered electron liquid with the help of the Keldysh sigma model. Whenever possible, we try to highlight those aspects of the analysis that are specific for the Keldysh approach. We conclude, that despite the differences related to working with Keldysh matrices instead of replicas, the RG-procedure in both schemes is rather similar. 

The relevance of this study goes beyond a mere confirmation of previously obtained results. We consider it as a step towards tackling problems that are sensitive to the kinetics of the electronic system at energy scales of the order of the temperature or below. Such problems are transparently treated within the Keldysh formalism. The renormalized Keldysh NL$\sigma$M allows to analyze the sub-temperature regime with effective parameters encoding the physics originating from the RG interval, i.e., from energies exceeding temperature. An important problem of this kind is the calculation of thermal transport.\cite{Schwieteheat}

This paper is organized as follows. In Sec.~\ref{sec:derivation} we describe the main steps of the derivation of the Keldysh NL$\sigma$M and cast it into a form that is convenient for the RG analysis. Due to the complex structure of the appearing fields and matrices in spin, Keldysh, time (frequency) and coordinate (momentum) spaces, the notation can at times be involved. We therefore include, from the very beginning, a compact summary of our notations as a reference point in Sec.~\ref{subsec:notations}. Section~\ref{sec:correlation functions} is concerned with the general structure of correlation functions for particle-number densities and spin densities in the diffusive regime. We perform their calculation in the Keldysh formalism emphasizing the important role of conservation laws. In Sec.~\ref{sec:renormalization} we present the RG analysis of the model. After introducing the general formalism, we discuss in detail the renormalization of the parameters (RG-charges) appearing in the model, and derive the set of coupled RG equations. In Sec.~\ref{sec:together} we return to the analysis of the correlation functions and calculate corrections to the static parts as well as vertex corrections that arise in connection with the source fields for particle-number and spin densities. This allows us to obtain the temperature dependence of the spin susceptibility in Sec.~\ref{subsec:Corrections to gamma}, and the electric conductivity in Sec.~\ref{subsec:conductivity}. Finally, we conclude in Sec.~\ref{sec:conclusion}.

\section{Keldysh sigma model for interacting electron systems}
\label{sec:derivation}
In this section, we present a derivation of the Keldysh NL$\sigma$M for the interacting electron liquid. We include the Coulomb interaction and Fermi-liquid type interactions in the singlet and triplet channels as well as source fields coupling to density and spin, see Sec.~\ref{subsec:derivation}. The resulting sigma model, which contains the Fermi liquid renormalizations, is presented in Sec.~\ref{subsec:sources}. In Sec.~\ref{subsec:modelforrg} we rewrite the sigma-model in a form that is convenient for the RG procedure that will be presented later in Sec.~\ref{sec:renormalization}. For the convenience of the reader, we first summarize our notations in Sec.~\ref{subsec:notations}.

\subsection{Notations}
\label{subsec:notations}
In the approach we use, the original Keldysh contour \cite{Keldish65} disappears from the explicit formulation of the theory which, instead, is reformulated in terms of matrices.\cite{Kamenev11}  The $2\times 2$ matrices in Keldysh space are decorated with a hat and labeled by a lower index, e.g., $\hat{\gamma}_2$ or $\hat{\sigma}_3$. For the Hubbard-Stratonovich (H-S) fields generating the electron-electron interactions the lower index is also related to the Keldysh space. We write, e.g., $\theta_k$, where $k=1,2$ indicates the so-called classical or quantum fields.

The Pauli matrices written without hats and labeled by the upper indices are used to describe interactions in the density/spin-density channels. They can be unified into the four-component vector $\vec{\sigma}=(\sigma^0,\sigma^1,\sigma^2,\sigma^3)^T$ or the three-component vector $\bfsigma=(\sigma^1,\sigma^2,\sigma^3)^T$. For the H-S fields, e.g. for $\theta^l$,  where $l=(0,1-3)$, the upper index indicates whether the field acts in the density channel (component 0) or spin-density channels (components $1-3$). Vector fields combine all four components, $\vec{\theta}=(\theta^{0},\theta^1,\theta^2,\theta^3)^T$, or three components, $\bftheta=(\theta^{1},\theta^{2},\theta^{3})^T$.
Usually, each of the components of these vectors itself is a two-component vector in the Keldysh space, e.g., $\theta^l_k$. In total, the vectors $\vec{\theta}$ and $\bftheta$ acquire eight or six components, respectively. Besides the H-S fields, the auxiliary potentials (fields) $\vec{\varphi}=(\varphi,\varphi^1,\varphi^2,\varphi^3)^T$, $\bfvarphi=(\varphi^1,\varphi^2,\varphi^3)^T$ are introduced to generate the correlation functions describing the density (singlet) and spin-density (triplet) channels.

We will use the symbols $\tr$ and $\Tr$ for traces. The symbol $\tr$ includes a trace in Keldysh space, an integration over frequencies, and a summation over spin degrees of freedom. The symbol $\Tr$, in addition to all above, includes an integration over the spatial coordinates of all the functions appearing under the trace.

Underscoring of matrices and fields denote multiplication by the matrices $\hat{u}$ from the left and right, e.g., $\underline{\hat{Q}}=\hat{u}\circ \hat{Q}\circ \hat{u}$; here the convolution is in the time domain. After the Fourier transform, the convolution converts into an algebraic product, $\underline{\hat{Q}}_{\eps,\eps'}=\hat{u}_{\eps}\hat{Q}_{\eps, \eps'}\hat{u}_{\eps'}$. The definition of the matrix $\hat{u}_{\eps}$ is given in Eq.~\eqref{eq:umatrix}; these matrices carry the information on the fermionic equilibrium distribution function $\mathcal{F}_\eps=\tanh(\eps/2T)$.

Finally, in order to lighten the notation we will in the following often write $\int_t=\int_{-\infty}^\infty dt$ and $\int_{x}=\int_{\bfr,t}$. Whenever the frequency integration is made explicit, we use the symbol $\int_\eps=\int d\eps/2\pi$. Furthermore, $\hat{\eps}$ acts trivially on a matrix in the frequency space as $\hat{\eps}\hat{Q}_{\eps\eps'}=\eps \hat{Q}_{\eps\eps'}$.

The term $\it {irreducible}$ correlation function in this paper means that only those diagrams should be considered, which cannot be separated into two disconnected parts by cutting a single Coulomb interaction line. In order to find the irreducible correlation function in the singlet channel, the long-range Coulomb interaction $V_{0}(\bfq)$ has to be separated from the rest of the interaction amplitudes. The argument $\bf q$ in $\it{any}$ amplitude of the electron-electron interaction indicates that this amplitude is reducible with respect to the Coulomb interaction.

\subsection{Derivation of the model}
\label{subsec:derivation}

Starting point for the derivation is the Keldysh partition function for the interacting electron liquid in the coherent state representation
\be
\mathcal{Z}=\int D[\psi^\dagger,\psi]\; \mbox{exp}(iS[\psi^\dagger,\psi]),
\ee
where the action $S$ is defined as
\be
S[\psi^\dagger,\psi]&=&\int_\mathcal{C}dt \;\mathcal{L}[\psi^\dagger,\psi]\\
\mathcal{L}[\psi^\dagger,\psi]&=&\int_\bfr \; {\psi}^\dagger_x i\partial_t \psi_x-K[\psi^\dagger,\psi] .
\ee
Here, $\mathcal{C}$ symbolizes the Keldysh contour,\cite{Schwinger61,Kadanoff62,Keldish65,Kamenev11} which consists of the forward (+) and backward ($-$) paths; $x=(\bfr,t)$ and $\psi_x=(\psi_{\uparrow}(x),\psi_\downarrow(x))^T$, $\psi^\dagger_x=(\psi^*_\uparrow(x),\psi^*_\downarrow(x))$ are vectors of Grassmann fields comprising the two spin components. $K$ is the grand canonical hamiltonian
\be
K=H-\mu N,\quad H=H_0+H_{int}.
\ee
The non-interacting part of the Hamiltonian is
\be
H_0=\int_\bfr\; {\psi}^\dagger_xh_0\psi_x,
\ee
where $h_0=-\nabla^2/2m^*+u_{dis}$.
Here, $u_{dis}(\bfr)$ is the disorder potential and $m^*$ is the (renormalized) mass.
The interaction Hamiltonian $H_{int}$ can be subdivided into singlet and triplet parts, $H_{int}=H_{int,\rho}+H_{int,\sigma}$, where
\be
H_{int,\rho}&=&\frac{1}{2}\int_{\bfr,\bfr'}\; n(\bfr,t)\;V_{\rho}(\bfr-\bfr')\;n(\bfr',t)\\
H_{int,\sigma}&=&2 \int_{\bfr,\bfr'} \;\bfs(\bfr,t)\;V_{\sigma}(\bfr-\bfr')\;\bfs(\bfr',t).
\ee
We introduced the particle-number density and spin densities
\be
n(x)={\psi}^\dagger_x\sigma^0\psi_x,\qquad \bfs(x)=\frac{1}{2}{\psi}^\dagger_x \bfsigma \psi_x .
\ee

The interactions in the singlet and triplet channels are described in terms of the amplitudes
\be
V_{\rho}(\bfq)= V_{0}(\bfq)+\frac{F_0^\rho}{2\nu},\quad V_{\sigma}=\frac{F_0^\sigma}{2\nu}.
\ee
Here, $F_0^\rho$ and $F_0^\sigma$ are the Fermi liquid parameters known from the phenomenological Fermi liquid theory\cite{AGD63,Lifshitz80} and $\nu$ is the single-particle density of states per spin direction. In $V_{\rho}(\bfq)$ the bare long-range part of the Coulomb interaction is separated from the short-range part. The latter determines the Fermi liquid renormalization of the polarization operator.

Next, we introduce fields on the forward and backward paths of the Keldysh contour, $\psi_{\pm}$, and group them into the vector
\be
\vec{\psi}=\left(\ba{cc}\psi_+\\\psi_-\ea\right).
\ee
The corresponding action reads
\be
S[\vec{\psi}^\dagger,\vec{\psi}]=\int_{-\infty}^{\infty} dt \left(\mathcal{L}[\psi_+^\dagger,\psi_+]-\mathcal{L}[\psi_-^\dagger,\psi_-]\right).\label{eq:S+-}
\ee
The interaction part can be decoupled with the help of a four-component H-S field for each of the $\pm$ paths, $\vartheta_{\pm}^l$, organized into a matrix
\be
\hat{\vartheta}^l=\left(\ba{cc}  \vartheta_+^{l}&0\\0&\vartheta_-^l\ea\right),\quad l=(0,1-3).
\ee
As a result, the partition function can be written as
\be
\mathcal{Z}=\int D[\vec{\vartheta}]D[\vec{\psi}^\dagger,\vec{\psi}]\mbox{exp}(iS[\vec{\psi}^\dagger,\vec{\psi},\vec{\vartheta}]),
\ee
where
\be
S[\vec{\psi}^\dagger,\vec{\psi},\vec{\vartheta}]=\int_x\; \vec{\psi}^\dagger_{x}\left(i\partial_t-h_0+\mu+\hat{\vartheta}^l\sigma^l\right)\hat{\sigma}_3\vec{\psi}_{x}\no\\
+\frac{1}{2}\int_{\bfr,\bfr',t} \;\vec{\vartheta}^T(\bfr,t)V^{-1}(\bfr-\bfr')\hat{\sigma}_3\vec{\vartheta}(\bfr',t).\quad
\ee
In the last formula, the sum over the repeated index $l$ from $0$ to $3$ is implied, while $\hat{\sigma}_3$ is the third Pauli matrix in the space of forward and backward fields. As we have already noted in Sec.~\ref{subsec:notations}, $\vec{\vartheta}$ has eight components: each of the $l$-components has two components in the Keldysh space. (The same will hold for $\vec{\theta}$ and $\vec{\varphi}$ introduced below.) We also introduced a matrix $V$ comprising the interaction potentials for the singlet and triplet channels
\be
V=\mbox{diag}(V_{\rho},V_{\sigma},V_{\sigma},V_{\sigma}).
\ee

It is convenient to change the basis and perform the Keldysh rotation\cite{Larkin75,Kamenev11}
defined by
\be
\vec{\Psi}^\dagger=\vec{\psi}^\dagger \hat{L}^{-1},\qquad\vec{\Psi}=\hat{L}\hat{\sigma}_3\vec{\psi},
\ee
where the rotation matrix $L$ is given by
\be
\hat{L}=\frac{1}{\sqrt{2}}\left(\ba{cc}1&-1\\1&1\ea\right),\quad \hat{L}^{-1}=\hat{L}^T=\hat{\sigma}_3\hat{L}\hat{\sigma}_3 .
\ee
Under the rotation $\hat{L}$, the field $\hat{\vartheta}$ transforms into $\hat{\theta}$ (the upper index $l$ is not shown),
\be
\hat{\theta}\equiv \hat{L} \hat{\vartheta} \hat{L}^{-1}=\left(\ba{cc}\theta_{cl}&\theta_q\\\theta_{q}&\theta_{cl}\ea\right).
\ee
As a result, we come to a description in terms of the classical (cl) and quantum (q) components of the bosonic fields
\be
\theta^i_{cl/q}=(\vartheta^i_+\pm\vartheta^i_-)/2.
\ee
With the help of two matrices in Keldysh space,
\be
\hat{\gamma}_1=\hat{\sigma}_0,\quad \hat{\gamma}_2=\hat{\sigma}_1,
\ee
one may write
\be
\hat{\theta}^l=\sum_{k=1,2}\theta_k^l\hat{\gamma}_k,
\ee
where $k=1$ denotes the classical component, while $k=2$ denotes the quantum component. As a result, the Keldysh action in the rotated basis reads
\be
S[\vec{\Psi}^\dagger,\vec{\Psi},\vec{\theta}]=\int_x \;\vec{\Psi}^\dagger_x\left(i\partial_t-h_0+\mu+\hat{\theta}^l\sigma^l\right)\vec{\Psi}_x\no\\
+\int_{\bfr,\bfr',t}  \;\vec{\theta}^T(\bfr,t)\hat{\gamma}_2V^{-1}(\bfr-\bfr') \vec{\theta}(\bfr',t).\quad
\ee
Working with classical and quantum fields is useful for the calculation of physical quantities like correlation functions.\cite{Kamenev11}

The first step in the derivation of the NL$\sigma$M is the averaging of the partition function over disorder configurations. For the sake of simplicity, we will work with a delta-correlated impurity potential. This choice corresponds to the statistical weight
\be
\left\langle \dots\right\rangle_{dis} =\mathcal{N}\int D[u_{dis}](\dots)\; \mbox{e}^{-\pi\nu \tau\int d\bfr u_{dis}^2(\bfr)}.
\ee
The normalization factor $\mathcal{N}$ is chosen so that $\left\langle 1\right\rangle_{dis}=1$. Averaging of the disorder-dependent part of the partition function gives
\be
&&\left\langle \mbox{e}^{-i\int_x \vec{\Psi}^\dagger_{x}u_{dis}(\bfr)\vec{\Psi}_x}\right\rangle_{dis}=\mbox{e}^{iS_{dis}},
\ee
where
\be
S_{dis}=\;\frac{i}{4\pi\nu \tau}\int_{\bfr,t,t'} (\vec{\Psi}^\dagger_{\bfr,t}\vec{\Psi}_{\bfr,t})(\vec{\Psi}^\dagger_{\bfr,t'}\vec{\Psi}_{\bfr,t'}).
\ee
Following further the standard route for the derivation of the NL$\sigma$M,\cite{Efetov99}
the four fermion term $S_{dis}$ is decoupled with a H-S field $\hat{Q}$ as
\be
\mbox{e}^{iS_{dis}}&=&\int D[\hat{Q}]\mbox{e}^{-\frac{1}{2\tau}\int_{\bfr,t,t'}   \vec{\Psi}^\dagger_{\bfr,t}\hat{Q}_{t,t'}(\bfr)\vec{\Psi}_{\bfr,t'}}\no\\
&&\qquad \times \mbox{e}^{-\frac{\pi\nu}{4\tau}\int_{\bfr,t,t'} \; {\rm tr}[\hat{Q}_{t,t'}(\bfr)\hat{Q}_{t',t}(\bfr)]}.\;
\ee
The matrix $\hat{Q}$ is Hermitian (note that the transposition involves the interchange of the time arguments).

To summarize, the Keldysh partition function has been presented in the form
\be
\mathcal{Z}=\int D[Q]D[\vec{\Psi}^\dagger,\vec{\Psi}]D[\vec{\theta}]\exp(iS[\vec{\Psi}^\dagger,\vec{\Psi},\vec{\theta},\hat{Q}]),
\ee
where
\be
&&S[\vec{\Psi}^\dagger,\vec{\Psi},\vec{\theta},Q]\label{eq:fermaction}\\
&=&\int_{x,x'} \;\vec{\Psi}^\dagger_{x}\left[\hat{G}_{0}^{-1}(x-x')+\delta_{\bfr-\bfr'}\frac{i}{2\tau} \hat{Q}_{t,t'}(\bfr)\right]\vec{\Psi}_{x'}\no\\
&+&\int_x\; \vec{\Psi}^\dagger_{x}\hat{\theta}^l(x)\sigma^l\vec{\Psi}_{x}+\frac{i\pi\nu}{4\tau}\int_{\bfr,t,t'} {\rm tr}[\hat{Q}_{t,t'}(\bfr)\hat{Q}_{t',t}(\bfr)]\no\\
&+&\int_{\bfr,\bfr',t}  \;\vec{\theta}^T(\bfr,t)\hat{\gamma}_2V^{-1}(\bfr-\bfr') \vec{\theta}(\bfr',t).\no
\ee
After the averaging, the matrix Green's function $\hat{G}(x,x')=-i\langle \vec{\Psi}_x\vec{\Psi}^\dagger_{x'}\rangle$ (averaging is with respect to $\Psi$, $Q$ and $\theta$) acquires the typical triangular structure
\be
\hat{G}=\left(\ba{cc} G^R&G^K\\0&G^A\ea\right),
\ee
where $G^R$, $G^A$, and $G^K$ are the retarded, advanced and Keldysh components, respectively. Needless to say, the free Green's function $\hat{G}_0$ has the same structure.

At this point it is convenient to introduce the auxiliary potentials $\varphi_{cl,q}^l(x)$ into the theory. To this end we replace
\be
\vec{\Psi}^\dagger\hat{\theta}^l\sigma^l\vec{\Psi}\rightarrow  \vec{\Psi}^\dagger\left(\hat{\theta}^l-\hat{\varphi}^l\right)\sigma^l\vec{\Psi}.\label{eq:sourceintr}
\ee
Here, $\varphi_{cl}(x)$ can be interpreted as a classical external potential, while $\varphi_{cl}^i$ ($i=1,2,3$) describes a magnetic coupling to the spin degrees of freedom. The corresponding quantum components do not have an immediate physical interpretation. They merely play the role of source fields used to generate correlation functions, see Sec.~\ref{sec:correlation functions}.

The main purpose of the manipulations presented in this section so far was to perform the disorder average and to cast the Keldysh partition function into a form that is convenient for further analysis. No approximations have been introduced. The resulting functional with action \eqref{eq:fermaction} is still very complicated. On the other hand, as is well known, perturbations of charge and spin relax diffusively at low temperatures. One may therefore seek to find a low-energy theory of the disordered system by integrating out the fast electronic degrees of freedom and focus on diffusion modes only. As described below, this eventually yields the low-energy field theory that describes the diffusion modes including effects of their re-scattering, the so-called NL$\sigma$M. For non-interacting electrons, the NL$\sigma$M was first introduced by Wegner.\cite{Wegner79}

In a system with time-reversal symmetry, the modes in the particle-particle channel (i.e. the Cooper channel) also have a diffusive form. Therefore, the two mentioned types of diffusion modes, known as Diffusons and Cooperons, should both be included in the effective description. Initially, the generalization of the sigma model description to the interacting electron liquid with the help of the replica approach concentrated on the charge and spin degrees of freedom.\cite{Finkelstein83} Subsequently, both the electron interaction in the Cooper channel and the Cooperon modes were also included into the RG analysis.\cite{Finkelstein84} Compared to the model presented in \eqref{eq:fermaction}, this generalization requires a further doubling of the size of vectors $\Psi$ and matrices $Q$ as to include the so-called time-reversal sector.\cite{Efetov99} For the sake of clarity, Cooperons and the interaction in the Cooper channel will be ignored in the present work. Physically, this corresponds to the effect of a weak perpendicular magnetic field. 

The next important step in the derivation of the NL$\sigma$M is to find a saddle point for the field $\hat{Q}$. In the presence of the e-e interaction, this is a highly nontrivial task. One possible route to deal with this problem is to use the saddle point of the non interacting theory (i.e., in the absence of $\theta$) as a first approximation, and then to analyze deviations with respect to this reference point. This is the strategy chosen by Finkel'stein\cite{Finkelstein83} in its original work and we also will follow this route here. (An alternative course was chosen in Ref. \onlinecite{Kamenev99}. There, a part of the effects of the electron interaction was accounted for by a modification of the equation determining the saddle point.)

Let us, therefore, write the saddle point equation for the matrix field $\hat{Q}$ in the absence of the e-e interaction
\be
\hat{Q}_{0;t,t'}(\bfr)=\frac{i}{\pi\nu}\left(\hat{G}_0^{-1}+\frac{i}{2\tau}\hat{Q}_0\right)^{-1}_{\bfr,\bfr,t,t'}.\label{eq:saddle}
\ee
In equilibrium, it can be solved by the ansatz $\hat{Q}_{0;t,t'}(\bfr)=\hat{\Lambda}_{t-t'}$, where
\be
\hat{\Lambda}_\eps =\left(\ba{cc}1&2\mathcal{F}_\eps\\0&-1\ea\right),
\ee
and $\mathcal{F}_\eps=\tanh(\eps/2T)$ is the fermionic equilibrium distribution function. It is sometimes important to remember that the saddle point $\hat{\Lambda}$ inherits the analytical structure of the Keldysh Green's function. In particular, the unities in the $11$ and $22$ components should be interpreted as retarded and advanced elements, i.e., slightly displaced in the time domain in accordance with the analytical properties of the Green's function. It is instructive to present $\hat{\Lambda}$ in the form $\hat{\Lambda}=\hat{u}\circ \hat{\sigma}_3\circ  \hat{u}$ (here, $\circ$ symbolizes a convolution), where
\be
\hat{u}_\eps=\hat{u}_\eps^{-1}\equiv\left(\ba{cc} 1&\mathcal{F}_\eps\\0&-1\ea\right). \label{eq:umatrix}
\ee
In order to discuss slow (in space and time) fluctuations around this saddle point, we parametrize the matrix $\hat{Q}$ as
\be
\hat{Q}=\hat{U}\circ\hat{\sigma}_3\circ \hat{\overline{U}},\label{eq:nonunderlineQ}
\ee
where $\hat{\overline{U}}=\hat{U}^{-1}$. We will also often use the matrix $\underline{\hat{Q}}$ defined as
\be
\underline{\hat{Q}}=\hat{u}\circ \hat{Q}\circ \hat{u}. \label{eq:underlineQ}
\ee
Recall in this connection that $\hat{\Lambda}=\underline{\hat\sigma}_3$. The so defined $\hat{Q}$ and $\underline{\hat{Q}}$ fulfill the constraint
\be
\hat{Q}\circ \hat{Q}=\hat{\underline{Q}}\circ \hat{\underline{Q}} =\hat{1}.\label{eq:constraint}
\ee

The frequency representation of the matrix $\hat{Q}$ is formed according to the prescription
\be
\hat{Q}_{\eps\eps'}(\bfr)&=&\int_{t,t'}\;\hat{Q}_{tt'}(\bfr)\;\mbox{e}^{i\eps t-i\eps't'}.
\ee
The matrices $\hat{\underline{Q}}$ and $\hat{U}$ transform as $\hat{Q}$ does, following the same prescription. Naturally, we will consider the Fourier transformed quantities $\hat{Q}_{\eps\eps'}$, $\hat{U}_{\eps\eps'}$, etc., as matrices in frequency space and write the parametrization presented in Eq.~\eqref{eq:nonunderlineQ} as $\hat{Q}=\hat{U}\hat{\sigma_3}\hat{\overline{U}}$, $\hat{U}\hat{\overline{U}}=\hat{1}$, so that $\hat{Q}^2=\hat{1}$.
When choosing this parametrization, we immediately restrict ourselves to the so-called "massless" manifold. Fluctuations that violate the constraint \eqref{eq:constraint} are massive and their dynamics is beyond our interest.\cite{Wegner79,Efetov80,Efetov99} The parametrization of Eq.~\eqref{eq:underlineQ} is very convenient for the RG procedure. For frequencies exceeding the temperature, matrices $\hat{u}$ are almost frequency-independent. One may therefore integrate out $\hat{U}$ until the moment when $\hat{u}$ introduces the information about temperature.

After integrating the fermionic fields $\Psi$, $\Psi^\dagger$, one can perform a gradient expansion in the slow fields $\hat{U}$ and $\hat{\overline{U}}$ and also expand in the fields $\vec{\theta}$ and sources $\vec{\varphi}$ (which are slowly varying by definition). The relevant steps have been described many times in the literature and we refer, e.g., to Refs.~\onlinecite{Efetov99} and \onlinecite{Kamenev11} for details. The result is the nonlinear sigma model in the form
\be
S&=&\frac{\pi\nu i}{4}\Tr\left[D(\nabla \hat{Q})^2+4i\left(\hat{\eps}+(\hat{\theta}^l-\hat{\varphi}^l)\sigma^l\right)\underline{\hat{Q}}\right]\no\\
&&+\int_{\bfr,\bfr',t}  \;\vec{\theta}^T(\bfr,t)\;\hat{\gamma}_{2}V^{-1}(\bfr-\bfr') \;\vec{\theta}(\bfr',t)\no\\
&&+2\nu\int_{x} \;(\vec{\theta}-\vec{\varphi})^T(x)\;\hat{\gamma}_2\;(\vec{\theta}-\vec{\varphi})(x), \label{eq:action}
\ee
where $\hat{\eps}$ acts trivially on a matrix in the frequency space as $\hat{\eps}\hat{Q}_{\eps\eps'}=\eps \hat{Q}_{\eps\eps'}$. Note that for non-interacting electrons (in contrast to the case of e-e interactions), owing to the trace operation $\Tr(\hat{\eps}\underline{\hat{Q}})=\Tr(\hat{\eps}\hat{Q})$, only the source term prevents one from removing the distribution function from the action. The last term in Eq.~\eqref{eq:action} arises as a result of integrating out fast electronic degrees of freedom with energies exceeding $1/\tau$. The interval of energies below $1/\tau$ down to temperature $T$ is dominated by diffusion modes, and it will be studied later in Secs.~\ref{sec:correlation functions} and \ref{sec:renormalization} on the basis of the NL$\sigma$M.

\subsection{NL$\sigma$M after Fermi liquid renormalizations}
\label{subsec:sources}
The last term in Eq.~\eqref{eq:action} allows us to obtain the Fermi liquid (FL) renormalizations in the NL$\sigma$M in a systematic 
way, including the renormalization of the source fields. A similar
treatment of the Fermi liquid corrections in the Keldysh formalism can be found, e.g., in Refs.~\onlinecite{Chtchelkatchev08} and \onlinecite{Zala01}. Upon integration in $\theta$, one finds the action of the Keldysh sigma model for interacting electrons in the form
\be
S=S'_0+S_{int}+S'_\varphi,\label{eq:Sgeneral}
\ee
where
\be
S'_0&=&\frac{\pi\nu i}{4}\int_\bfr \tr\left[D(\nabla \hat{Q})^2+4i\hat{\eps}\hat{Q}\right],\label{eq:Sp0}\\
S_{int}&=&-\frac{\pi^2\nu}{8}\int_{\bfr\bfr't} \tr[\hat{\gamma}_i\underline{\hat{Q}}_{tt}(\bfr)]\hat{\gamma}_2^{ij}\tilde{\Gamma}_\rho(\bfr-\bfr') \tr[\hat{\gamma}_j \underline{\hat{Q}}_{tt}(\bfr')]\no\\
&&-\frac{\pi^2\nu}{8}\int_{\bfr t} \tr[\hat{\gamma}_i\bfsigma \underline{\hat{Q}}_{tt}(\bfr)]\hat{\gamma}_2^{ij}\Gamma_\sigma\tr[\hat{\gamma}_j\bfsigma \underline{\hat{Q}}_{tt}(\bfr)],\label{eq:Sint}
\ee
and $S'_{\varphi}=S'_{\varphi Q}+S'_{\varphi\varphi}$ with
\be
S'_{\varphi Q}&=&\pi\nu \int_\bfr \;\tr\left[\hat{\varphi}^l_{FL}(\bfr)\sigma^l \underline{\hat{Q}}(\bfr)\right],\label{eq:SpphiQ}\\
S'_{\varphi\varphi}&=&2\nu\int_{\bfr t}\; {\vec{\varphi}}^{T}_{FL}(\bfr,t)\hat{\gamma}_2\vec{\varphi}(\bfr,t).\label{eq:Spphiphi}
\ee
Notation $S'$ indicates that the corresponding terms in the action are not yet written in the final form suitable for the RG analysis, and will be treated further in Sec.~\ref{subsec:modelforrg}.

As a result of the integration in $\vec{\theta}$, the source fields $\vec{\varphi}=(\varphi,\bfvarphi^T)^T$ acquire static vertex corrections describing the FL renormalizations and screening. Namely, we get for the singlet component
\be
\varphi_{FL}(\bfq,t)&=&\frac{\varphi(\bfq,t)}{1+F_0^\rho+2\nu V_0(\bfq)},\label{eq:phipstatic0}
\ee
and for the triplet components
\be
\bfvarphi^{i}_{FL}&=&\frac{\bfvarphi^i}{1+F_0^\sigma},\quad i=1,2,3.\label{eq:phipstatick}
\ee
Furthermore, the interaction amplitudes in the singlet and triplet channels, symbolized by $\tilde{\Gamma}_\rho$ and $\Gamma_\sigma$ respectively, acquire the desired form
\be
\tilde{\Gamma}_\rho(\bfq)&=&\frac{2\nu V_0(\bfq)+F_0^\rho}{1+(2\nu V_0(\bfq)+F_0^\rho)},\quad
\Gamma_\sigma=\frac{F_0^\sigma}{1+F_0^\sigma}.
\ee
For future purposes it will be convenient to decompose the interaction in the singlet channel into two parts.\cite{Finkelstein83,Finkelstein90} One of them is the statically screened Coulomb interaction $\Gamma_0(\bfq)$, while the other one is the short-range interaction $\Gamma_\rho$ which acts within the polarization operator along with $\Gamma_\sigma$,
\be
\tilde{\Gamma}_\rho(\bfq)&=&2\Gamma_0(\bfq)+\Gamma_\rho,\label{eq:tildeG}
\ee
where
\be
\Gamma_0(\bfq)=\frac{\nu}{(1+F_0^\rho)^2}\frac{1}{V^{-1}_0(\bfq)+\frac{\partial n}{\partial \mu}},\quad\Gamma_\rho=\frac{F_0^\rho}{1+F_0^\rho}.
\ee
We also obtained the FL renormalization for $\frac{\partial n}{\partial\mu}$, the quantity that determines the value of the polarization operator in the static limit
\be
\frac{\partial n}{\partial\mu}=\frac{2\nu}{1+F_0^\rho}.
\ee

This concludes the derivation of the Keldysh sigma model which, in principle, can be used as a starting point for the RG analysis of the disordered electron liquid. In the next section, we will nevertheless cast the NL$\sigma$M in an equivalent form that will turn out to be more suitable for the renormalization group analysis.

\subsection{NL$\sigma$M: Preparation for the RG-procedure}
\label{subsec:modelforrg}

As a preparation for the RG analysis, we will now present the model in a slightly modified form. We write the action as
\be
S=S_0+S_{int}+S_\varphi,
\ee
and comment on the individual terms next.

The second (i.e., the frequency) term in the expression for $S_0'$, Eq.~\eqref{eq:Sp0}, acquires logarithmic corrections at low temperatures in the presence of the electron interactions. In other words, not only $D$, but also the dynamics of the diffusion modes is modified in the course of the renormalization of the NL$\sigma$M. Following Refs.~\onlinecite{Finkelstein83,Finkelstein83a,Baranov99} we will introduce the parameter $z$ into the model in order to account for these changes.
As a result, $S_0$ takes the form
\be
S_0&=&\frac{\pi\nu i}{4}\int_\bfr \tr\left[D(\nabla \hat{Q})^2+4iz\hat{\eps}\hat{Q}\right].\label{eq:S0}
\ee

For technical reasons, it is convenient to rewrite the interaction term, Eq.~\eqref{eq:Sint}, in a different form. Instead of organizing the short-range part of the interaction amplitudes into the singlet and triplet channel amplitudes, $\Gamma_\rho$ and $\Gamma_\sigma$, we will pass to a representation that separates small-angle and large-angle scattering, described by $\Gamma_1$ and $\Gamma_2$, respectively. The RG-analysis takes a simpler form in this representation.\cite{Finkelstein83,Finkelstein90} To this end we rewrite the interaction terms with the help of the identity
\be
\Gamma_\sigma\vec{\bfsigma}_{\alpha\beta}\vec{\bfsigma}_{\gamma\delta}=
2\Gamma_\sigma\delta_{\alpha\delta}\delta_{\beta\gamma}-\Gamma_\sigma\delta_{\alpha\beta}\delta_{\gamma\delta},
\ee
where $\alpha$, $\beta$, $\gamma$ and $\delta$ are spin indices. The interaction amplitudes $\Gamma_1$ and $\Gamma_2$ are defined as
\be
\Gamma_1&=&\frac{1}{2}\left(\Gamma_\rho-\Gamma_\sigma\right),\quad \Gamma_2=-\Gamma_\sigma.
\ee
The amplitude $\Gamma_2$ describes large angle scattering, while $\Gamma_1$ describes small angle scattering. It is therefore convenient to define a new amplitude $\Gamma(\bfq)$, which comprises both $\Gamma_1$ and the screened Coulomb interaction $\Gamma_0(\bfq)$:
\be
\Gamma(\bfq)&=&\Gamma_0(\bfq)+\Gamma_1.
\ee
In terms of the new amplitudes one finds the relation $\tilde{\Gamma}_\rho(\bfq)=2\Gamma(\bfq)-\Gamma_2$, cf.~Eq.~\eqref{eq:tildeG}.
Note that in the limit of small $\bfq$, the effective amplitude in the $\rho$ channel can be expressed in terms of $\Gamma_1$ and $\Gamma_2$ as follows
\be
\tilde{\Gamma}_\rho(\bfq\rightarrow 0)=\frac{1}{1+F_0^\rho}+2\Gamma_1-\Gamma_2.\label{eq:gammrhodefinition}
\ee
Returning to the action, the interaction term can be (identically) rewritten as
\be
&&S_{int}=\no\\
&&-\frac{\pi^2 \nu}{4}\int_{\bfr\bfr't}\tr[\hat{\gamma}_i\underline{\hat{Q}_{\alpha\alpha;tt}}(\bfr)]\gamma_2^{ij}\Gamma(\bfr-\bfr')\tr[\hat{\gamma}_j\underline{\hat{Q}_{\beta\beta;tt}}(\bfr')]\no\\
&&+\frac{\pi^2 \nu}{4}\int_{\bfr t}\tr[\hat{\gamma}_i\underline{\hat{Q}_{\alpha\beta;tt}}(\bfr)]\gamma_2^{ij}\Gamma_2\tr [\hat{\gamma}_j\underline{\hat{Q}_{\beta\alpha;tt}}(\bfr)].\label{eq:Sintiden}
\ee

In order to obtain a more tractable form for the interaction part of the action, let us introduce a set of H-S fields: real $\phi_0(x)$, $\phi_1(x)$ and Hermitian $\phi_{2,\alpha\beta}(x)$, each with classical and quantum components, which we characterize by their correlations
\be
&&\langle \phi_0^i(x)\phi_0^j(x')\rangle=\frac{i}{2\nu}\Gamma_0(\bfr-\bfr')\delta(t-t')\gamma_2^{ij},\label{eq:phi0phi0}\\
&&\langle \phi^i_1(x)\phi_1^j(x')\rangle=\frac{i}{2\nu}\Gamma_1\delta(x-x')\gamma_2^{ij},\label{eq:phi1phi1}\\
&&\langle \phi^i_{2,\alpha\beta}(x)\phi_{2,\gamma\delta}^j(x')\rangle=-\frac{i}{2\nu}\Gamma_2\delta_{\alpha\delta}\delta_{\beta\gamma}\delta(x-x')\gamma_2^{ij}.\qquad\label{eq:phi2phi2}
\ee
This definition allows us to cast $S_{int}$ in a compact form
\be
S_{int}=\frac{i(\pi\nu)^2}{2}\sum_{n=0}^2\int_{\bfr\bfr'}\langle \tr[\hat{\phi}_n(\bfr)\underline{\hat{Q}}(\bfr)]\tr[\hat{\phi}_n(\bfr')\underline{\hat{Q}}(\bfr')]\rangle.\label{eq:Sintphi}
\ee
Here, the frequency representation of the fields $\phi_n$ has been introduced in the matrix form, $\hat{\phi}_{n;\eps\eps'}$, according to the convention:
\be
\hat{\phi}_{n;\eps\eps'}(\bfr)&=&\int_t\;\hat{\phi}_n(\bfr,t)\;\mbox{e}^{i(\eps-\eps')t}.
\ee
We will sometimes use the notation $\underline{\hat{\phi}}=\hat{u}\circ \hat{\phi}\circ \hat{u}$ in analogy to Eq.~\eqref{eq:underlineQ}, so that $\tr[\hat{\phi}_n(\bfr)\underline{\hat{Q}}(\bfr)]=\tr[\underline{\hat{\phi}_n}(\bfr)\hat{Q}(\bfr)]$.

We had to split the interaction in the singlet channel into $\phi_0$ and $\phi_1$, because for the calculation of the irreducible density-density correlation function (i.e., the polarization operator) one needs to consider the Coulomb and the short-range parts of the interaction separately. (Recall that the term irreducible in this context means that only those contributions should be considered, which cannot be separated into two disconnected parts by cutting a single Coulomb interaction line.) We encounter this problem considering the source terms associated with the singlet channel, see Eq.~\eqref{eq:phipstatic0}.
Source fields were introduced because they allow generating correlation functions by functional differentiation of the Keldysh partition function, for details see Sec.~\ref{sec:correlation functions} below. The potential related to the singlet channel, $\varphi$, can be used to obtain the density-density correlation function which, in turn, is related to electric conductivity through the Einstein relation, see Sec.~\ref{subsec:conductivity}.  It is important to note, however, that only the knowledge of the \emph{irreducible} density-density correlation function is required for that purpose [for a detailed discussion of this point we refer to Ref. \onlinecite{Pines66}]. For this reason we will not work with the source term $S'_{\varphi}$, but with a slightly modified one, $S_{\varphi}$, for which the dependence on $V_0(\bfq)$ is removed. Note that the triplet part is unaffected by this change.

Finally, we write $S_{\varphi}=S_{\varphi Q}+S_{\varphi\varphi}$, where
\be
S_{\varphi Q}&=&\pi\nu \int_\bfr \;\tr\left[\left(\gamma_\triangleleft^\rho \hat{\varphi}(\bfr)+\gamma_{\triangleleft}^\sigma\hat{\bfvarphi}(\bfr)\bfsigma \right)\underline{\hat{Q}}(\bfr)\right]\label{eq:SourcephiQ} \\
S_{\varphi\varphi}&=&2\nu\int_{\bfr t}\; {\vec{\varphi}}^{T}(\bfr,t)\hat{\gamma}_2\mbox{diag}(\gamma^\rho_\bullet,\gamma^\sigma_\bullet,
\gamma^\sigma_\bullet,\gamma^\sigma_\bullet)\vec{\varphi}(\bfr,t).\label{eq:Sphiphi}\qquad
\ee
Here, the constants $\gamma_\triangleleft^{\rho/\sigma}$ and $\gamma_\bullet^{\rho/\sigma}$ have been introduced. $\gamma_\triangleleft^{\rho/\sigma}$ characterize the (triangular) vertices and $\gamma_\bullet^{\rho/\sigma}$ the static part of the correlation function. By comparison with Eqs.~\eqref{eq:SpphiQ}-\eqref{eq:phipstatick} and keeping in mind the previous remarks one finds that the initial values for the renormalization procedure read
\be
\gamma_\triangleleft^\rho=\gamma_\bullet^\rho=\frac{1}{1+F_0^\rho},\quad \gamma_\triangleleft^\sigma=\gamma_\bullet^\sigma=\frac{1}{1+F_0^\sigma}.\label{eq:initial}
\ee
As one can see, $\gamma_\triangleleft^{\rho/\sigma}=\gamma_\bullet^{\rho/\sigma}$ initially coincide. It is a priori not obvious, however, whether this important relation remains true under renormalization, and this is why the different constants have been introduced.

To summarize, the nonlinear sigma model contains several parameters ("charges") that may in principle acquire logarithmic corrections at low temperatures, $D$, $z$, $\Gamma_1$ and $\Gamma_2$, $\gamma^{\rho/\sigma}_\triangleleft$ and $\gamma_\bullet^{\rho/\sigma}$. Let us state the initial values, which follow directly from the derivation presented in Sec.~\ref{sec:derivation}, namely
\be
D=v_F^2\tau/2, \quad z=1;
\ee
\be
\Gamma_1=\frac{1}{2}\left(\frac{F^\rho_0}{1+F^\rho_0}-\frac{F_0^\sigma}{1+F_0^\sigma}\right),\quad\Gamma_2
=-\frac{F_0^\sigma}{1+F_0^\sigma}\label{eq:GammaF},
\ee
and the values for $\gamma_\bullet^{\rho/\sigma}$, $\gamma_\triangleleft^{\rho/\sigma}$ are written in Eq.~\eqref{eq:initial}.

\section{Correlation functions}
\label{sec:correlation functions}
In this section we first recall how retarded correlation functions can be generated from the Keldysh partition function by taking derivatives with respect to the so-called quantum and classical components of suitably chosen source fields. Next, we discuss the general structure of the correlation functions for particle-number densities and spin densities in the diffusive regime. The conservation laws for the total number of particles and for spin impose important constraints on the structure of these correlation functions.

\subsection{Generalities}
\label{subsec:Generalities}
We are interested in the retarded correlation functions, which are defined as a commutator of operators:
\be
\chi^R_{oo}(x_1-x_2)=-i\theta(t_1-t_2)\left\langle \left[\hat{o}(x_1),\hat{o}(x_2)\right]\right\rangle_T.\label{eq:chidef}
\ee
In order to be in line with common notation, we use hats to denote operators in this section up to Eq.~\eqref{eq:chiRoo}. Afterwards, the hat symbol will again be reserved for Keldysh matrices only. In Eq.~\eqref{eq:chidef}, $\hat{o}$ can be either the operator of the density $\hat{n}$ or of a component of the spin density operator $\hat{\bfs}$,
\be
\hat{n}(x)&=&\sum_{\alpha\beta}\hat{\psi}_{\alpha}^\dagger(x)\sigma^0_{\alpha\beta}\hat{\psi}_{\beta}(x),\\
\hat{\bfs}(x)&=&\frac{1}{2}\sum_{\alpha\beta} \hat{\psi}^\dagger_\alpha(x)\bfsigma_{\alpha\beta}\hat{\psi}_\beta(x).
\ee
In Eq.~\eqref{eq:chidef}, $\theta(t-t')$ is the Heaviside function and thermal averaging is with respect to the grand canonical ensemble,
\be
\left\langle \dots\right\rangle_T=\tr\left[\hat{\rho}\dots\right],\quad  \hat{\rho}=\frac{\mbox{e}^{-\hat{K}/T}}{\tr(\mbox{e}^{-\hat{K}/T})},
\ee
where $\hat{K}=\hat{H}-\mu \hat{N}$, $\hat{H}$ is the Hamiltonian, and $\hat{N}$ the number operator. The field operators $\hat{\psi}$ and $\hat{\psi}^\dagger$ are written in the Heisenberg representation with respect to $\hat{K}$.

Using the time ordered product $T[\hat{o}(t_1)\hat{o}(t_2)]=\theta(t_1-t_2)\hat{o}(t_1)\hat{o}(t_2)+\theta(t_2-t_1)\hat{o}(t_2)\hat{o}(t_1)$ and anti-time ordered product $\tilde{T}[\hat{o}(t_2)\hat{o}(t_1)]=\theta(t_1-t_2)\hat{o}(t_2)\hat{o}(t_1)+\theta(t_2-t_1)\hat{o}(t_1)\hat{o}(t_2)$, one may present the correlation function as
\be
\chi^R_{oo}(x_1-x_2)&=&-\frac{i}{2}\Big\langle T[\hat{o}(x_1)\hat{o}(x_2)]-\tilde{T}[\hat{o}(x_2)\hat{o}(x_1)]\no\\
&&+\hat{o}(x_1)\hat{o}(x_2)-\hat{o}(x_2)\hat{o}(x_1)\Big\rangle_T.\label{eq:chiRoo}
\ee
In the Keldysh formalism, this expression can conveniently be represented with the help of the functional integral, namely
\be
\chi^R_{oo}(x_1-x_2)&=&-\frac{i}{2}\big\langle o_+(x_1) o_+(x_2)- o_-(x_1)o_-(x_2)\no\\
&&\quad+o_-(x_1)o_+(x_2)-o_-(x_2)o_+(x_1)\big\rangle\no\\
&=&-\frac{i}{2}\left\langle [o_++o_-](x_1)[o_+-o_-](x_2)\right\rangle,\quad
\ee
where $o_\pm$ are now the corresponding (bosonic) fields on forward and backward paths of the Keldysh contour and averaging is with respect to the action $S$ (compare Eq.~\eqref{eq:S+-}). Introducing the classical and quantum components of the densities $o$ as $o_{cl/q}=\frac{1}{2}(o_+\pm o_-)$, one may write the correlation function in the form
\be
\chi^R_{oo}(x_1-x_2)=-2i\left\langle o_{cl}(x_1)o_{q}(x_2)\right\rangle.
\ee

The source term that has been introduced into the action in Eq.~\eqref{eq:sourceintr} can be re-written as follows:
\be
S_{source}&=&-\vec{\Psi}^\dagger\hat{\varphi}^i\sigma^i\vec{\Psi}\\
&=&-2(\varphi_2 n_{cl}+\varphi_1 n_q+2\bfvarphi_2\bfs_{cl}+2\bfvarphi_1\bfs_q)\no
\ee
Therefore, the correlation functions for the density $n$ and the spin components $\bfs^i$ can conveniently be written as
\be
\chi^R_{nn}(x_1-x_2)&=&\frac{i}{2}\left.\frac{\delta^2 \mathcal{Z}}{\delta \varphi_2(x_1)\delta \varphi_1(x_2)}\right|_{\varphi_1=\varphi_2=0},\label{eq:chinngeneral}\\
\chi^R_{s^is^j}(x_1-x_2)&=&\frac{i}{8}\left.\frac{\delta^2 \mathcal{Z}}{\delta \bfvarphi^i_2(x_1)\delta \bfvarphi^j_1(x_2)}\right|_{\bfvarphi_1=\bfvarphi_2=0}.\label{eq:chissgeneral}
\ee
This is rather intuitive, as
\be
\langle n_{cl}(x)\rangle=\frac{i}{2}\frac{\delta Z}{\delta \varphi_2(x)}, \quad \langle {\bf{s}}^i_{cl}(x)\rangle=\frac{i}{4}\frac{\delta Z}{\delta\bfvarphi^i_2(x)}
\ee
are the average particle-number and spin densities in the presence of the external (classical) potentials and, hence, the correlation functions describe the corresponding responses.

A very important observation can be made directly from the definition of the correlation function, Eq.~\eqref{eq:chidef}. To this end, let us first define the Fourier transform of the retarded correlation functions as $\chi_{oo}^R(x_1-x_2)=\int_\bfq \chi_{oo}^R(\bfq,t_1-t_2)\exp(i\bfq(\bfr_1-\bfr_2))$. Since for any given time the operators of the total density and spin $\int_\bfr \hat{n}(x)$ and $\int_\bfr \hat{\bfs}(x)$ commute with $\hat{K}$, as the total number of particles and the total spin are conserved, the correlation function $\chi(\bfq,t_1-t_2)$ vanishes in the limit $\bfq\rightarrow 0$. (This fact imposes an important constraint on the RG flow of the various charges in the model.) When further introducing the Fourier transform with respect to time, $\chi_{oo}^R(\bfq,t_1-t_2)=\int_\omega \chi^R_{oo}(\bfq,\omega)\exp(-i\omega(t_1-t_2))$, it should be appreciated that the limits $\bfq\rightarrow 0$ and $\omega\rightarrow 0$ do not commute with each other. In particular, if the limit $\omega\rightarrow 0$ is taken first, the correlation functions do not vanish, but their values are related to the corresponding thermodynamic susceptibilities.

\subsection{Correlation functions from the sigma model}
\label{subsec:contractions}
We will now discuss the density-density and spin-spin correlation functions in the framework of the NL$\sigma$M. The discussion will be restricted to the so-called ladder approximation, i.e., to an approximation, for which no internal momentum and frequency integrations over diffusion modes are carried out. In fact, those integrations give rise to logarithmic corrections (arising from the interval $T<\eps<1/\tau$), which is the essence of the RG-scheme. The logarithmic corrections may be absorbed into the various charges of the model, while the form of the model is unchanged. The results for the correlation functions obtained in the ladder approximation are therefore applicable at different scales (or temperatures) once the appearing charges are replaced by their renormalized values. As already mentioned before, the conservation laws for the number of particles and the total spin impose certain constraints on the relation between different RG charges which must be obeyed at each step of the renormalization procedure. This observation serves~\cite{Finkelstein84,DiCastro04,Finkelstein10} as an important check for the correctness of the obtained RG equations.

In short, we now find the correlation functions for density and spin in the Gaussian approximation with respect to fluctuations, i.e., with respect to diffusion modes. We thereby assume that all non-Gaussian integrations that lead to RG-corrections have been already performed. As a preparation, let us start with the parametrization of the matrix $\hat{U}$. A convenient choice of the parametrization is
\be
\hat{U}=\mbox{e}^{-\hat{P}/2},\quad \hat{\overline{U}}=\mbox{e}^{\hat{P}/2},\label{eq:U}
\ee
with the additional constraint $\{\hat{P},\hat{\sigma}_3\}=0$ in order to avoid overcounting of the relevant degrees of freedom. The chosen parametrization is not the only possible one. In fact, it gives rise to a non-trivial Jacobian, which, however, does not become relevant for the one-loop calculation discussed in this manuscript. Other parametrizations exist; for an instructive discussion within the context of the Keldysh NL$\sigma$M we refer to Ref.~\onlinecite{Ivanov06}. Returning to the exponential parametrization, Eq.~\eqref{eq:U}, note that $\hat{Q}=\hat{\sigma}_3\mbox{exp}(\hat{P})$. Further, the matrices $\hat{P}$ can be written as
\be
\hat{P}_{\eps\eps'}(\bfr)=\left(\ba{cc} 0& d_{cl;\eps\eps'}(\bfr)\\d_{q;\eps\eps'}(\bfr)&0\ea\right),
\ee
where $d_{cl/q}$ are hermitian matrices both in the frequency domain and in spin space, $[d^{\alpha\beta}_{cl/q;\eps\eps'}]^*=d^{\beta\alpha}_{cl/q;\eps'\eps}$. Expanding $S_0+S_{int}$ (see Eqs.~\eqref{eq:S0} and \eqref{eq:Sintphi}) up to second order in the generators $\hat{P}$, one obtains
\be
&&S=-\frac{i\pi\nu}{4}\int\tr[D(\nabla \hat{P})^2-2iz\hat{\eps}\hat{\sigma}_3\hat{P}^2]\label{eq:Psq}\\
&&+\frac{i}{2}(\pi\nu)^2\sum_{n=0}^2\int_{\bfr\bfr'}\langle \tr[\underline{\hat{\phi}}_n(\bfr)\hat{\sigma}_3\hat{P}(\bfr)]\tr[\underline{\hat{\phi}}_n(\bfr')\hat{\sigma}_3\hat{P}(\bfr')]\rangle.\no
\ee
Recall that for the frequency representation of the fields $\phi_n$, the matrix form $\hat{\phi}_{n;\eps\eps'}$ has been introduced.

By inverting the corresponding quadratic form, i.e., in the Gaussian approximation, this action gives rise to certain correlations for the components of $\hat{P}$. The result is most easily obtained after separation into singlet and triplet channels. Defining
\be
d_{cl/q;\eps\eps'}^{l}=\frac{1}{2}\sum_{\alpha\beta}\sigma_{\beta\alpha}^ld^{\alpha\beta}_{cl/q;\eps\eps'},\qquad l=(0,1-3),
\ee
one obtains for the correlation functions describing diffusion of the particle-hole pairs in the singlet (indicated by $0$) and triplet (indicated by $i,j\in\{1,2,3\}$) channels, respectively:
\be
&&\left\langle d^{0}_{cl;\eps_1\eps_2}(\bfq) d^0_{q;\eps_3\eps_4}(-\bfq)\right\rangle=-\frac{1}{\pi\nu}\mathcal{D}(\bfq,\omega)\times\label{eq:dsinglet} \\
&&\left(\delta_{\eps_1,\eps_4}\delta_{\eps_2,\eps_3}-
\delta_{\omega,\eps_4-\eps_3}i\pi\Delta_{\eps_1\eps_2}
\tilde{\Gamma}_\rho(\bfq)\tilde{\mathcal{D}}_1(\bfq,\omega)\right),\no
\ee
and
\be
&&\left\langle d^{i}_{cl;\eps_1\eps_2}(\bfq) d^j_{q;\eps_3\eps_4}(-\bfq)\right\rangle=-\frac{1}{\pi\nu}\delta^{ij}\mathcal{D}(\bfq,\omega)\times\label{eq:dtriplet} \\
&&\Big(\delta_{\eps_1,\eps_4}\delta_{\eps_2,\eps_3}-\delta_{\omega,\eps_4-\eps_3}i\pi\Delta_{\eps_1\eps_2}\Gamma_\sigma \mathcal{D}_2(\bfq,\omega)\Big),\no
\ee
where $\omega=\eps_1-\eps_2$, $\Delta_{\eps,\eps'}=\mathcal{F}_\eps-\mathcal{F}_{\eps'}$ and $\delta_{\eps,\eps'}=2\pi \delta(\eps-\eps')$. Obviously, on the level of the Gaussian fluctuations, the singlet and triplet channels do not interfere with each other. Note that three types of diffusons have been introduced\cite{Finkelstein83,Finkelstein90} in the above correlation functions:
\be
\mathcal{D}(\bfq,\omega)&=&\frac{1}{D\bfq^2-iz\omega}\label{eq:Dzalone}
\ee
\be
\tilde{\mathcal{D}}_1(\bfq,\omega)&=&\frac{1}{D\bfq^2-i\tilde{z}_{1}\omega}\label{eq:Dzone}
\ee
\be
\mathcal{D}_2(\bfq,\omega)&=&\frac{1}{D\bfq^2-iz_2\omega}\label{eq:Dztwo},
\ee
where $\tilde{z}_1(\bfq)=z-2\Gamma(\bfq)+\Gamma_2=z-\tilde{\Gamma}_\rho(\bf q)$, and $z_2=z+\Gamma_2=z-\Gamma_\sigma$.
We will see soon that actually $\tilde{z}_1(\bfq)\approx 0$ and, therefore,  $\tilde{\mathcal{D}_1}$ does not depend on $\omega$.\cite{Finkelstein83,Baranov99}

Transforming back to the original representation in terms of spin projections, one finds
\be
&&\left\langle d^{\alpha\beta}_{cl;\eps_1\eps_2}(\bfq)d^{\gamma\delta }_{q;\eps_3\eps_4}(-\bfq)\right\rangle\label{eq:basic_contraction}\\
&&=-\frac{2}{\pi\nu}\left[\delta_{\alpha\delta}\delta_{\beta\gamma}\delta_{\eps_1,\eps_4}\delta_{\eps_2,\eps_3}\mathcal{D}(\bfq,\omega)\right.\no\\
&&+\delta_{\alpha\delta}\delta_{\beta\gamma}\delta_{\omega,\eps_4-\eps_3}i\pi\Delta_{\eps_1,\eps_2}\mathcal{D}(\bfq,\omega)\Gamma_2\mathcal{D}_{2}(\bfq,\omega)\no\\
&&\left.-\delta_{\alpha\beta}\delta_{\gamma\delta}\delta_{\omega,\eps_4-\eps_3} i\pi \Delta_{\eps_1,\eps_2}\mathcal{D}_{2}(\bfq,\omega)\Gamma(\bfq)\tilde{\mathcal{D}}_{1}(\bfq,\omega)\right].\no
\ee
In order to demonstrate the general structure of the correlation functions for conserved quantities, we will be interested in the irreducible correlation function in the singlet channel, $\hat{\bar{\chi}}_{nn}\equiv \hat{\chi}_{nn}|_{irr}$.  For that, the ladder which is irreducible with respect to the Coulomb interaction is required. It can be found simply by excluding $\Gamma_0(\bf q)$, so that the expression for the irreducible average
$\left\langle d^0_{cl;\eps_1\eps_2}(\bfq) d^0_{q;\eps_3\eps_4}(-\bfq)\right\rangle_{irr}$
coincides with the one stated in Eq.~\eqref{eq:dsinglet} up to the replacement $\tilde{\Gamma}(\bfq)\rightarrow  \Gamma_\rho$ and $\tilde{\mathcal{D}}_1\rightarrow \mathcal{D}_1$, where
\be
\mathcal{D}_1(\bfq,\omega)=\frac{1}{D\bfq^2-iz_1\omega}
\ee
with
\be
z_1=z-2\Gamma_1+\Gamma_2=z-\Gamma_\rho.
\ee

\begin{figure}
\includegraphics[width=8.5cm]{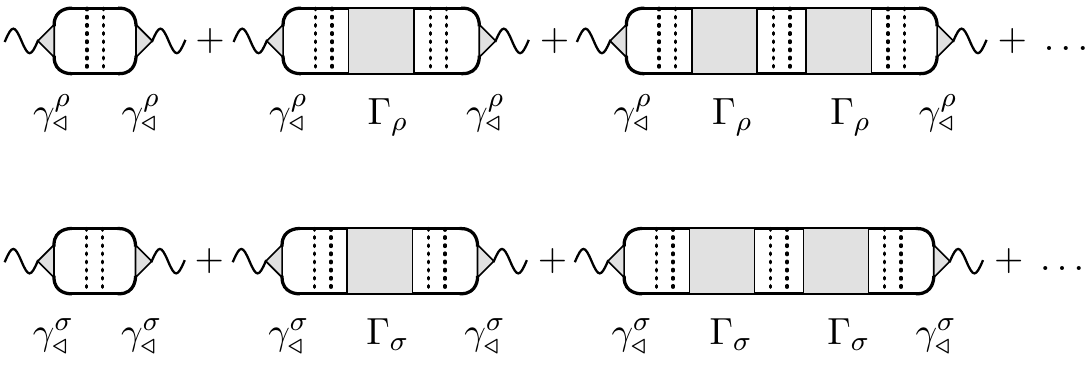}
\caption{Dynamical correlation functions $\bar{\chi}_{nn}^{\rm dyn,R}$ (top) and $\chi_{s_ks_k}^{\rm dyn,R}$ (bottom).}
\label{fig:Dynrhosigma}
\end{figure}

With this preparation, the correlation functions in the ladder approximation can be calculated. In view of Eqs.~\eqref{eq:chinngeneral} and \eqref{eq:chissgeneral}, we may integrate out the $\hat{P}$ modes and keep resulting terms only up to quadratic order in $\varphi$. Therefore, we calculate the dressed term
\be
S_{\varphi\varphi;d}=S_{\varphi\varphi}+\frac{i}{2}\left\langle\!\left\langle S_{\varphi Q}^2\right\rangle\!\right\rangle_{irr},\label{eq:Stild}
\ee
where for the second term both appearing matrices $\hat{Q}$ are replaced by $\hat{\sigma}_3\hat{P}$, and the averaging is with respect to the action \eqref{eq:Psq} for which the contraction rules obtained above can be used. In Eq.~\eqref{eq:Stild}, $\left\langle\!\left\langle \dots\right\rangle\!\right\rangle$ denotes the connected average. One may anticipate that $S_{\varphi\varphi;d}$ has the following form,
\be
S_{\varphi\varphi;d}=-\int_{xx'} \vec{\varphi}^T(x)\hat{X}(x-x') \vec{\varphi}(x'),
\ee
where
$\hat{X}=\mbox{diag}(\hat{\bar{\chi}}_{nn},4\hat{\chi}_{s^xs^x},4\hat{\chi}_{s^ys^y},4\hat{\chi}_{s^zs^z})$, and the $2\times 2$ blocks $\hat{\chi}_{oo}$ have a structure that is typical for correlation functions in the Keldysh formalism. Indeed,
\be
\hat{\chi}_{oo}=\left(\ba{cc} 0&\chi_{oo}^A\\\chi_{oo}^R&\chi_{oo}^K\ea\right)\label{eq:Kstructure},
\ee
where $\chi_{oo}^A(\omega)=\chi_{oo}^R(-\omega)$, and
\be
\chi_{oo}^K(\omega)=\mathcal{B}_\omega\left(\chi_{oo}^R(\omega)-\chi_{oo}^A(\omega)\right).
\ee
Furthermore, the two terms in Eq.~\eqref{eq:Stild} for $S_{\varphi\varphi;d}$ give rise to the static (st) and dynamical (dyn) parts of the correlation functions, respectively. As can directly be read off from Eq~\eqref{eq:Sphiphi}, the contribution from $S_{\varphi\varphi}$ is
\be
\bar{\chi}_{nn}^{{\rm st}, R}=-2\nu \gamma_\bullet^\rho,\quad \chi_{s^is^i}^{{\rm st}, R}=-2\nu\gamma_\bullet^\sigma\label{eq:static parts},
\ee
while for the dynamical part one finds
\be
&&\frac{i}{2}\left\langle\!\left\langle S_{\varphi Q}^2\right\rangle\!\right\rangle_{irr}\\
&=&\frac{i(\pi\nu)^2}{2}\left\langle\!\!\!\left\langle \left[\int_\bfr \tr\left[(\gamma_\triangleleft^\rho \underline{\hat{\varphi}}(\bfr)+\gamma_{\triangleleft}^\sigma\underline{\hat{\bfvarphi}}(\bfr)\bfsigma) \hat{\sigma}_3\hat{P}(\bfr)\right]\right]^2 \right\rangle\!\!\!\right\rangle_{irr}\no\\
&=&-\int_{xx'} \vec{\varphi}^T(x)\hat{X}^{\rm dyn}(x-x') \vec{\varphi}(x'),\no
\ee
where $\hat{X}^{\rm dyn}=\mbox{diag}(\hat{\chi}^{\rm dyn}_{nn},4\hat{\chi}^{\rm dyn}_{s^xs^x},4\hat{\chi}_{s^ys^y}^{\rm dyn},4\hat{\chi}_{s^zs^z}^{\rm dyn})$. The components of $\hat{\chi}^{\rm dyn}$ have again the structure indicated in Eq.~\eqref{eq:Kstructure}, and
\be
\bar{\chi}_{nn}^{{\rm dyn,R}}(\bfq,\omega)&=&-2\nu(\gamma^\triangleleft_{\rho})^2\;i\omega\mathcal{D}_1(\bfq,\omega),\quad\label{eq:chidyn1}\\
\chi_{s^is^i}^{{\rm dyn,R}}(\bfq,\omega)&=&-2\nu(\gamma^\triangleleft_{\sigma})^2\;i\omega\mathcal{D}_2(\bfq,\omega).\label{eq:chidyn2}
\ee
For a diagrammatic illustration see Fig.~\ref{fig:Dynrhosigma}.

In order to obtain this result, the following relation has been used
\be
&&1-\mathcal{F}_{\eps+\frac{\omega}{2}}\mathcal{F}_{\eps-\frac{\omega}{2}}=\mathcal{B}_{\omega}\left(\mathcal{F}_{\eps+\frac{\omega}{2}}-\mathcal{F}_{\eps-\frac{\omega}{2}}\right),
\ee
where
\be
\mathcal{B}_{\omega}=\coth\left(\frac{\omega}{2T}\right)
\ee
is the bosonic equilibrium correlation function. A second important identity is
\be
\pi \int_\eps\left(\mathcal{F}_{\eps+\frac{\omega}{2}}-\mathcal{F}_{\eps-\frac{\omega}{2}}\right)=\omega.\label{eq:windowint}
\ee
The total correlation function is then found by adding the static and the dynamical parts,
\be
\chi_{oo}^R(\bfq,\omega)=\chi_{oo}^{{\rm st},R}+\chi_{oo}^{{\rm dyn},R}(\bfq,\omega),
\ee
with the result
\be
\bar{\chi}^R_{nn}(\bfq,\omega)&=&-2\nu\gamma^\rho_\bullet\frac{D\bfq^2-i\omega \left(z_1-\frac{(\gamma_\triangleleft^\rho)^2}{\gamma_\bullet^\rho}\right)}{D\bfq^2-i z_1\omega},\label{eq:diffcorrelationfunctions}\\
\chi^{R}_{s^is^i}(\bfq,\omega)&=&-2\nu\gamma_\bullet^{\sigma}\frac{D\bfq^2-i\omega \left(z_2-\frac{(\gamma_\triangleleft^\sigma)^2}{\gamma_\bullet^\sigma}\right)}{D\bfq^2-iz_2\omega}.\label{eq:diffcorrelationfunctionsS}
\ee

As discussed in Sec.~\ref{subsec:Generalities}, conservation of charge and spin demands that
\be
\chi^R_{oo}(\bfq=0,\omega\rightarrow 0)=0.
\ee
In order to fulfill these conditions, the following relations must hold in view of Eqs.~\eqref{eq:diffcorrelationfunctions} and \eqref{eq:diffcorrelationfunctionsS},
\be
z_1=\frac{(\gamma_\triangleleft^\rho)^2}{\gamma_\bullet^\rho},\quad z_2=\frac{(\gamma_\triangleleft^\sigma)^2}{\gamma_\bullet^\sigma},\label{eq:twoconditions}
\ee
where the first relation is related to charge conservation\cite{Finkelstein83,Finkelstein90} and the second one to the conservation of spin.\cite{Finkelstein84,Finkelstein10} One may readily check that for the bare values of $z_1$, $\gamma_\bullet^{\rho/\sigma}$ and $\gamma_\triangleleft^{\rho/\sigma}$, these relations are fulfilled. Below, we will discuss the renormalization of the NL$\sigma$M for interacting electrons. In the RG scheme, the parameters $z$, $\Gamma_1$, $\Gamma_2$ (which determine $z_1$ and $z_2$) as well as $\gamma_\bullet^{\sigma}$ and $\gamma^{\sigma}_\triangleleft$ acquire logarithmic corrections and thereby become scale-dependent. It will be an important check of the theory that the two conditions displayed in Eq.~\eqref{eq:twoconditions} still hold {\it after} renormalization. Indeed, we will find that
\be
\gamma_\triangleleft^\rho=\gamma_\bullet^\rho=\frac{1}{1+F_0^\rho}\label{eq:gammatrirho}
\ee
are not renormalized, and that the relation $z_1=1/(1+F_0^\rho)$ holds under the RG flow. Therefore, the first relation in Eq.~\eqref{eq:twoconditions} is fulfilled. As a byproduct, it follows from these relations that $2\Gamma_0({\bf q})=1/(1+F_0^\rho)$ for small enough $q$ when $V_0^{-1}({\bf q}){\partial \mu}/{\partial n} \ll 1$. Therefore, $\tilde {z}_1({\bfq})=0$ in this limit and, hence, $\tilde{\mathcal{D}}_1(\bfq,\omega)=1/D{\bf q}^2$.

Further, we will find that
\be
\gamma_\triangleleft^\sigma=\gamma_\bullet^\sigma=z_2,\label{eq:gammatrisigma}
\ee
and the relation for the conservation of spin also holds, so that
\be
\bar{\chi}^R_{nn}(\bfq,\omega)&=&-\frac{\partial n}{\partial \mu}\frac{D\bfq^2}{D\bfq^2-i\frac{\omega}{1+F_0^\rho}}\label{eq:chibarfinaln}\\
\chi^R_{s^is^j}(\bfq,\omega)&=&-2\nu z_2\frac{D\bfq^2}{D\bfq^2-iz_2\omega}\delta^{ij}\label{eq:chibarfinals}.
\ee
The correlation functions $\bar{\chi}^R_{nn}(\bfq,\omega)$ and $\chi^R_{ss}(\bfq,\omega)$ have a universal
form, which is typical for diffusive correlation functions of the densities of a conserved quantity.
In a separate publication, we show that the same structure, compare Eqs.~\eqref{eq:chibarfinaln} and \eqref{eq:chibarfinals}, also holds for the heat density - heat density correlation function reflecting energy conservation.~\cite{Schwieteheat}

Finally, a comment is in order. The vanishing of the correlation function $\chi_{nn} (\bfq,\omega)$ in the limit $\bfq\rightarrow 0$ does not request it to be irreducible. However, the obtained universal form for the diffusive correlation functions will be lost for the reducible correlation function because of plasmons.
Recall that the irreducible correlation function $\bar{\chi}_{nn}$ is, in fact, the polarization operator.
Furthermore, we need to know only the irreducible function $\bar\chi_{nn} (\bfq,\omega)$ in order to extract the conductivity using the Einstein relation.~\cite{Finkelstein83,Finkelstein10}

\section{Renormalization}
\label{sec:renormalization}
The renormalization group approach for the problem at hand follows a general philosophy that is common to many problems in condensed matter physics. For the RG procedure, the fields in the action are separated into fast and slow modes. Subsequently, the fast modes are integrated out with logarithmic accuracy, leading to an effective action for the slow modes with scale-dependent parameters, i.e., RG charges.
A remark about the RG procedure in the Keldysh technique is in order: for any theory in which a quenched disorder average is performed, diagrams
that can be cut into separate parts by cutting only impurity lines should not appear. In the original model of Ref.~\onlinecite{Finkelstein83} the so-called replica trick was used in order to make sure that such contributions vanish. When using the Keldysh approach, the vanishing is effected in a somewhat different way. Generally speaking, the most important observation about the vanishing of unphysical terms in the Keldysh technique is that the frequency integral over a product of several retarded or advanced functions (but not a mixture of them) vanishes. This argument will frequently be used later on. The argument, however, does not carry over to the case when a single retarded or advanced function is connected to the rest by impurity lines only. This special case is discussed in connection with Fig.~\ref{fig:Sint2} in Sec.~\ref{subsec:Keldyshnotations}. [An alternative to the replica and Keldysh approaches exists, the so-called supersymmetry technique.\cite{Efetov99} It is a very powerful tool for noninteracting systems. Its application to interacting systems, however, is a formidable challenge, and progress in this direction is so far limited.\cite{Schwiete05}]

In order to lighten notations, starting from Sec.~\ref{subsec:Keldyshnotations} we will leave out the hats symbolizing matrices in Keldysh space.

\subsection{Generalities}
For the NL$\sigma$M the separation into fast and slow modes should be done in such a way that the nonlinear constraint $\hat{Q}^2=1$ is preserved\cite{Polyakov75}
\be
\hat{Q}=\hat{U}\hat{Q}_0\hat{\overline{U}},\quad \hat{Q}_0=\hat{U}_0\hat{\sigma}_3 \hat{\overline{U}}_0, \quad \hat{U}_0\hat{\overline{U}}_0=\hat{U}\hat{\overline{U}}=\hat{1}.\quad\label{eq:Qonion}
\ee
Here, $\hat{Q}_0$ contains the fast variables, $\hat{U}$ and $\hat{\overline{U}}$ represent the slow degrees of freedom. It is also convenient to introduce the slow field $\hat{Q}_s$ as
\be
\hat{Q}_s=\hat{U}\hat{\sigma}_3\hat{\overline{U}}.
\ee
When inserting $\hat{Q}$ in the form specified in Eq.~\eqref{eq:Qonion} into the action $S_0$, one obtains
\be
S_0&=&\frac{\pi \nu i}{4}\Tr\left[D(\nabla \hat{Q}_0)^2+D[\hat{Q}_0,\hat{\Phi}]^2\right.\no\\
&&\left.\qquad+2D\hat{\Phi}[\hat{Q}_0,\nabla \hat{Q}_0]+4iz\hat{\eps} \hat{U} \hat{Q}_0 \hat{\overline{U}}\right],
\ee
where $\hat{\Phi}=\hat{\overline{U}}\nabla \hat{U}=-\nabla \hat{\overline{U}} \hat{U}$. Using this notation, the interaction reads
\be
S_{int}=\frac{i(\pi\nu)^2}{2}\sum_{n=0}^2\left\langle \Tr\left[\underline{\hat{\phi}_n}\hat{U}\hat{Q}_0 \hat{\overline{U}}\right]\Tr\left[\underline{\hat{\phi}_n}\hat{U}\hat{Q}_0\hat{\overline{U}}\right]\right\rangle.\quad
\ee
For the RG-procedure, a particular parametrization for the fast degrees of freedom needs to be chosen. In accord with the previous Section, we will work with the exponential parametrization $\hat{U}_0=\exp(-\hat{P}/2)$, $\hat{Q}_0=\hat{\sigma}_3 \exp(\hat{P})$, $\{\hat{\sigma}_3, \hat P\}=0$. It turns out to be sufficient to expand up to second order in $\hat{P}$. We left out terms linear in $\hat{P}$. Such terms describe the decay (or fusion) of a fast mode into slow modes. These processes do not not play any role in the RG-analysis.
Then the result of the expansion reads
\be
S_0&=&\frac{\pi\nu i}{4}\;\Tr\left[D(\nabla \hat{Q}_s)^2+4iz\hat{\eps} \hat{Q}_s\right]\\
&+&\frac{\pi\nu i}{2} \;\Tr\left[ D(\hat{\sigma}_3 P\hat{\Phi})^2+D\hat{P}^2(\hat{\Phi}\hat{\sigma}_3)^2\right.\no\\
&&\left.\qquad\qquad +D\hat{\Phi}[\nabla \hat{P},\hat{P}]+iz \hat{\eps} \hat{U}\hat{\sigma}_3 \hat{P}^2 \hat{\overline{U}}\right]\no\\
&-&\frac{\pi\nu i}{4}\;\Tr\left[ D(\nabla \hat{P})^2\right],\no
\ee
\be
S_{int}&=&\frac{i}{2}(\pi\nu)^2\sum_{n=0}^2\left\langle \Tr\left[\underline{\hat{\phi}_n}\hat{Q}_s\right]\Tr\left[\underline{\hat{\phi}_n}\hat{Q}_s\right]\right\rangle\\
&+&\frac{i}{2}(\pi\nu)^2\sum_{n=0}^2\left\langle \Tr\left[\underline{\hat{\phi}_n} \hat{Q}_s\right]\tr\left[\underline{\hat{\phi}_n}\hat{U}\hat{\sigma}_3 \hat{P}^2\hat{\overline{U}}\right]\right\rangle\no\\
&+&\frac{i}{2}(\pi\nu)^2\sum_{n=0}^2\left\langle \Tr\left[\underline{\hat{\phi}_n}\hat{U}\hat{\sigma}_3 \hat{P}\hat{\overline{U}}\right]\left[\underline{\hat{\phi}_n}\hat{U}\hat{\sigma}_3\hat{P}\hat{\overline{U}}\right]\right\rangle.\no
\ee

So far, the separation into fast and slow degrees was purely formal. Let us now qualify this distinction:
\begin{enumerate}

\item Frequencies in the interval $\lambda \tau^{-1}<|\eps|<\tau^{-1}$, $0<\lambda<1$ and momenta in the shell $\lambda\tau^{-1}<Dk^2/z<\tau^{-1}$ are referred to as fast.

\item If at least one of the frequencies $\eps$ or $\eps'$ for the slow field $\hat{U}_{\eps\eps'}$ is fast, it has to be set equal to the unit matrix.

\item In the fast variables $\hat{P}_{\eps\eps'}$ at least one of the frequencies $\eps$, $\eps'$ or the momentum should be fast.

\end{enumerate}
For the frequency term in the action, one should explicitly distinguish fast and slow frequencies, i.e., $\hat{\eps}_f$ and $\hat{\eps}_s$. Then
\be
\Tr\left[z\hat{\eps} \hat{U}\hat{\sigma}_3\hat{P}^2\hat{\overline{U}}\right]=\Tr\left[z\hat{\eps}_s\hat{U}\hat{\sigma}_3 \hat{P}^2\hat{\overline{U}}\right]+\Tr\left[z \hat{\eps}_f\hat{\sigma}_3 \hat{P}^2\right].\no\\
\ee
We will now present a list of all the terms that are relevant for the one-loop RG-analysis. The following terms contain only slow modes
\be
S_{D}&=&\frac{i\pi\nu D}{4}\Tr\left[(\nabla \hat{Q}_s)^2\right]\label{eq:SD}\\
S_{z}&=&-\pi\nu z \Tr\left[\hat{\eps}_s \hat{Q}_s\right]\label{eq:Sz}\\
S_{\Gamma}&=&\frac{i}{2}(\pi\nu)^2\left\langle \Tr\left[\underline{\hat{\phi}_n}\hat{Q}_s\right]\Tr\left[\underline{\hat{\phi}_n}\hat{Q}_s\right]\right\rangle\label{eq:SGamma}\\
S_{\gamma_\triangleleft}&=&{\pi\nu}\Tr\left[\left(\gamma_\triangleleft^\rho \underline{\hat{\varphi}}+\gamma_{\triangleleft}^\sigma \underline{\hat{\bfvarphi}}\bfsigma\right)\hat{Q}_s\right]\label{eq:Sgammatriangle}\\
S_{\gamma_\bullet}&=&2\nu\int_x\vec{\varphi}^T(x)\hat{\gamma}_2\mbox{diag}(\gamma_\bullet^\rho,\gamma_\bullet^\sigma,\gamma_\bullet^\sigma,\gamma_\bullet^\sigma)\vec{\varphi}(x)\label{eq:Sgammabullet}.
\ee
Terms $S_{\gamma_\triangleleft}$ and $S_{\gamma_\bullet}$ arise from the source term $S_{\varphi}$. In fact, $S_{\gamma_\bullet}$ is identical to $S_{\varphi\varphi}$; the present notation is used to emphasize the dependence on the parameters $\gamma^{\rho/\sigma}_\bullet$.

Next, we come to the terms containing fast modes. The terms originating from $S_0$ read
\be
S_{f,0}&=&-\frac{i\pi\nu}{4}\Tr\left[D(\nabla \hat{P})^2-2iz\hat{\eps}_f \hat{\sigma}_3 \hat{P}^2\right] \\
S_1&=&-\frac{\pi \nu i}{2}\Tr\left[ D\hat{\Phi}[\hat{P},\nabla \hat{P}]\right]\\
S_2&=&\frac{\pi\nu i}{2} \Tr\left[D\hat{P}^2(\hat{\Phi}\hat{\sigma}_3)^2+D(\hat{\sigma}_3 \hat{P}\hat{\Phi})^2\right]\\
S_{\eps}&=&-\frac{\pi\nu }{2} \Tr\left[z\hat{\eps}_s\hat{U}\hat{\sigma}_3 \hat{P}^2 \hat{\overline{U}}\right].
\ee
Here, $S_2$ has two parts, which we label as $S_{2a}$ and $S_{2b}$ in the order of appearance.

The interaction part of the action $S_{int}$ gives rise to the following terms
\be
S_{int,1}&=&\frac{i}{2}(\pi\nu)^2 \sum_{n=0}^2\left\langle \Tr\left[\underline{\hat{\phi}_n}\hat{U}\hat{\sigma}_3\hat{P}\hat{\overline{U}}\right]\Tr\left[\underline{\hat{\phi}_n}\hat{U}\hat{\sigma}_3 \hat{P}\hat{\overline{U}}\right]\right\rangle\no\\
S_{int,2}&=&\frac{i}{2}(\pi\nu)^2\sum_{n=0}^2\left\langle \Tr\left[\underline{\hat{\phi}_n} Q_s\right]\Tr\left[\underline{\hat{\phi}_n}\hat{U}\hat{\sigma}_3 \hat{P}^2\hat{\overline{U}}\right]\right\rangle.\no\\
&&
\ee
Note that the labeling of these two terms refers to their different structure with respect to $\hat{P}$, and is not related to the fields $\phi_1$ and $\phi_2$.

Finally, the source term $S_{\varphi Q}$, see \eqref{eq:SourcephiQ}, generates a term
\be
S_{\varphi,2}=\frac{\pi\nu}{2}\Tr\left[\left(\gamma_\triangleleft^\rho \underline{\hat{\varphi}}+\gamma_{\triangleleft}^\sigma\underline{\hat{\bfvarphi}}\bfsigma \right)\hat{U}\sigma_3\hat{P}^2\hat{\overline{U}}\right],
\ee
where the labeling is chosen in analogy to $S_{int,2}$.

\begin{figure}
\includegraphics[height=7cm]{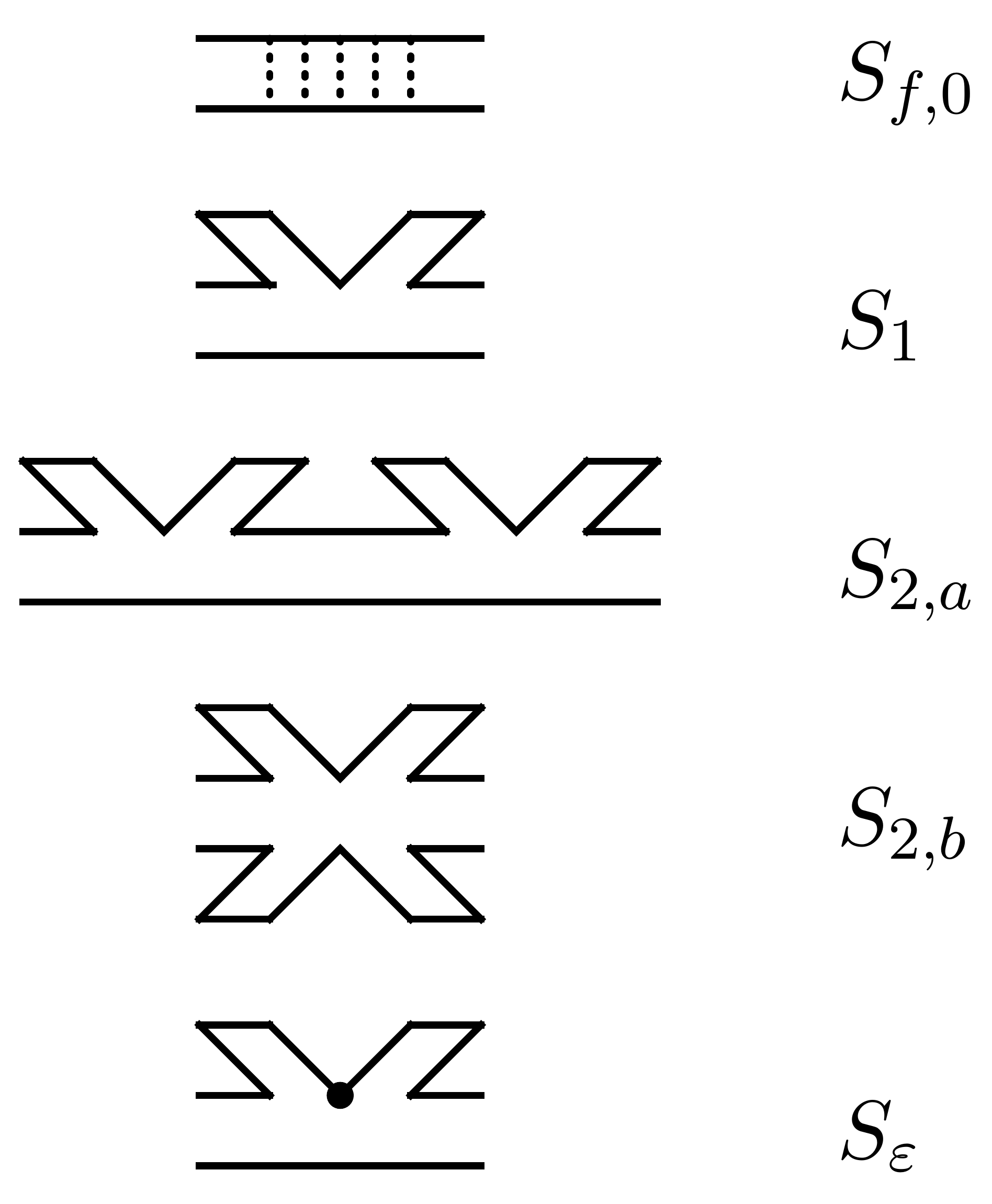}
\caption{The elements of the RG-procedure originating from the noninteracting part of the action. Open ends imply $P$. Closed sleeves correspond to $U$ or $\overline U$.
When separated by an angle, a gradient acts on one of them. A slow frequency $\eps_s$ stands in the vertex marked by a dot.
}
\label{fig:elementsS0}
\end{figure}

\begin{figure}
\includegraphics[height=7cm]{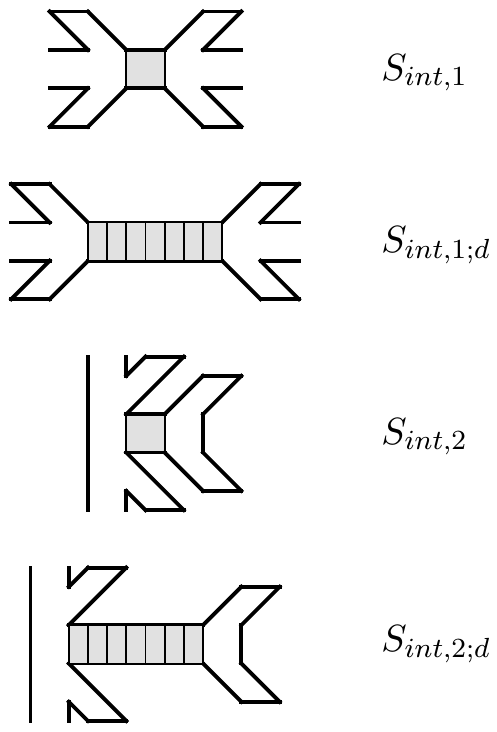}
\caption{The elements of the RG-procedure originating from the interaction part of the action. A shaded square implies one of the interaction amplitudes. A ladder means that the interaction was dressed by ladder diagrams. Such terms are indicated by the subscript "d".
}
\label{fig:elementsSint}
\end{figure}

\begin{figure}
\includegraphics[height=1.8cm]{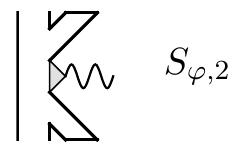}
\caption{Source term
}
\label{fig:elementsSphi}
\end{figure}

The terms containing fast modes are conveniently represented in a diagrammatic language as depicted in Figs.~\ref{fig:elementsS0}, \ref{fig:elementsSint} and \ref{fig:elementsSphi}.

We want to integrate out fast modes $\hat{P}$ in the Gaussian approximation, and in this way generate a new effective action. Besides the slow part of the action, compare Eqs.~\eqref{eq:SD} to \eqref{eq:Sgammabullet}, corrections arise from the term
\be
\Delta{S}=-i\ln\left[\int D[\hat{P}] \;\mbox{e}^{iS_1+iS_2+iS_{\eps} +iS_{int}+iS_{\varphi,2}}\;\mbox{e}^{iS_{f,0}}\right].\quad
\ee
In general, if there are  $N$ different parts in the action in which slow and fast modes couple to each other, one finds
\be
&&\Delta{S}=-i\ln\left(\int D[\hat{P}]\;\left(\mbox{e}^{i\sum_{i=1}^{N}S_i}\right)\mbox{e}^{iS_{f,0}}\right)\\
&=&\sum_{i=1}^N \left\langle S_i\right\rangle+\frac{i}{2}\sum_{ij=1}^N\left\langle \!\left\langle S_i S_j\right\rangle\!\right\rangle-\frac{1}{6}\sum_{ijk=1}^N\left\langle\!\left\langle S_iS_jS_k\right\rangle\!\right\rangle+\dots\no
\ee
Here, the connected average means that contractions between different terms must be taken as
\be
\left\langle\!\left\langle AB\right\rangle\!\right\rangle=\left\langle AB\right\rangle -\left\langle A\right\rangle \left\langle B\right\rangle,
\ee
and so on.

When integrating out fast modes, two cases should be distinguished. If \emph{at least} one of the frequencies of the $\hat{P}$-matrix is slow, then the contractions should be performed using $S_{f,0}$ alone. One can formulate two contraction rules for this case. Rule (i) applies when the two contracted $\hat{P}$s stand under different traces
\be
&&\left\langle \tr\left[\hat{A}\hat{P}_{\eps_1\eps_2}(\bfr_1)\right]\tr\left[\hat{B}\hat{P}_{\eps_3\eps_4}(\bfr_2)\right]\right\rangle\\
&=&-\frac{2}{\pi\nu}\tr\left[ \hat{A}^\perp\hat{\Pi}_{\eps_1\eps_2}(\bfr_1-\bfr_2)\hat{B}^\perp\right]\delta_{\eps_1,\eps_4}\delta_{\eps_2,\eps_3},\no
\ee
where we denote $\hat{A}^\perp=\frac{1}{2}(\hat{A}-\hat{\sigma}_3\hat{A}\hat{\sigma}_3)$, and
\be
\hat{\Pi}_{\eps+\frac{\omega}{2}\eps-\frac{\omega}{2}}(\bfq)=\left(\ba{cc}\mathcal{D}(\bfq,\omega)&0\\0&\overline{\mathcal{D}}(\bfq,\omega)\ea\right)
\ee
contains a retarded diffuson $\mathcal{D}$ and an advanced one, $\overline{\mathcal{D}}(\omega)=\mathcal{D}(-\omega)$. A second contraction rule (ii) applies when two contracted $\hat {P}$s appear within one trace. It reads as follows
\be
&&\left\langle \tr\left[AP_{\eps_1\eps_2}(\bfr_1)BP_{\eps_3\eps_4}(\bfr_2)\right]\right\rangle\\
&=&-\frac{1}{\pi\nu}\left(\tr[A\hat{\Pi}_{\eps_1\eps_2}(\bfr_1-\bfr_2)]\tr[B]\right.\no\\
&&\left.\qquad-\tr[A\hat{\sigma}_3 \hat{\Pi}_{\eps_1\eps_2}(\bfr_1-\bfr_2)]\tr[B\hat{\sigma}_3]  \right)\delta_{\eps_1\eps_4}\delta_{\eps_2,\eps_3}.\no
\ee

In the second case, when both frequencies of the $\hat{P}$ matrix are fast, the free Gaussian action of the fast modes besides $S_{f,0}$ also contains a part originating from $S_{int,1}$. In the case in question, it takes the form $S_{int,1}\rightarrow S_{f,int}$, where
\be
S_{f,int}=\frac{i}{2}(\pi\nu)^2 \sum_{n=0}^2\left\langle \Tr\left[\underline{\hat{\phi}_n}\hat{\sigma}_3\hat{P}\right]\Tr\left[\underline{\hat{\phi}_n}\hat{\sigma}_3 \hat{P}\right]\right\rangle.
\ee
Correspondingly, one should take the contraction with the full quadratic form
\be
S_f=S_{f,0}+S_{f,int}.
\ee
The relevant contraction rule for the components of $\hat{P}$ has already been stated in Eqs.~\eqref{eq:dsinglet}, \eqref{eq:dtriplet} and \eqref{eq:basic_contraction}. As is clear from the discussion presented in connection with these formulas in Sec.~\ref{subsec:contractions}, the extension of the quadratic form corresponds to "dressed" diffusons, which include not only impurity scattering but also a rescattering in the singlet and triplet channels as described by the ampitudes $\Gamma_\rho$ and $\Gamma_\sigma$. An example when this extension becomes important is the dressing of the interaction  which will be discussed next.

\subsection{Dressed interaction}
\begin{figure}[tb]
\includegraphics[width=8cm]{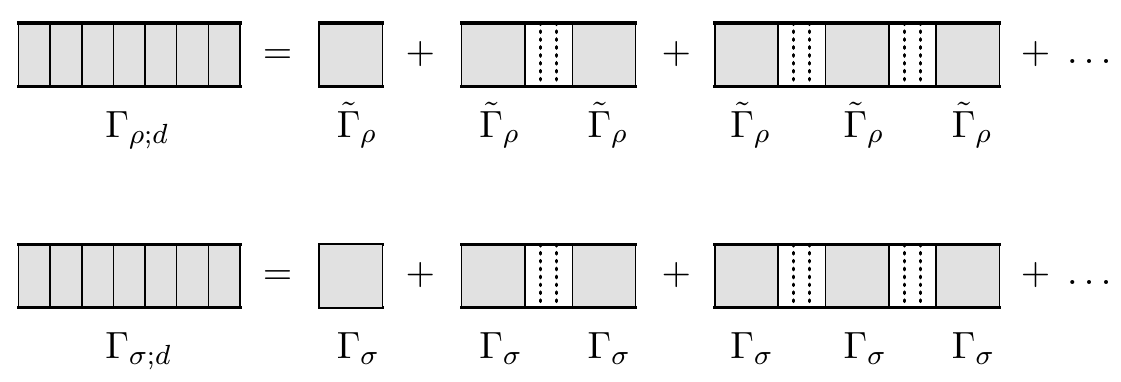}
\caption{Dressed interactions.
}
\label{fig:Dressedintc}
\end{figure}

Suppose that a certain average contains the interaction part of the action, $S_{int}$. Besides $S_{int}$, one may as well insert in its place the second cumulant $\frac{i}{2}\langle \!\langle S_{int}^2\rangle\!\rangle$, where for each of the interaction terms one $\hat{Q}_{\eps\eps'}$ will be replaced by the fast $\hat{\sigma}_3 \hat{P}_{\eps\eps'}$ with both frequencies fast, so that adjacent $\hat{U}$, $\hat{\overline{U}}$ should be substituted by $1$. The contraction of such fast $\hat{P}$s has to be taken with respect to $S_f$. This case may occur because the interaction fixes only the difference of frequencies $\eps-\eps'$ rather than the two frequencies individually. It means that $S_{int}$ should be replaced by its dressed (extended) counterpart
\be
S_{int;d}=S_{int} +\frac{i}{2}\left\langle\!\left\langle S_{int}^2\right\rangle\!\right\rangle,
\ee
where specifically
\be
\frac{i}{2}\left\langle\!\left\langle S_{int}^2\right\rangle\!\right\rangle=-\frac{i}{2}(\pi\nu)^2\left\langle\!\!\left\langle\langle \Tr[\underline{{\phi}_n}\hat{Q}]\Tr[\underline{{\phi}_n}\hat{\sigma}_3 \hat{P}]\rangle^2_\phi\right\rangle\!\!\right\rangle_{S_f}\;\;\;
\ee
and we indicated by the labels $\phi$ and $S_f$ which kind of average should be used.
For the calculation of this object a separation into singlet and triplet channel is useful,  in close analogy to the calculation of the correlation functions demonstrated before, see Fig.~\ref{fig:Dynrhosigma}. The calculation gives
\be
&&S_{int;d}\\
&&=-\frac{\pi^2\nu}{8}\int_{\bfr\bfr',\eps_i} \tr[\hat{\gamma}_i\sigma^0 \underline{\hat{Q}_{\eps_1\eps_2}}(\bfr)]\tr[\hat{\gamma}_j\sigma^0\underline{\hat{Q}_{\eps_3\eps_4}}(\bfr')]\no\\
&&\times \hat{\Gamma}^{ij}_{\rho;d}(\bfr-\bfr',\eps_1-\eps_2)\delta_{\eps_1-\eps_2,\eps_4-\eps_3}\no\\
&&-\frac{\pi^2\nu}{8}\int_{\bfr\bfr',\eps_i}\tr[\hat{\gamma}_i\bfsigma \underline{\hat{Q}_{\eps_1\eps_2}}(\bfr)]\tr[\hat{\gamma}_j\bfsigma\underline{\hat{Q}_{\eps_3\eps_4}}(\bfr')]\no\\
&&\times \hat{\Gamma}^{ij}_{\sigma;d}(\bfr-\bfr',\eps_1-\eps_2)\delta_{\eps_1-\eps_2,\eps_4-\eps_3}.\no
\ee
The dressed $(d)$ interaction can be obtained by the substitutions $\hat{\gamma}_2^{ij}\Gamma_\rho(\bfq)\rightarrow\hat{\Gamma}^{ij}_{\rho;d}(\bfq,\omega)$ and $\hat{\gamma}_2^{ij}\Gamma_\sigma(\bfq)\rightarrow\hat{\Gamma}^{ij}_{\sigma;d}(\bfq,\omega)$, where the interaction matrices $\hat{\Gamma}_{\rho/\sigma;d}^{ij}$ have a Keldysh space structure (compare with Eq.~\eqref{eq:Kstructure}). As a result, one gets
\be
\hat{\Gamma}_{\mu;d}(\bfq,\omega)&=&\left(\ba{cc} \Gamma_{\mu;d}^K(\bfq,\omega)&\Gamma_{\mu;d}^R(\bfq,\omega)\\\Gamma_{\mu;d}^A(\bfq,\omega)&0\ea\right),\; \mu=\{\rho,\sigma\},\no\label{eq:Gammahat}\\&&
\ee
where
\be \Gamma^A_{\mu;d}(\bfq,\omega)=\Gamma^R_{\mu;d}(\bfq,-\omega),\\
\Gamma^K_{\mu;d}(\bfq,\omega)=\mathcal{B}_\omega(\Gamma^R_{\mu;d}(\bfq,\omega)-\Gamma^A_{\mu;d}(\bfq,\omega)),
\ee
and
\be
\Gamma^R_{\rho;d}(\bfq,\omega)&=&\tilde{\Gamma}_{\rho}(\bfq)\Big(1-i\omega\tilde{\Gamma}_\rho(\bfq) \tilde{\mathcal{D}}_1(\bfq,\omega)\Big)\\
\Gamma^R_{\sigma;d}(\bfq,\omega)&=&\Gamma_{\sigma}\Big(1-i\omega \Gamma_\sigma\mathcal{D}_2(\bfq,\omega)\Big).
\ee
Obviously, the difference between the dressed and bare amplitudes is in the dynamic properties; in the static limit the amplitudes are equal. A diagrammatic illustration of dressing is shown in Fig.~\ref{fig:Dressedintc}.

Clearly, $\Gamma^R_{\rho/\sigma;d}$ describe rescattering in the singlet and triplet channels with intermediate sections composed of a pair of retarded and advanced Green's functions (sometimes referred to as $RA$ sections). Each $RA$ section gives rise to a window function $\Delta_{\eps+\omega/2,\eps-\omega/2}$ which, when integrated in $\eps$, produces a factor of $\omega$ (compare relation \eqref{eq:windowint}). This is why the coefficients of the frequencies of the diffusion modes $\tilde{\mathcal{D}}_1$ and $\mathcal{D}_2$ are modified by the interaction amplitudes, see Eqs~\eqref{eq:Dzalone}-\eqref{eq:Dztwo}. An important difference to the calculation of the correlation function is that in the present case the interaction may be reducible with respect to the Coulomb interaction, and $\tilde{\Gamma}_\rho({\bf q})$ and $\tilde{\mathcal{D}}_1$ appear in the singlet channel.

A somewhat simplified way to express the same result is
\be
\Gamma^R_{\rho;d}(\bfq,\omega)=\tilde{\Gamma}_\rho({\bf q})\frac{\tilde{\mathcal{D}}_1}{\mathcal{D}},\quad \Gamma^R_{\sigma;d}(\bfq,\omega)=\Gamma_\sigma\frac{\mathcal{D}_2}{\mathcal{D}}.
\ee
In order to obtain $\Gamma_d$ and $\Gamma_{2;d}$, one may use the relations $\Gamma^R_d=\frac{1}{2}\left(\Gamma^R_{\rho;d}-\Gamma^R_{\sigma;d}\right)$ and $\Gamma^R_{2;d}=-\Gamma^R_{\sigma;d}$ to find
\be
\Gamma^R_d(\bfq,\omega)=\Gamma({\bf q})\frac{\tilde{\mathcal{D}}_1\mathcal{D}_2}{\mathcal{D}^2},\quad \Gamma^R_{2;d}(\bfq,\omega)=\Gamma_2\frac{\mathcal{D}_2}{\mathcal{D}}.\label{eq:dress}
\ee
Needless to say, $\Gamma_d^R$ and $\Gamma_{2;d}^R$ are components of interaction matrices $\hat{\Gamma}_d$ and $\hat{\Gamma}_{2;d}$ with a structure as indicated in Eq.~\eqref{eq:Gammahat}.

If a model for a disordered Fermi liquid with short range interactions is considered, one may use the replacement $\Gamma({\bf q})\rightarrow \Gamma_1$, $\tilde{\mathcal{D}}_1\rightarrow \mathcal{D}_1$ in the final expressions. For the Coulomb case, it is useful to single out the screened Coulomb interaction explicitly. To this end, one may use the identity
\be
\Gamma^R_{d}=\Gamma({\bf q})\frac{\tilde{\mathcal{D}}_1\mathcal{D}_2}{\mathcal{D}^2}=\Gamma_0({\bf q})\frac{\tilde{\mathcal{D}}_1\mathcal{D}_1}{\mathcal{D}^2}+
\Gamma_1\frac{\mathcal{D}_1\mathcal{D}_2}{\mathcal{D}^2}.
\ee
After defining
\be
\tilde{\Gamma}_{0;d}^R=\Gamma_0({\bf q})\frac{\tilde{\mathcal{D}}_1}{\mathcal{D}_1},
\ee
one may single out the Coulomb interaction
\be
\Gamma^R_{d}=\Gamma^R_{0;d}+\Gamma^R_{1;d},
\ee
where
\be
\Gamma^R_{0;d}=\tilde{\Gamma}_{0;d}\frac{\mathcal{D}_1^2}{\mathcal{D}^2_0},\quad
\Gamma^R_{1;d}=\Gamma_1\frac{\mathcal{D}_1\mathcal{D}_2}{\mathcal{D}_0^2}\label{eq:Gamma01}.
\ee
Note that the entire dependence on the Coulomb interaction is delegated to $\tilde{\Gamma}_{0;d}$. Furthermore, with the use of the identities $\partial_\mu n=2\nu/(1+F_0^\rho)$ as well as $z_1=1/(1+F_0^\rho)$, one can obtain $\tilde{\Gamma}_{0;d}$ in the form
\be
\tilde{\Gamma}^R_{0;d}(\bfq,\omega)=\frac{\nu}{(1+F_0^\rho)^2}\left[V^{-1}_0(\bfq)+\frac{\partial n}{\partial \mu}\frac{D\bfq^2}{D\bfq^2-iz_1\omega}\right]^{-1}.
\ee
We observe that $\tilde{\Gamma}^R_{0;d}$ is the dynamically screened Coulomb interaction. Following this decomposition of the dressed interaction, we elevate relations \eqref{eq:phi0phi0}, \eqref{eq:phi1phi1} and \eqref{eq:phi2phi2} to
\be
&&\langle \phi^i_0(x)\phi_0^j(x')\rangle=\frac{i}{2\nu}\hat{\Gamma}^{ij}_{0;d}(x-x'),\\
&&\langle \phi^i_1(x)\phi_1^j(x')\rangle=\frac{i}{2\nu}\hat{\Gamma}^{ij}_{1;d}(x-x'),\\
&&\langle \phi^i_{2,\alpha\beta}(x)\phi_{2,\gamma\delta}^j(x')\rangle=-\frac{i}{2\nu}\hat{\Gamma}^{ij}_{2;d}(x-x')\delta_{\alpha\delta}\delta_{\beta\gamma},\qquad
\ee
whenever the dressed interaction is used. We remind that $\Gamma_{0;d}$ and $\Gamma_{1;d}$ are defined in Eq.~\eqref{eq:Gamma01} and $\Gamma_{2;d}$ in Eq.~\eqref{eq:dress}. The appearing interaction matrices have the typical Keldysh structure, compare Eq.~\eqref{eq:Gammahat}.
When dressing is not needed (such as for external vertices defined below), the static limit may be taken and $\Gamma^R_{n;d}\rightarrow \Gamma_{n}$ for $n=0-2$.

\subsection{Renormalization of the diffusion coefficient}
\label{subsec:Keldyshnotations}

\begin{figure}[tb]
\includegraphics[height=6.5cm]{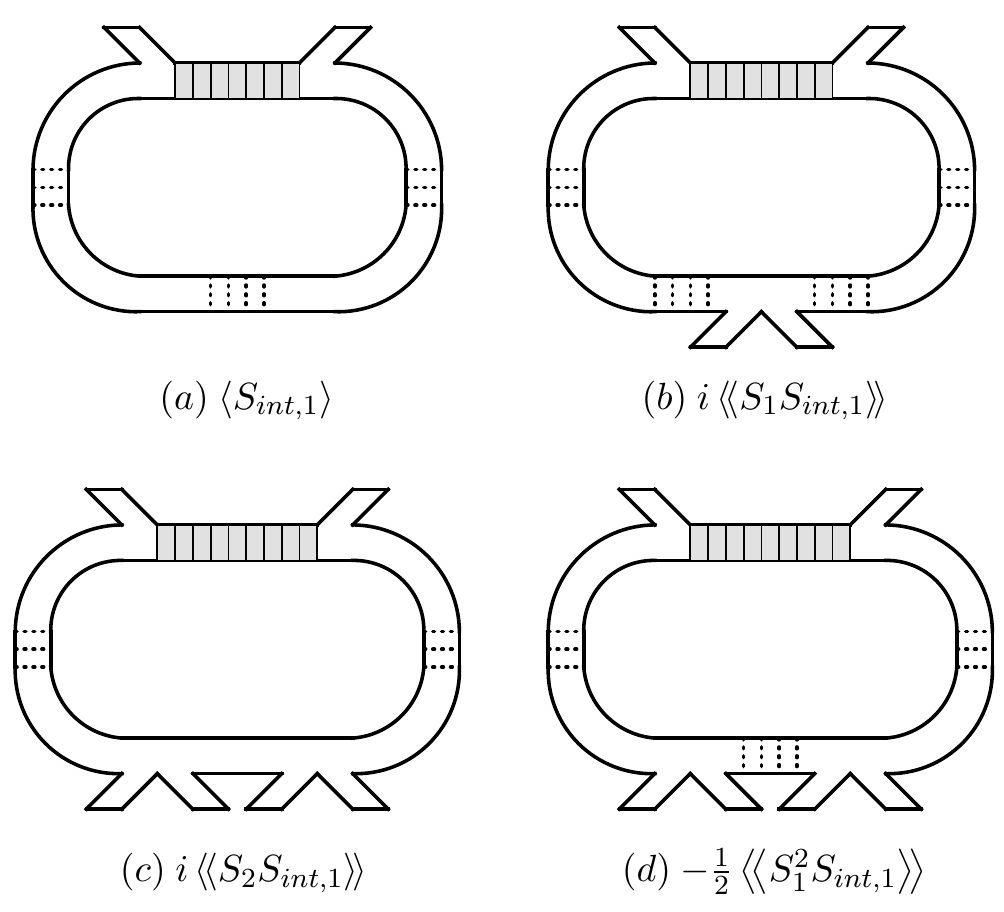}
\caption{The four different terms contributing to $\Delta{S_D}$. For the terms ($a$) and ($b$) a gradient expansion is needed.}
\label{fig:fourforSD}
\end{figure}

In this section, we discuss the renormalization of the diffusive term $S_D$ in the one loop approximation. This term contains two slow momenta (spatial gradients). It means that we can use $S_1$ at most twice or $S_2$ once. Additionally, gradients can be generated by Taylor expansion of the slow fields $U$, $\bar{U}$. As a result, one should consider
\be
\Delta{S_D}&=&\left\langle S_{int}\right\rangle+i\left\langle\!\left\langle S_1S_{int}\right\rangle\!\right\rangle+i\left\langle\!\left\langle S_2S_{int}\right\rangle\!\right\rangle\\
&&-\frac{1}{2}\left\langle\!\left\langle S_1^2S_{int}\right\rangle\!\right\rangle.\no
\ee
We will discuss these terms one by one and use the opportunity to highlight some aspects that are specific for the RG procedure in the Keldysh formalism. For a diagrammatic illustration of the four terms, see Fig.~\ref{fig:fourforSD}. Recall that for notational simplicity, we will from now on leave out hats for matrices in Keldysh space.

\subsubsection{$\left\langle S_{int}\right\rangle$}
\begin{figure}[tb]
\includegraphics[height=2cm]{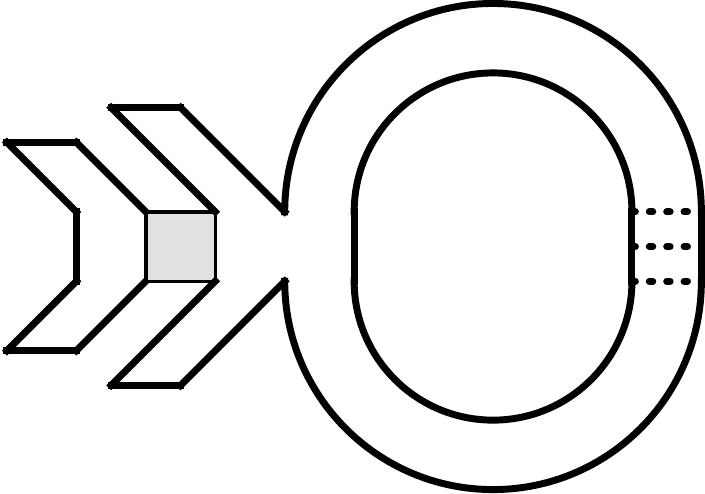}
\caption{Diagram for $\left\langle S_{int,2}\right\rangle$. This term vanishes as discussed in the text.
}
\label{fig:Sint2}
\end{figure}

$S_{int}$ consists of two parts, $S_{int,1}$ and $S_{int,2}$. First consider $\left\langle S_{int,2}\right\rangle$. The corresponding expression contains the following average
\be
\int_{\eps_2} \left\langle {P}_{\eps_1\eps_2}{P}_{\eps_2\eps_3}\right\rangle\propto \left(\ba{cc} \mathcal{D}(\omega)&0\\0&\overline{\mathcal{D}}(\omega)\ea\right)
\ee
The diagram for $\langle S_{int,2}\rangle$ is displayed in Fig.~\ref{fig:Sint2}. It is immediately obvious that this diagram can be cut into separate parts by cutting only impurity lines. As is well known, such diagrams should not appear for any theory in which a quenched disorder average is performed.
The so-called replica method was invented \cite{Edwards75} to eliminate such contributions. Indeed, the internal Green's function allows for a free summation over the replica index, and therefore the diagram vanishes in the zero-replica limit. In the Keldysh technique, the vanishing of unphysical terms mostly occurs because the frequency integral over a product of several retarded or advanced functions (but not a mixture of them) vanishes. This argument, however, does not carry over to the case of a single retarded or advanced function as is relevant for the discussed term. In this case one needs to argue that the contribution of the unphysical diagram to the calculation of any physical quantity will always contain the frequency integral of the \emph{sum} of one retarded and one advanced function, and it is simple to see that their sum vanishes. In the example at hand, the retarded and advanced diffuson appear as separate elements of the matrix $M_{\eps_1\eps_3}=\int_{\eps_2} \left\langle {P}_{\eps_1\eps_2}{P}_{\eps_2\eps_3}\right\rangle$.
Whenever physical quantities are calculated, \emph{all} modes have to be integrated out, which implies that \emph{eventually} the sum of retarded and advanced functions will appear. Anticipating this fact, diagrams as encountered for $\langle S_{int,2}\rangle$ may safely be dropped; Fig.~\ref{fig:Sint2} illustrates this important point.

\begin{figure}[tb]
\includegraphics[height=5cm]{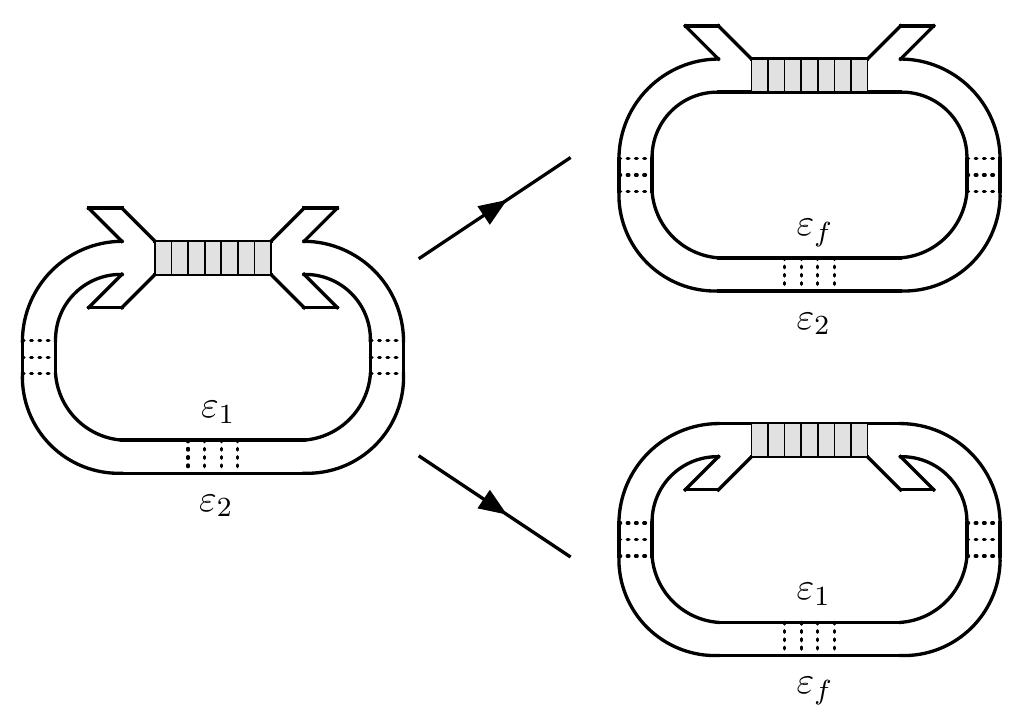}
\caption{This figure illustrates the two choices of $\eps_f$ for the average $\left\langle S_{int,1}\right\rangle$, where $\eps_f$ symbolizes the fast frequency. In fact, both choices are equivalent and in this way one comes from Eq.~\eqref{eq:Sint1average1} to Eq.~\eqref{eq:Sint1average2}.
}
\label{fig:choices}
\end{figure}

For the other term, $\langle S_{int,1}\rangle$, see Fig.~\ref{fig:fourforSD}(a), one finds
\be
&&\left\langle S_{int,1}\right\rangle\label{eq:Sint1average1}\\
&=& i\pi\nu  \sum_{n=0}^2 \int_{\bfr_1\bfr_2,\eps_1\eps_2}\left\langle\tr\left[\left(\bar{U}(\bfr_1)\underline{\phi_{n}}(\bfr_1)U(\bfr_1)\right)^\perp_{\eps_1\eps_2}\right.\right.\no\\
&&\left.\left.\left(\bar{U}(\bfr_2)\underline{\phi_n}(\bfr_2)U(\bfr_2)\right)^\perp_{\eps_2\eps_1}\Pi_{\eps_1\eps_2}(\bfr_1-\bfr_2)\right]\right\rangle.\no
\ee
Here and in the following we denote $M^\perp=(M-\sigma_3 M\sigma_3)/2$, and $M^\parallel=(M+\sigma_3 M\sigma_3)/2$, so that $M=M^\parallel+M^\perp$. $M^\parallel$ is the diagonal part of $M$ in Keldysh space, and $M^\perp$ the off-diagonal part; $[M^\parallel,\sigma_3]=0$, $\{M^\perp,\sigma_3\}=0$.
We will mostly work in such a way that contractions in ${P}$ are performed first, while the choice of fast and slow frequencies for the $P$-matrices is made a posteriori. This is a straightforward procedure since the frequency arguments of $P$ always reappear explicitly as arguments of the diffusion propagators $\Pi$.
For the renormalization of the diffusion coefficient, in the discussed contribution precisely one frequency argument of the $P$-matrices is fast. This might be either $\eps_1$ or $\eps_2$, see Fig.~\ref{fig:choices} for an illustration. Due to the identity $X^\perp \hat{\Pi}_{\eps_1\eps_2}=\hat{\Pi}_{\eps_2\eps_1} X^\perp$ for any matrix $X$ in Keldysh space both possibilities are equivalent. For definiteness, we choose here $\eps_2$ as fast and write $\eps_2=\eps_f$.
This leads to the intermediate result
\be
\left\langle S_{int,1}\right\rangle&=&2i\pi\nu \sum_{n=0}^2\int_{\bfr_1\bfr_2,\eps_1\eps_f}\left\langle \tr\left[(\bar{U}(\bfr_1)\underline{\phi_n}(\bfr_1))^\perp_{\eps_1\eps_f}\right.\right.\no\\
&&\left.\left.\times (\underline{\phi_n}(\bfr_2)U(\bfr_2))^\perp_{\eps_f\eps_1}\Pi_{\eps_1\eps_f}(\bfr_1-\bfr_2)\right]\right\rangle.\label{eq:Sint1average2}
\ee
We do not evaluate this term right now, but first proceed with the other terms.

\subsubsection{$i\left\langle \!\left\langle S_1 S_{int}\right\rangle\!\right\rangle$}
The relevant contribution comes from $S_{int,1}$ only. One finds
\be
&&i\left\langle \!\left\langle S_1 S_{int,1}\right\rangle\!\right\rangle=-2i\pi\nu \sum_{n=0}^2\int_{\bfr_i,\eps_i} D(\nabla_{\bfr_3''}-\nabla_{\bfr_3'})\\
&& \left\langle \tr\left[\left(\bar{U}(\bfr_1)\underline{\phi_n}(\bfr_1)U(\bfr_1)\right)^\perp_{\eps_1\eps_2}\left(\bar{U}(\bfr_2)\underline{\phi_n}(\bfr_2)U(\bfr_2)\right)^\perp_{\eps_2\eps_3}\right.\right.\no\\
&&\times \left.\left.\left.\Phi_{\eps_3\eps_1}^\parallel(\bfr_3)\hat{\Pi}_{\eps_1\eps_2}(\bfr_1-\bfr_3')\hat{\Pi}_{\eps_3\eps_2}(\bfr_3''-\bfr_2)\right]\right\rangle\right|_{\bfr_3''=\bfr_3'=\bfr_3},\no
\ee
where $\eps_1$ and $\eps_3$ are necessarily slow, because of $\Phi_{\eps_1\eps_3}$. Recall that $\Phi=\overline{U}\nabla U=-\nabla \overline{U} U$, and  it is clear that it can only have two slow indices or vanish. Therefore $\eps_2$ needs to be fast and
\be
&&i\left\langle \!\left\langle S_1 S_{int,1}\right\rangle\!\right\rangle=-2i \pi\nu \sum_{n=0}^2\int_{\bfr_i,\eps_i} D(\nabla_{\bfr_3''}-\nabla_{\bfr_3'})\label{eq:S1Sint1final}\\
&& \left\langle \tr\left[\left(\bar{U}(\bfr_1)\underline{\phi_n}(\bfr_1)\right)^\perp_{\eps_1\eps_f}\left(\underline{\phi_n}(\bfr_2)U(\bfr_2)\right)^\perp_{\eps_f\eps_3}\right.\right.\no\\
&&\times \left.\left.\left.\Phi_{\eps_3\eps_1}^\parallel(\bfr_3)\Pi_{\eps_1\eps_f}(\bfr_1-\bfr_3')\Pi_{\eps_3\eps_f}(\bfr_3''-\bfr_2)\right]\right\rangle\right|_{\bfr_3''=\bfr_3'=\bfr_3}.\no
\ee
Fig.~\ref{fig:fourforSD}(b) illustrates the structure of this term. One may already notice the structural similarity to Eq.~\eqref{eq:Sint1average2}; the same observation also holds for the remaining contributions to $\Delta S_{D}$. This is why the further evaluation is postponed until all four terms have been discussed.

\subsubsection{$i\left\langle\!\left\langle S_2S_{int}\right\rangle\!\right\rangle$}

$S_2$ contains two terms, $S_{2a}$ and $S_{2b}$. In $S_{2b}$, all frequencies of the $P$ matrices are forced to be slow due to the presence of $\Phi$ and this does not lead to an RG-contribution to the diffusion coefficient. The relevant contribution comes from a combination of $S_{2a}$ and $S_{int,1}$:
\be
&& i\left\langle\!\left\langle S_{2a}S_{int,1}\right\rangle\!\right\rangle=2i\pi\nu D\sum_{n=0}^2\int_{\bfr_i,\eps_i}\label{eq:S2aSint1final} \\
&&\left\langle\tr\left[\left(\bar{U}(\bfr_1) \underline{\phi_n}(\bfr_1)\right)^\perp_{\eps_1\eps_f}\left(\underline{\phi_n}(\bfr_2)U(\bfr_2)\right)^\perp_{\eps_f\eps_3}\right.\right.\no\\
&&\left.\left.\times (\Phi(\bfr_3)\Lambda\Phi(\bfr_3)\Lambda)_{\eps_3\eps_1}^\parallel\Pi_{\eps_1\eps_f}(\bfr_1-\bfr_3)\Pi_{\eps_3\eps_f}(\bfr_3-\bfr_2)\right]\right\rangle.\no
\ee
For an illustration of this contribution see Fig.~\ref{fig:fourforSD}(c). The expression will be evaluated further together with the other contributions to $\Delta S_D$.

\subsubsection{$-\frac{1}{2}\left\langle\!\left\langle S_1^2S_{int}\right\rangle\!\right\rangle$}

\begin{figure}[tb]
\includegraphics[height=2.5cm]{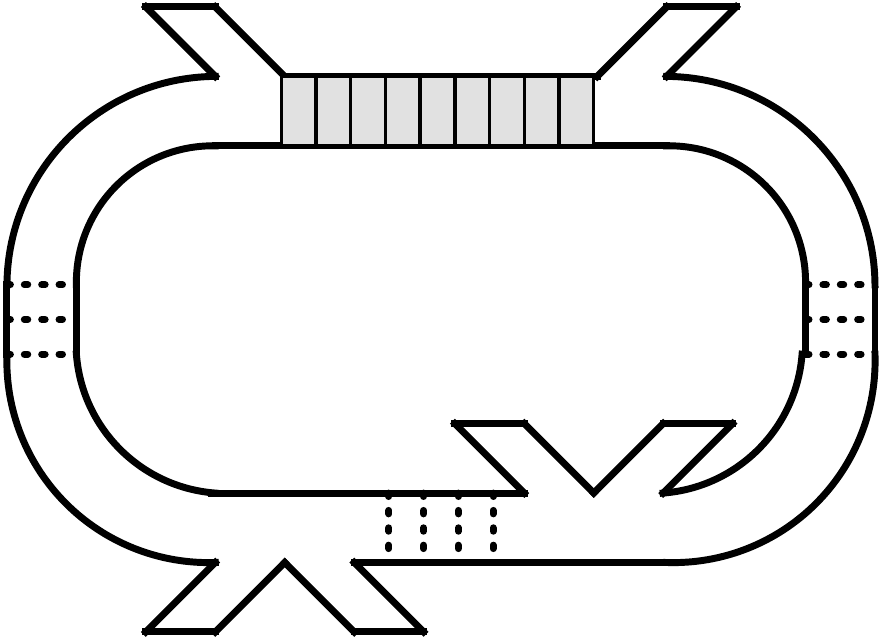}
\caption{Terms of the kind displayed in this figure arise when evaluating the average $-\frac{1}{2}\left\langle\!\left\langle S_1^2S_{int,1}\right\rangle\!\right\rangle$, compare the first and third terms in Eq.~\eqref{eq:S1S1Sint1}. All frequencies involved are bound to be small. This makes the contributions of this type irrelevant.
}
\label{fig:S1S1Sintsmall}
\end{figure}

Similarly to the previously discussed terms, the dominant contribution comes from $S_{int,1}$.
The contractions can be performed in several ways, as indicated below
\begin{widetext}

\be
-\frac{1}{2}\left\langle\!\left\langle S_1^2S_{int,1}\right\rangle\!\right\rangle\label{eq:S1S1Sint1}
&=&\frac{i(\pi\nu)^4}{16} D^2 \sum_{n=0}^2\Big(2\Tr[\Phi
\bcontraction{}{P}{\overset{\leftrightarrow}{\nabla} P]\Tr[\Phi}{P}
\contraction{P\overset{\leftrightarrow}{\nabla}}{P}{]\Tr[\Phi P\overset{\leftrightarrow}{\nabla} P]\langle\Tr[\underline{\phi_n}U\sigma_3}{P}
\bcontraction{P\overset{\leftrightarrow}{\nabla} P]\Tr[\Phi P\overset{\leftrightarrow}{\nabla} }{P}{]\langle\Tr[\underline{\phi_n}U\sigma_3P\bar{U}]\Tr[\underline{\phi_n} U\sigma_3}{P}
P\overset{\leftrightarrow}{\nabla} P]\Tr[\Phi P\overset{\leftrightarrow}{\nabla} P]\langle\Tr[\underline{\phi_n}U\sigma_3P\bar{U}]\Tr[\underline{\phi_n} U\sigma_3 P\bar{U}]\rangle\\
&&\qquad\qquad\qquad+4\Tr[\Phi
\bcontraction{}{P}{\overset{\leftrightarrow}{\nabla} P]\Tr[\Phi P\overset{\leftrightarrow}{\nabla} }{P}
\contraction{P\overset{\leftrightarrow}{\nabla} }{P}{]\Tr[\Phi P\overset{\leftrightarrow}{\nabla} P]\langle\Tr[\underline{\phi_n}U\sigma_3}{P}
\bcontraction[1.5ex]{P\overset{\leftrightarrow}{\nabla} P]\Tr[\Phi}{P}{\overset{\leftrightarrow}{\nabla}P]\langle\Tr[\underline{\phi_n}U\sigma_3P\bar{U}]\Tr[\underline{\phi_n} U\sigma_3 }{P}
P\overset{\leftrightarrow}{\nabla} P]\Tr[\Phi P\overset{\leftrightarrow}{\nabla} P]\langle\Tr[\underline{\phi_n}U\sigma_3P\bar{U}]\Tr[\underline{\phi_n} U\sigma_3 P\bar{U}]\rangle\no\\
&&\qquad\qquad\qquad+2\Tr[\Phi
\bcontraction{}{P}{\overset{\leftrightarrow}{\nabla} P]\Tr[\Phi P\overset{\leftrightarrow}{\nabla} P]\langle\Tr[\underline{\phi_n}U\sigma_3}{P}
\bcontraction[1.5ex]{P\overset{\leftrightarrow}{\nabla}}{P} {]\Tr[\Phi P\overset{\leftrightarrow}{\nabla}}{P}
\contraction{P\overset{\leftrightarrow}{\nabla} P]\Tr[\Phi }{P}{\overset{\leftrightarrow}{\nabla} P]\langle\Tr[\underline{\phi_n}U\sigma_3P\bar{U}]\Tr[\underline{\phi_n} U\sigma_3 }{P}
P\overset{\leftrightarrow}{\nabla} P]\Tr[\Phi P\overset{\leftrightarrow}{\nabla} P]\langle\Tr[\underline{\phi_n}U\sigma_3P\bar{U}]\Tr[\underline{\phi_n} U\sigma_3 P\bar{U}]\rangle\Big).\no
\ee
\end{widetext}
For the first and last terms, all frequencies of $P$ are fixed to be slow by the presence of two $\Phi$-fields. This is why terms of this kind are irrelevant for the RG; see Fig.~\ref{fig:S1S1Sintsmall} for an illustration. Out of the three terms, the relevant one is the second which reduces to the contribution displayed in Fig.~\ref{fig:fourforSD}(d). It gives
\be
&&-\frac{1}{2}\left\langle\!\left\langle S_1^2S_{int}\right\rangle\!\right\rangle=\label{eq:S1S1Sintfinal}\\
&&2i \pi\nu D^2 \sum_{n=0}^2\int_{\bfr_i,\eps_i} (\nabla_{\bfr_3''}-\nabla_{\bfr_2'})(\nabla_{\bfr_4''}-\nabla_{\bfr_4'})\no\\
&&\times \tr\left[(\bar{U}(\bfr_1) \underline{\phi_n}(\bfr_1))^\perp_{\eps_1\eps_f}(\underline{\phi_n}(\bfr_2)U(\bfr_2))^\perp_{\eps_f\eps_3}\Pi_{\eps_3\eps_f}(\bfr_1,\bfr_3'')\right.\no\\
&&\left.\quad\times\Phi^\parallel_{\eps_3\eps_4}(z)\Pi_{\eps_4\eps_f}(\bfr_3',\bfr_4'')\Phi^\parallel_{\eps_4\eps_1}(\bfr_4)\Pi_{\eps_1\eps_f}(\bfr_4',\bfr_2)\right].\no
\ee

\subsubsection{The correction $\Delta D$}
In the previous sections, expressions were obtained for the four different contributions to the RG-corrections to $S_D$. They can be found in Eqs. \eqref{eq:Sint1average2}, \eqref{eq:S1Sint1final}, \eqref{eq:S2aSint1final} and \eqref{eq:S1S1Sintfinal}. As is obvious from these formulas, and also from the diagrammatic representation in Fig.~\ref{fig:fourforSD}, the following block is common to all four terms
\be
&&\sum_{n=0}^2\left\langle \left(\bar{U}(\bfr_1)\underline{\phi_n}(\bfr_1)\right)^\perp_{\eps_1\eps_f}\left(\underline{\phi_n}(\bfr_2)U(\bfr_2)\right)^\perp_{\eps_f\eps_3}\right\rangle\quad \label{eq:block}\\
&=&\frac{i}{2\nu}\int_{\eps_5}\hat{\mathcal{V}}_{\eps_5\eps_f}^{ij}(\bfr_1-\bfr_2)\no\\
&&\times \left(\overline{\mathcal{U}}_{\eps_1\eps_5}(\bfr_1)\gamma_iu_{\eps_f}\right)^\perp\left(u_{\eps_f}\gamma_j\mathcal{U}_{\eps_5\eps_3}(\bfr_2)\right)^\perp,\no
\ee
where $\mathcal{U}=uU$, $\bar{\mathcal{U}}=\bar{U}u$ and $\mathcal{V}=\Gamma_d-2\Gamma_{2;d}$.

The gradient expansion of $U$ and $\bar{U}$ mentioned at the beginning of the calculation is necessary for $\langle\!\langle S_{int}\rangle\!\rangle$ and $i\langle\!\langle S_{1}S_{int}\rangle\!\rangle$ only, since the expressions for $i\langle\!\langle S_{2}S_{int}\rangle\!\rangle$ and $-\frac{1}{2}\langle\!\langle S_1^2S_{int}\rangle\!\rangle$ already contain two slow gradients (via $\Phi$). Since $\eps_f$ is fast and all other frequencies are slow, we can neglect, with the logarithmic accuracy, the slow frequencies $\eps_i$ compared to $\eps_f$ in the RG-integrals. Putting these remarks into effect, one finds
\begin{widetext}
\be
\Delta S_D&=&\left\langle S_{int}\right\rangle+i\left\langle\!\left\langle S_1S_{int}\right\rangle\!\right\rangle+i\left\langle\!\left\langle S_2S_{int}\right\rangle\!\right\rangle-\frac{1}{2}\left\langle\!\left\langle S_1^2S_{int}\right\rangle\!\right\rangle\label{eq:DeltaSDintermediate}\\
&=&-\pi\int_{\bfr,\bfp,\eps_i} \tr\left[\left\{-\delta_{\eps_3\eps_1}D\nabla_{\bfr'}\nabla_{\bfr''}-\Phi^\parallel_{\eps_3\eps_1}(\bfr)D\nabla_{\bfr''}+\Phi^\parallel_{\eps_3\eps_1}(\bfr)D\nabla_{\bfr'}+D(\Phi_{\eps_3\eps_4}(\bfr)\sigma_3\Phi_{\eps_4\eps_1}(\bfr)\sigma_3)^\parallel\right\}\right. \no\\
&&\qquad\qquad\left.\left. \left(\bar{\mathcal{U}}_{\eps_1\eps_5}(\bfr')\gamma_i u_{\eps_f}\right)^\perp \hat{\mathcal{V}}_{-\eps_f}^{ij}(\bfp) \Pi^2_{\eps_f}(\bfp) \left(u_{\eps_f}\gamma_j\mathcal{U}_{\eps_5\eps_3}(\bfr'')\right)^\perp\right]\right|_{\bfr''=\bfr'=\bfr}\no\\
&&-\frac{4}{d}\pi\int_{\bfr,\bfp,\eps_i} \tr\left[\left\{\delta_{\eps_3\eps_1}D\nabla_{\bfr'}\nabla_{\bfr''}+\Phi^\parallel_{\eps_3\eps_1}(\bfr)D\nabla_{\bfr''}-\Phi^\parallel_{\eps_3\eps_1}(\bfr)D\nabla_{\bfr'}-D\Phi^\parallel_{\eps_3\eps_4}(\bfr)\Phi^\parallel_{\eps_4\eps_1}(\bfr)\right\}\right.\no\\
&&\qquad  \left.\left.\qquad\qquad \left(\bar{\mathcal{U}}_{\eps_1\eps_5}(\bfr')\gamma_iu_{\eps_f}\right)^\perp \hat{\mathcal{V}}_{-\eps_f}^{ij}(\bfp)D\bfp^2\Pi^3_{\eps_f}(\bfp) \left(u_{\eps_f}\gamma_j\mathcal{U}_{\eps_5\eps_3}(\bfr'')\right)^\perp\right]\right|_{\bfr''=\bfr'=\bfr}\no,
\ee
\end{widetext}
where $d$ is the dimension. An additional term, which does not contain any gradients, was left out here. Fortunately, such terms need to cancel once all corrections are considered, as they would make the diffuson massive. (We have checked this cancellation by a perturbative calculation.) In order to further evaluate this expression, we study the quantity
\be
\tilde{\mathcal{R}}^{m}_{ab}(\bfp)=\int_{\eps_f} \;[\gamma_i u_{\eps_f}]^{a} \mathcal{V}_{-\eps_f}^{ij}(\bfp)\Pi^m_{\eps_f}(\bfp)[u_{\eps_f}\gamma_{j}]^b,\quad
\ee
where ${a,b}\in\{\parallel,\perp\}$, $m=2,3$ is the power with which the diffusons enter the expressions, and  $\tilde{\mathcal{R}}^{m}_{ab}(\bfp)$
is a matrix in Keldysh space.
For example, $\tilde{\mathcal{R}}^{m}_{\parallel\parallel}(\bfp)$ is a diagonal matrix with entries
\be
\tilde{\mathcal{R}}^{m}_{\parallel\parallel}(\bfp)_{11}&=&\int_{\eps_f}\left(\mathcal{B}_{\eps_f}\mathcal{V}^R_{\eps_f}+(F_{\eps_f}-\mathcal{B}_{\eps_f})\mathcal{V}^A_{\eps_f}\right)\mathcal{D}_{\eps_f}^m,\\
\tilde{\mathcal{R}}^{m}_{\parallel\parallel}(\bfp)_{22}&=&\int_{\eps_f}\left(-\mathcal{B}_{\eps_f}\mathcal{V}^A_{\eps_f}+(\mathcal{B}_{\eps_f}-F_{\eps_f})\mathcal{V}^R_{\eps_f}\right)\overline{\mathcal{D}}_{\eps_f}^m. \qquad
\ee
For the RG calculation in $2d$, these integrals need to be found with logarithmic accuracy only. To this end note that for the purpose of the RG analysis, we may set
\be
\mathcal{F}_{\eps_f}\approx \mathcal{B}_{\eps_f}\approx \mbox{sign}(\eps_f).\label{eq:apprsigmaf}
\ee
Due to the frequent occurrence of the sign-factor, let us introduce the notation
\be
\sigma_f=\mbox{sign}(\eps_f).
\ee
As a consequence
\be
\tilde{\mathcal{R}}^{m}_{\parallel\parallel}(\bfp)\approx \int_{\eps_f}\sigma_f\mathcal{D}_{\eps_f}^m\mathcal{V}^R_{\eps_f}.\label{eq:diagonalR}
\ee
In a similar way one finds $\tilde{\mathcal{R}}^{m}_{\perp\perp}(\bfp)=\tilde{\mathcal{R}}^{m}_{\parallel\parallel}(\bfp)$.

Next, consider the off-diagonal matrix $\tilde{\mathcal{R}}^{m}_{\parallel\perp}$ with entries
\be
\tilde{\mathcal{R}}^{m}_{\parallel\perp}(\bfp)_{12}&=&\int_{\eps_f} \left(\mathcal{F}_{\eps_f}\mathcal{B}_{\eps_f}\mathcal{V}^R_{\eps_f}+\mathcal{V}^A_{\eps_f}(1-\mathcal{F}_{\eps_f}\mathcal{B}_{\eps_f})\right)\mathcal{D}_{\eps_f}^m,\no\\
\tilde{\mathcal{R}}^{m}_{\parallel\perp}(\bfp)_{21}&=&\int_{\eps_f}\overline{\mathcal{D}}^n_{\eps_f} \mathcal{V}^A_{\eps_f}.
\ee
Employing again the approximations of Eq.~\eqref{eq:apprsigmaf}, we see that both components reduce to integrals over a product of only retarded or only advanced functions. A similar structure, obviously, holds for $\tilde{\mathcal{R}}^{m}_{\perp\parallel}(\bfp)$. In perturbative calculations such terms vanish after integration in frequency a discussed earlier. In the RG procedure it is a little bit more complicated. After integration in momentum, such terms are odd functions in frequency. Thus, although the integration over the fast frequency is performed within limited intervals, the sum over the positive and negative frequency-intervals vanishes. It is useful in this connection to compare the expressions for the diagonal and off-diagonal matrices $\tilde{\mathcal{R}}$. The diagonal ones, see Eq.~\eqref{eq:diagonalR}, contain an additional factor $\sigma_f$ which makes the $\eps_f$-integrals finite.

Therefore, we need to keep only the $\parallel\parallel$ and $\perp\perp$ components. Coming back to $\Delta S_D$ as given in Eq.~\eqref{eq:DeltaSDintermediate}, one obtains
\be
&&\int_{\eps_5,\eps_f}\left(\bar{\mathcal{U}}_{\eps_1\eps_5}\gamma_i u_{\eps_f}\right)^\perp \mathcal{V}_{-\eps_f}^{ij}\Pi^n_{\eps_f} \left(u_{\eps_5}\gamma_j\mathcal{U}_{\eps_f\eps_3}\right)^\perp\no\\
&=&\sum_{a,b=\perp\parallel}\;\left[\bar{\mathcal{U}}^a\;\tilde{\mathcal{R}}^{m}_{a'b'}\;\mathcal{U}^b\right]_{\eps_1\eps_3}\no\\
&=&\left(\bar{\mathcal{U}}^\parallel\mathcal{U}^\parallel+\bar{\mathcal{U}}^\perp\mathcal{U}^\perp\right)_{\eps_1\eps_3}
\int_{\eps_f}\sigma_f\mathcal{D}_{\eps_f}^m\mathcal{V}^R_{\eps_f},\label{eq:aux}
\ee
where in the second line we denoted ${\perp}'=\parallel$, $\parallel'=\perp$ for $a'$ and $b'$, and used the obvious fact that the off-diagonal part of the product $C=AB$ is given by $C^{\perp}=\sum_{a=\perp,\parallel} A^aB_{a'}$. As only the parallel component of the total matrix considered in Eq.~\eqref{eq:aux} enters the trace in Eq.~\eqref{eq:DeltaSDintermediate}, we may effectively replace
\be
\bar{\mathcal{U}}^\parallel(\bfr')\mathcal{U}^\parallel(\bfr'')+\bar{\mathcal{U}}^\perp(\bfr')\mathcal{U}^\perp(\bfr'')\rightarrow \bar{U}(\bfr')U(\bfr'').
\ee
It was used that the matrices $u$ cancel.
Let us further introduce the notation
\be
\mathcal{R}_{2}&=&\int_\bfp\tilde{\mathcal{R}}^{2}_{\parallel\parallel}(\bfp)
=\int_{\bfp,\eps_f}\sigma_f\mathcal{D}_{\eps_f}^2(\bfp)\mathcal{V}^R_{\eps_f}(\bfp),\\
\mathcal{R}_3&=&\int_\bfp D\bfp^2\tilde{\mathcal{R}}^{3}_{\parallel\parallel}(\bfp)=\int_{\bfp,\eps_f} \sigma_f D\bfp^2\mathcal{D}_{\eps_f}^3(\bfp) \mathcal{V}^R_{\eps_f}(\bfp).\no
\ee
The expression for the renormalization of the diffusion constant reads
\be
\Delta S_D\label{eq:DeltaS_D2+3}
&=&-\pi \mathcal{R}_{2}\Tr\left[-D\nabla U \nabla \overline{U}-D\nabla U\Phi^\parallel \overline{U}\right.\no\\
&&\left.\qquad+DU\Phi^\parallel \nabla \overline{U}+D[(\Phi^{\parallel})^2-(\Phi^{\perp})^2]\right]\no\\
&&-\frac{4\pi }{d} \mathcal{R}_{3}\Tr\left[D\nabla U \nabla \overline{U}+D\nabla U\Phi^\parallel \overline{U}\right.\no\\
&&\left.\qquad -DU\Phi^\parallel \nabla \overline{U}-D\Phi^\parallel\Phi^\parallel\right]\no\\
&=&\frac{4\pi }{d} \mathcal{R}_{3}\Tr\left[D(\Phi^{\perp})^2\right].
\ee
We see that the two-diffuson contributions cancel out (as it may be expected from general arguments\cite{Finkelstein90,Finkelstein10}), and the remaining term comes from the three-diffuson term only. Using $\Tr[(\Phi^\perp)^2]=-\frac{1}{4}\Tr[(\nabla Q_s)^2]$, one finds
\be
&&\Delta S_D=-\frac{\pi}{d}\int \tr[D(\nabla Q_s)^2] \times\\
&&\int_{\bfp,\eps_f} \sigma_f D\bfp^2\mathcal{D}_{\eps_f}^3(\bfp) \left[\Gamma^R_{d}(\bfp,\eps_f)-2\Gamma^R_{2,d}(\bfp,\eps_f)\right],\no
\ee
This leads to the following result for the correction to the diffusion coefficient
\be
\Delta D&=&\frac{4i D}{d\nu}\int_{\bfp,\eps_f}\sigma_f D\bfp^2 \mathcal{D}^3_{\eps_f}(\bfp)\no\\
&&\qquad\times  \left[\Gamma^R_{d}(\bfp,\eps_f)-2\Gamma_{2,d}^R(\bfp,\eps_f)\right].\label{eq:DeltaDprelim}
\ee
The factor $d$ in the denominator results from an averaging over the direction of momentum. The logarithmic integral will be evaluated in Sec.~\ref{subsec:logarithmic integrals} below.

Finally, the situation with the abandoned terms, where all frequencies were forced to be slow, is worth commenting. See Fig.~\ref{fig:S1S1Sintsmall} as an example. Such terms have a hybrid structure, as they resemble at the same time the $S_D$-term and the interaction term of the action: they contain gradients \emph{and} mix frequencies. The remaining momentum integrals are not logarithmic, and are determined by the lower cutoff $\lambda \tau^{-1}$ of the RG-interval. Compared to the electron-electron interaction terms, the discussed terms contain a small parameter $\rho Dk^2/(\lambda\tau^{-1})$, which is not compensated by a large logarithm. Here, the small parameter $\rho$ is the only small parameter introduced for the RG analysis:
\be
\rho=\frac{1}{(2\pi)^2\nu D}\label{eq:rhodefinition}.
\ee
It has the meaning of the sheet resistance measured in dimensional units; note an extra factor $\pi$ as compared to the quantum resistance.

\subsection{Renormalization of $z$}

There are two corrections to $S_z$,
\be
\Delta S_z=i\left\langle\!\left\langle S_{\eps}S_{int}\right\rangle\!\right\rangle+\left\langle S_{int}\right\rangle.
\ee
Below we present some details of the calculation. As it turns out, the dominant contributions arise from those terms for which $S_{int}$ is replaced by $S_{int,1}$.

\subsubsection{$i\left\langle\!\left\langle S_{\eps}S_{int,1}\right\rangle\!\right\rangle$}

\begin{figure}[tb]
\includegraphics[height=2.5cm]{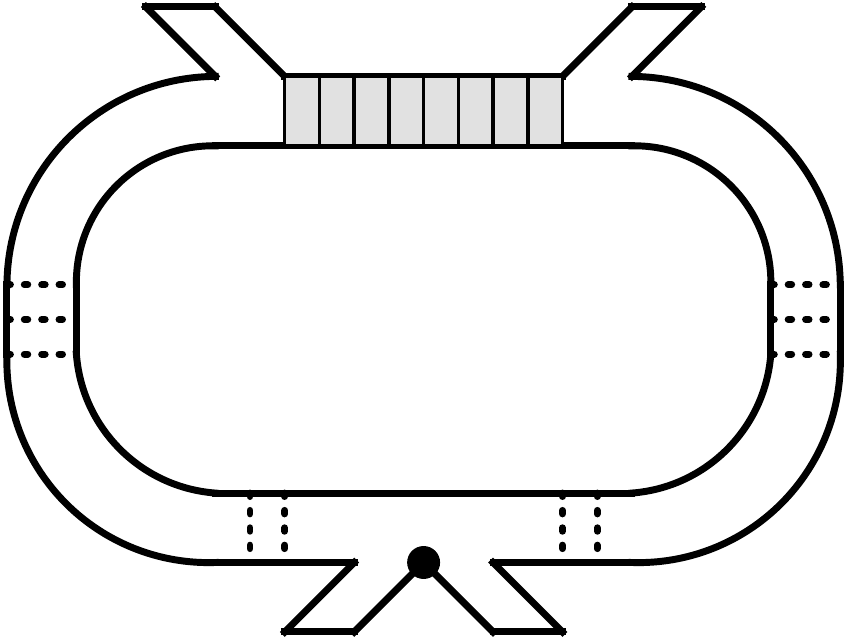}
\caption{Diagrammatic representation for $i\left\langle\!\left\langle S_{\eps} S_{int,1}\right\rangle\!\right\rangle$. This term contributes to $\Delta z$.}
\label{fig:SESint}
\end{figure}

After evaluating the relevant contractions in the $P$-matrices, one obtains the expression
\be
&&i\left\langle\!\left\langle S_{\eps}S_{int,1}\right\rangle\!\right\rangle=2\pi\nu\sum_{n=0}^2\int_{\bfr_i,\eps_i}\\
&&\left\langle \tr\left[(\overline{U}(\bfr_1)\underline{\phi_n}(\bfr_1)U(\bfr_1)\sigma_3)^\perp_{\eps_1\eps_2}\Pi_{\eps_2\eps_3}(\bfr_3-\bfr_2)\right.\right.\no\\
&&\left.\quad\times\Pi_{\eps_2\eps_1}(\bfr_3-\bfr_1)(\overline{U}(\bfr_2)\underline{\phi_n}(\bfr_2)U(\bfr_2))^\perp_{\eps_2\eps_3}\right.\no\\
&&\left.\quad\times(\overline{U}({\bfr_3}) z \eps_sU({\bfr_3}))_{\eps_3\eps_1}^\parallel ]\right\rangle.\no
\ee
The frequencies $\eps_1$ and $\eps_3$ are bound to be slow due to presence of $\eps_s$, while $\eps_2$ is fast. This observation directly leads to the result
\be
i\left\langle\!\left\langle S_{\eps}S_{int,1}\right\rangle\!\right\rangle&=&-\pi i\mathcal{R}_{2}\Tr\left[ z \eps_s Q_s \right].\label{eq:SESint1final}
\ee
The corresponding diagram is displayed in Fig.~\ref{fig:SESint}.

\subsubsection{$\left\langle S_{int,1}\right\rangle$}
\begin{figure}
\includegraphics[height=2.5cm]{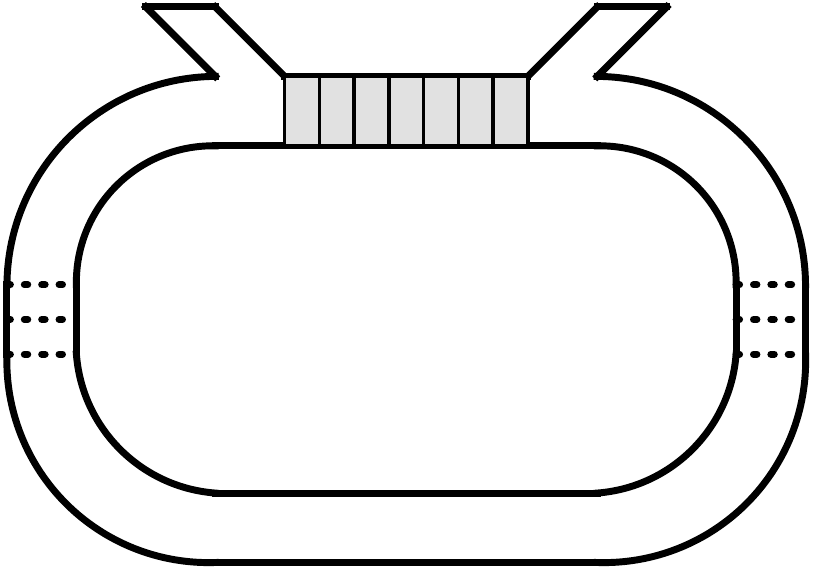}
\caption{The average $\left\langle S_{int,1}\right\rangle$ as relevant for the calculation of $\Delta z$. Expansion in the slow frequency is needed to be performed.}
\label{fig:SintforDz}
\end{figure}

This term is somewhat special, as it contains a contribution from the boundaries of the frequency integration interval. Starting point is formula \eqref{eq:Sint1average2}, see also Fig.~\ref{fig:SintforDz}, where (unlike previously) $\bfr_2$ may directly be set equal to $\bfr_1$, but an expansion in slow frequencies is performed. In order to see how it works, it is convenient to first perform the average in $\phi$
\be
\left\langle S_{int,1}\right\rangle
&\approx&-\pi \Tr\left[(\overline{\mathcal{U}}_{\eps_1\eps_5}\gamma_i u_{\eps_f})^\perp\hat{\Pi}_{\eps_f-\eps_1}(\bfp)\mathcal{V}^{ij}_{\eps_5-\eps_f}(\bfp)\right.\no\\
&&\left.\quad \times (u_{\eps_f} \gamma_j\mathcal{U}_{\eps_5\eps_1})^\perp\right].
\ee
An expansion in slow frequencies could be either in $\eps_5$ or in $\eps_1$. When expanding in $\eps_1$, the matrices $\mathcal{U}$, $\bar{\mathcal{U}}$ cancel following the previous arguments. Therefore, one should consider an expansion in $\eps_5$ and study
\be
\tilde{\mathcal{R}}_{ab}^{1}(\bf{p};\eps_5)&=&\int_{\eps_f}[\gamma_{i}u_{\eps_f}]^a \left[\mathcal{V}_{\eps_5-\eps_f}(\bfp)-\mathcal{V}_{-\eps_f}(\bfp)\right]^{ij}\no\\
&&\quad\times \Pi_{\eps_f}(\bfp)[u_{\eps_f} \gamma_j]^{b},\quad a,b\in\{\perp,\parallel\}.\qquad
\ee
Only the $\perp\perp$ and $\parallel\parallel$ components give a logarithmic contribution. Further, it should be noted that an expansion of the  distribution function in $\eps_5$ is not necessary since such terms would be exponentially suppressed in the RG regime. Defining
\be
\mathcal{R}_{1}=\frac{1}{z}\int_{\eps_f,p}\sigma_f \mathcal{D}_{\eps_f}(\bfp)\partial_{\eps_f}V^R_{\eps_f}(\bfp),
\ee
one obtains
\be
\tilde{\mathcal{R}}^{\parallel\parallel}_{1,\eps_5}\approx-\tilde{\mathcal{R}}_{1,\eps_5}^{\perp\perp} \approx -z\eps_5\mathcal{R}_{1}\sigma_3,
\ee
and further on
\be
\left\langle S_{int,1}\right\rangle&\approx&-\pi \mathcal {R}_{1} \Tr\left[z\eps Q_s\right].\label{eq:SintforDzfinal}
\ee
The integral $\mathcal{R}_1$ may be rearranged with the use of a partial integration in $\eps_f$:
\be
\mathcal{R}_{1}&=&\frac{1}{2\pi z}\int_{\bfp}\sigma_f\mathcal{D}(\bfp)\left.\mathcal{V}_{\eps_f}^R(\bfp)\right|_{bound}-i\mathcal{R}_{2},\no\\
\ee
where the index $bound$ indicates that expression should be evaluated at the boundaries of the frequency integration interval.

\subsubsection{The correction $\Delta z$}
When combining the two contributions, Eqs.~\eqref{eq:SESint1final} and  $\eqref{eq:SintforDzfinal}$, a partial cancellation occurs and only the boundary terms remain. For the total correction to $z$ one reads off
\be
\Delta z=\frac{1}{2\pi\nu}\int_{\bfp}\sigma_f\mathcal{D}_{\eps_f}(\bfp)\left.\mathcal{V}_{\eps_f}^R(\bfp)\right|_{bound}\label{eq:Deltaz}.
\ee
It is important to note that once the integrand is evaluated at the two boundaries, i.e., the upper and lower limits of the frequency integral, the momentum integral is convergent and yields a logarithmic correction.

\subsection{Renormalization of the interaction amplitudes}
The interaction term $S_{\Gamma}$ contains three interaction amplitudes, $\Gamma_0(\bfq)$, $\Gamma_1$ and $\Gamma_2$.  The amplitude $\Gamma_2$
differs by the spin structure from the other two and, therefore, corrections to either of these two classes are easily identified. The amplitudes $\Gamma_0(\bfq)$ and $\Gamma_1$ have the same spin structure, but they differ in another aspect. Recall that $\Gamma_0(\bfq)$ is the statically screened long-range Coulomb interaction, while $\Gamma_1$ is short-range as it is directly related to the Fermi liquid amplitudes. A correction to $\Gamma_0(\bfq)$ could arise only from diagrams, for which the Coulomb interaction is not
part of the logarithmic integration. Such type of diagrams can be generated with the help of $S_{int,2}$ and closely resemble vertex corrections for a scalar vertex. Importantly, such corrections, although they arise from individual diagrams, eventually cancel, once all contributions are summed up. Indeed, it turns out that the cancellation occurs between certain pairs of diagrams. The calculation will, therefore, be organized in such a way that these pair diagrams are treated together. As already indicated, the cancellation of the corrections to $\Gamma_0(\bfq)$ also reflects itself in the fact that the scalar triangular vertex $\gamma_\triangleleft^\rho$ remains unrenormalized. This will be demonstrated explicitly below in Secs.~\ref{subsec:Corrections to gamma} and \ref{subsec:vertex corrections}. In contrast, the correction to the amplitude $\Gamma_1$, which is short-range in character, is finite.

Generally, the RG-equations at the one-loop level sum the series of logarithmic corrections of the kind $(\rho \ln 1/T\tau)^n$, where $\rho$, the small parameter of the RG expansion, has been introduced in Eq.~\eqref{eq:rhodefinition}. Corrections to the interaction amplitude may contain a product of several interaction amplitudes, with some of them being dressed. Even on the level of the one-loop approximation, it is a priori not clear whether the number of diagrams that needs to be considered in order to derive such a system of equations is finite. As has first been demonstrated by Finkel'stein in Ref.~\onlinecite{Finkelstein83}, it is fortunately the case and the product of at most four (dressed and undressed) interaction amplitudes is involved in the calculation. The main guiding rule here is that the order of the RG-equation is determined by the number of momentum integrations: each integration generates the small parameter $\rho$. There cannot be too many dressed amplitudes, because otherwise it is impossible to arrange them without an additional momentum integration.

In order to structure the calculation, we will present the correction to $S_\Gamma$ as the sum of $6$ individual contributions. Apart from the first one, all of them consist of pairs of diagrams. These pairs arise as a result of a different choice of the fast frequency for the logarithmic integration. The above mentioned cancellation of corrections to $\Gamma_0(\bfq)$ takes place between the two partner diagrams forming a pair [whenever such correction appears]. For the corrections to $\Gamma_1$ and $\Gamma_2$ the cancellation is not complete, and these corrections remain finite. We write
\be
\Delta S_{\Gamma}=\sum_{i=0}^5 (\Delta S_{\Gamma})_i,
\ee
where
\be
(\Delta S_{\Gamma})_0&=&\left\langle S_{int,1}\right\rangle\label{eq:DeltaSGamma0}\\
(\Delta S_{\Gamma})_1&=&\frac{i}{2}\left\langle \!\left\langle S_{int,1}^2\right\rangle\!\right\rangle\label{eq:DeltaSGamma1}\\
(\Delta S_{\Gamma})_2&=&i\left\langle\!\left\langle S_{int,1}S_{int,2}\right\rangle\!\right\rangle\label{eq:DeltaSGamma2}\\
(\Delta S_{\Gamma})_3&=&-\frac{1}{2}\left\langle\!\left\langle S_{int,1}^2S_{int,2}\right\rangle\!\right\rangle\\
(\Delta S_{\Gamma})_4&=&-\frac{1}{2}\left\langle\!\left\langle S_{int,1}S^2_{int,2}\right\rangle\!\right\rangle\\
(\Delta S_{\Gamma})_5&=&-\frac{i}{4}\left\langle\!\left\langle S_{int,1}^2S_{int,2}^2\right\rangle\!\right\rangle.\label{eq:DeltaSGamma5}
\ee
We will present details of the calculation of the first two contributions, the other ones can be considered in a similar way, but we will only state the results and display the corresponding diagrams. As already mentioned, the calculation of vertex corrections presented in Secs.~\ref{subsec:Corrections to gamma} and \ref{subsec:vertex corrections} have a close similarity to some of the diagrams that are important here. The interested reader may find additional information there, in particular about the cancelations for pair diagrams.

\subsubsection{$\left\langle S_{int,1}\right\rangle$}
\begin{figure}
\includegraphics[height=2.5cm]{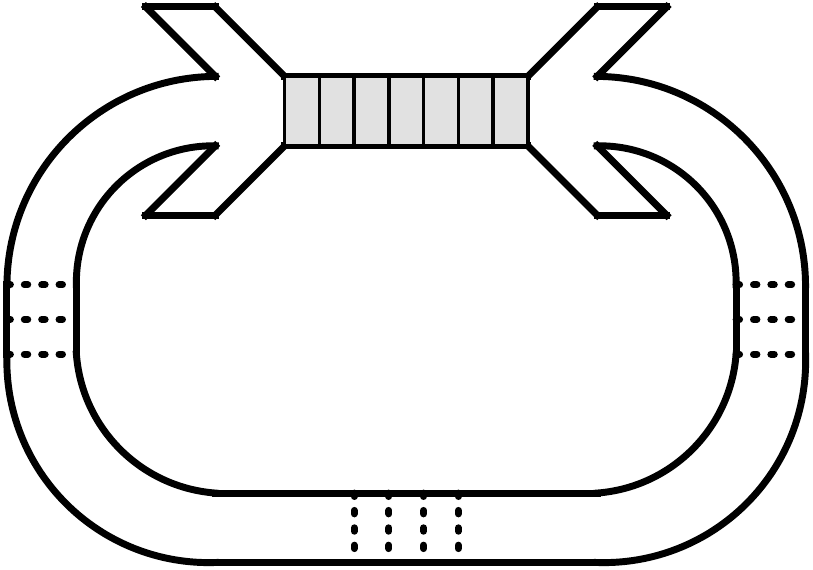}
\caption{$\left\langle S_{int,1}\right\rangle$ as relevant for the renormalization of the interaction amplitudes. In this case, all frequencies involved are slow, the logarithmic correction arises from an integration over fast momenta.}
\label{fig:Fig12}
\end{figure}

This term has been considered before and we may use formula \eqref{eq:Sint1average1} for $\left\langle S_{int,1}\right\rangle$ as our starting point. In the present context, we consider the case that the two frequency arguments $\eps_1$ and $\eps_2$ are slow, while the momentum entering $\Pi$ is fast, see Fig.~\ref{fig:Fig12}. Therefore we can approximate it by just $\Pi(\bfp,0)$, i.e., for the range of momenta $\bfp$ that are of interest, the frequency dependence may be neglected. In this approximation. $\Pi(\bfp,0)\approx \mathcal{D}(\bfp,0)\equiv \mathcal{D}(\bfp)$ becomes proportional to the unit matrix in Keldysh space and additionally the summation in $\eps_1$ and $\eps_2$ may be performed. As no expansion in slow momenta is required, we may put $\bfr_2\rightarrow \bfr_1$ for the arguments of the slow modes:
\be
\left\langle S_{int,1}\right\rangle&=& i\pi\nu \sum_{n=0}^2\int_{\bfr\bfr'}\left\langle\tr[\left(\bar{U}(\bfr)\underline{\phi_n}(\bfr)U(\bfr)\right)^\perp\right.\\
&&\left.\quad \times \left(\bar{U}(\bfr)\underline{\phi_n}(\bfr')U(\bfr')\right)^\perp]\right\rangle \mathcal{D}(\bfr-\bfr')\no\\
&=&\frac{i\pi\nu}{2}\sum_{n=0}^2\int_{\bfr,\bfr'} \tr[\underline{\phi_n}(\bfr)\underline{\phi_n}(\bfr')\no\\
&&\quad -Q_s(\bfr)\underline{\phi_n}(\bfr)Q_s(\bfr)\underline{\phi_n}(\bfr')]\mathcal{D}(\bfr-\bfr').\no
\ee
The first term in the last equation is just a constant and can be dropped. After performing the average in $\phi$ one obtains
\be
\left\langle S_{int,1}\right\rangle
&=&\frac{\pi}{4}\int_{\eps_i} \Tr\left[\underline{Q_{s,\alpha\beta;\eps_2\eps_1}}\gamma_i\underline{Q_{s,\beta\alpha;\eps_4,\eps_3}}\gamma_j\right]\\
&&\times \int_{\bfr}\hat{\Gamma}^{ij}(\bfp)\mathcal{D}(\bfp)\delta_{\eps_1-\eps_4,\eps_2-\eps_3}\no\\
&&-\frac{\pi}{4}\int \tr\left[\underline{Q_{s,\alpha\alpha;\eps_2\eps_1}}\gamma_i\underline{Q_{s,\beta\beta;\eps_4,\eps_3}}\gamma_j\right]\no\\
&&\times \int_{\bfr}\hat{\Gamma}^{ij}_2\mathcal{D}(\bfp)\delta_{\eps_1-\eps_4,\eps_2-\eps_3}.\no
\ee

As the frequency arguments of $\Gamma$, $\Gamma_2$ are slow (while the momenta are fast), no dressing of the interaction line was included, and the static amplitudes can be used. As was already noted before, in such a case $\hat{\Gamma}$ and $\hat{\Gamma}_2$ are off-diagonal matrices in Keldysh space and take the simple form
\be
\hat{\Gamma}=\left(\ba{cc}0&\Gamma\\
\Gamma&0\ea\right),\quad \hat{\Gamma}_2=\left(\ba{cc} 0&\Gamma_2\\\Gamma_2&0\ea\right).
\ee
We can use the relation (recall that $\gamma_1$ is the unit matrix)
\be
&&\Tr [\underline{Q_1} \underline{Q_2}\gamma_2]+\Tr[\underline{Q_1}\gamma_2 \underline{Q_2}]\no\\
&=&\Tr[ \underline{Q_1}] \Tr[\gamma_2 \underline{Q_2}]+\Tr[\gamma_2 \underline{Q_1}] \Tr[\underline{Q_2}],
\ee
where all appearing $Q$-matrices have fixed frequency arguments and spin indices. The result is
\be
\left\langle S_{int,1}\right\rangle&=&\frac{\pi}{4}\int_{\bfr,\eps_i} \tr\left[\gamma_i\underline{Q_{s,\alpha\beta;\eps_2\eps_1}}(\bfr)\right]\gamma_2^{ij}\times\\
&& \tr\left[\gamma_j\underline{Q_{s,\beta\alpha;\eps_4,\eps_3}}(\bfr)\right]\delta_{\eps_1-\eps_4,\eps_2-\eps_3}\int_{\bfp}\Gamma(\bfp)\mathcal{D}(\bfp)\no\\
&-&\frac{\pi}{4}\int_{\bfr,\eps_i} \tr\left[\gamma_i \underline{Q_{s,\alpha\alpha;\eps_2\eps_1}}(\bfr)\right]\gamma_2^{ij}\times\no\\
&&\times \tr \left[\gamma_j\underline{Q_{s,\beta\beta;\eps_4\eps_3}}(\bfr)\right]\delta_{\eps_1-\eps_4,\eps_2-\eps_3}\int_{\bfp}\Gamma_2\mathcal{D}(\bfp).\no
\ee
Comparing to the original interaction term, Eq.~\eqref{eq:Sintiden}, one finds that the structure of the $\Gamma_1$ and $\Gamma_2$ terms are reproduced, leading to the resulting corrections from $(\Delta S_{\Gamma})_0$:
\be
(\Delta \Gamma_1)_0&=&\frac{1}{\pi\nu}\Gamma_2\int_{\bfp}\mathcal{D}(\bfp),\label{eq:onezero}\\
(\Delta \Gamma_2)_0&=&\frac{1}{\pi\nu}\int_\bfp\Gamma(\bfp)\mathcal{D}(\bfp)\label{eq:twozero}.
\ee
\subsubsection{Pairs of diagrams}

\begin{figure}
\includegraphics[width=8cm]{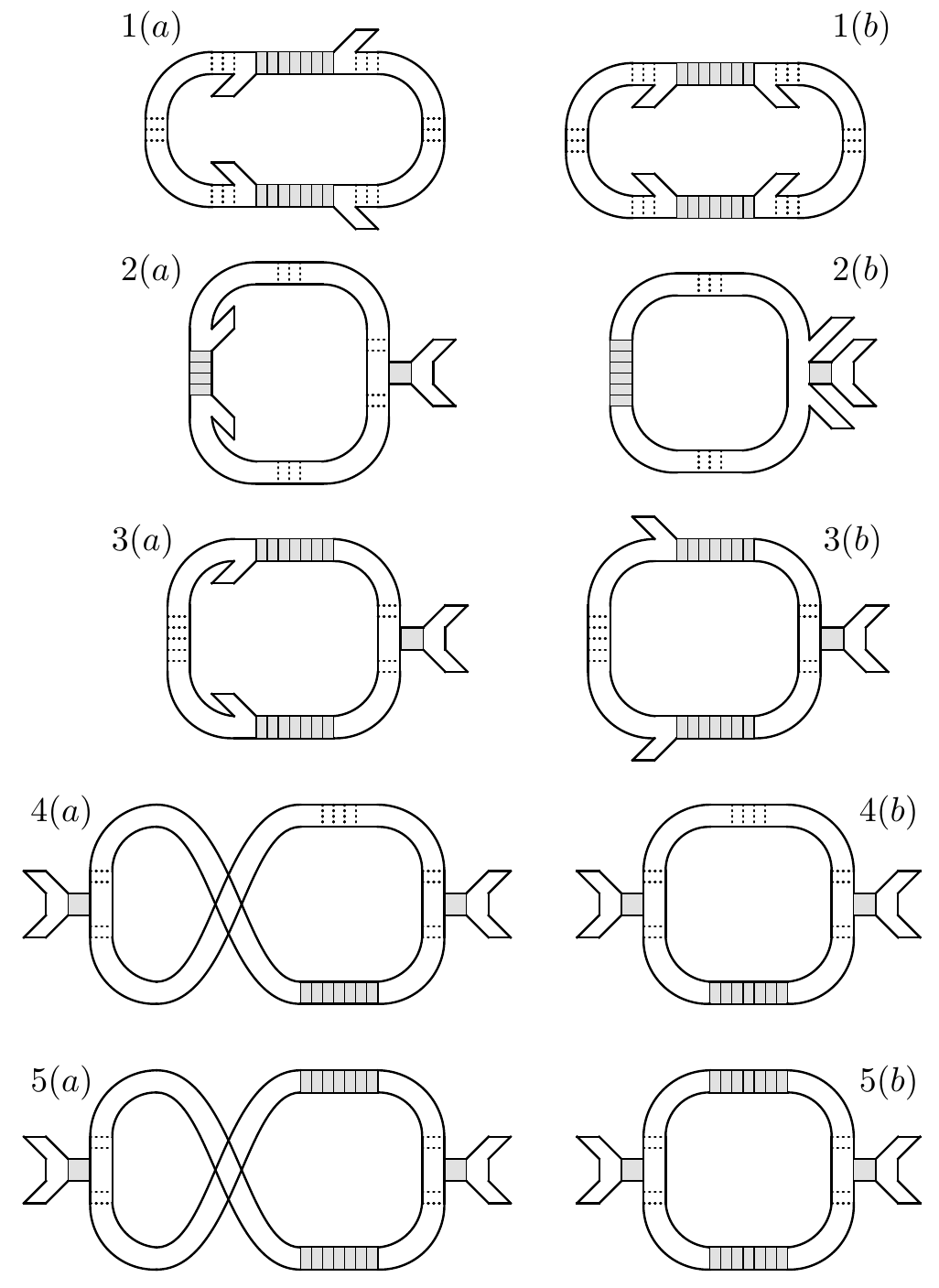}
\caption{The pairs of diagrams related to $(\Delta S_\Gamma)_i$, $i=1-5$. Diagrams labeled as $(a)$ give rise to the corrections $(\Delta \Gamma_1)_i$ and diagrams labeled as ($b$) to the corrections $(\Delta \Gamma_2)_i$. Only those contributions remain, for which all interaction amplitudes are of the $\Gamma_2$-type. All other contributions, which contain the amplitudes $\Gamma_0$ or $\Gamma_1$ at least once, cancel between the two diagrams forming a pair. An important consequence is that the amplitude $\Gamma_0$ remains unrenormalized.}
\label{fig:Summary_Gamma}
\end{figure}

As we have already mentioned, pairs of diagrams arise as a result of a different choice of the fast frequency $\eps_f$ for the logarithmic integration. These pairs of diagrams are displayed as two columns in Fig.~\ref{fig:Summary_Gamma}.
As an illustration, we discuss in detail one pair of diagrams, labeled as 1(a) and 1(b). This pair gives rise to the correction $(\Delta S_\Gamma)_1$, and originates from
\be
&&\frac{i}{2}\left\langle \!\left\langle S_{int,1}^2\right\rangle\!\right\rangle=\\
&&-\frac{i(\pi\nu)^4}{8}\left\langle\!\!\left\langle \left\langle \Tr[\underline{\phi_2}U\sigma_3 P \overline{U}]\Tr[\underline{\phi
_2}U\sigma_3 P\overline{U}]\right\rangle_{\phi_2}\right.\right.\no\\
&&\left.\left.\times \langle \Tr[\underline{\phi'_2}U\sigma_3 P \overline{U}]\Tr[\underline{\phi'
_2}U\sigma_3 P\overline{U}]\rangle_{\phi'_2}\right\rangle\!\!\right\rangle.\no
\ee
Note that $\phi'_2$ has the same correlation as $\phi_2$ (As it will become clear later, only the $\phi_2$-contractions have to be considered in all diagrams presented in Fig.~\ref{fig:Summary_Gamma}. Otherwise, the contributions are canceled out {\emph {within}} each of the pairs.)

We perform the contractions, and introduce a symmetry factor two:
\be
&&\frac{i}{2}\left\langle \!\left\langle S_{int,1}^2\right\rangle\!\right\rangle=-i(\pi\nu)^2\int_{\bfr_i,\eps_i}\\
&&\langle\tr[(\overline{U}\underline{\phi_2}U)^\perp_{\eps_1\eps_2}(\bfr_1)\Pi_{\eps_2\eps_1}(\bfr_1-\bfr_4)(\overline{U}\underline{\phi'_2}U)^\perp_{\eps_2\eps_1}(\bfr_4)]\no\\
&&\tr[(\underline{U}\underline{\phi'_2}U)_{\eps_3\eps_4}^\perp(\bfr_3) \Pi_{\eps_4\eps_3}(\bfr_3-\bfr_2)(\overline{U}\underline{\phi_2}U)^\perp_{\eps_4
\eps_3}(\bfr_2)]\rangle_{\phi_2\phi_2'}.\no
\ee
The different ways in which the occurring frequencies can be chosen as being fast are as follows:
\begin{enumerate}[(a)]
\item $(\eps_2,\eps_3)$ fast or equivalently $(\eps_1,\eps_4)$ fast $\rightarrow (\Delta \Gamma_1)_1$
\item $(\eps_2,\eps_4)$ fast or equivalently $(\eps_1,\eps_3)$ fast $\rightarrow (\Delta \Gamma_2)_1$.
\end{enumerate}
These two possibilities lead to the diagrams displayed in Figs.~\ref{fig:Summary_Gamma} as 1(a) and 1(b), respectively. For case (a), a correction to $\Gamma_1$ arises; for case (b) a correction to $\Gamma_2$.

(a) Let $(\eps_2,\eps_3)$ be fast: We account for the equivalent choices by a factor of two, neglect slow frequencies in the diffusion propagators, and take the slow $U$ modes at coinciding points. In this way one obtains
\be
&&\Delta S=\\
&&\frac{-i(\pi\nu)^2}{2}\int\left\langle \tr[(\underline{\phi'_2}(\bfr_4)Q(\bfr_1)\underline{\phi_2}(\bfr_1)\sigma_3)_{\eps_2\eps_2}\Pi_{\eps_2}(\bfr_1-\bfr_4)\right.\no\\
&&\qquad -(\underline{\phi'_2}(\bfr_4)\underline{\phi_2}(\bfr_1))_{\eps_2\eps_2}\Pi_{\eps_2}(\bfr_1-\bfr_4)]\no\\
&&\times \tr[(\underline{\phi'_2}(\bfr_3)Q(\bfr_2)\underline{\phi_2}(\bfr_2)\sigma_3)_{\eps_3\eps_3}\Pi_{\eps_3}(\bfr_2-\bfr_3)\no\\
&&\left.\qquad-(\underline{\phi'_2}(\bfr_3)\underline{\phi_2}(\bfr_2))_{\eps_3\eps_3}\Pi_{\eps_3}(\bfr_2-\bfr_3)]\right\rangle.\no
\ee
The term of interest is the one containing two $Q$'s and for this term one obtains
\be
\Delta S&=&\frac{\pi^2 i}{8}\int_{\bfr,\bfp,\eps_i} \Gamma^{ij}_{2;d}(\bfp,\eps_f)\Gamma_{2;d}^{kl}(\bfp,-\eps_f)\delta_{\eps_1-\eps_2,\eps_4-\eps_3}\no\\
&&\times\tr\left[(\gamma_k\Lambda_{\eps_f}\underline{\Pi_{\eps_f}}(\bfp)\gamma_i)\underline{Q_{\alpha\alpha;\eps_1\eps_2}}(\bfr)\right]\no\\
&&\times\tr\left[(\gamma_l\Lambda_{-{\eps_f}}\underline{\Pi_{-\eps_f}}(\bfp)\gamma_j)\underline{Q_{y,\beta\beta;\eps_3\eps_4}}(\bfr)\right],\label{eq:DeltaSintermediate}
\ee
where we remind that $\Lambda_\eps=u_\eps \sigma_3u_{\eps}$ and we defined $\underline{\Pi_\eps}=u_\eps\Pi_\eps u_\eps$. After a somewhat tedious but straightforward calculation one may show that the following expression emerges
\be
&&\Delta S=\frac{-i\pi^2}{2}\int_{\bfp,\eps_f} \;\sigma_f [\mathcal{D}^R_{\eps_f}(\bfp) \Gamma_{2;d}^R(\bfp,\eps_f)]^2\label{eq:DeltaSGamma1text}\\
&&\times\int_{\bfr,\eps_i} \tr[ \gamma_1\underline{Q_{\alpha\alpha;\eps_1\eps_2}}(\bfr)]\tr[\gamma_2\underline{Q_{\beta\beta;\eps_3\eps_4}}(\bfr)]\delta_{\eps_1-\eps_2,\eps_4-\eps_3}.\no\label{eq:DeltaSfinala}
\ee
We see that the typical structure of the $\Gamma_1$-type interaction term is reproduced.

(b) Now, let $(\eps_2,\eps_4)$ be fast: In a similar way we find that we should evaluate the following expression
\be
&&\Delta S=\\
&&\frac{-i(\pi\nu)^2}{2}\int \langle \tr[(\underline{\phi'_{2}}(\bfr_4)Q(\bfr_1)\underline{\phi_{2}}(\bfr_1)\sigma_3)_{\eps_2\eps_2}{\Pi}_{\eps_2}(\bfr_1-\bfr_4)]\no\\
&&\times
\tr[(\underline{\phi_{2}}(\bfr_2)Q(\bfr_2)\underline{\phi'_{2}}(\bfr_3)\sigma_3)_{\eps_4\eps_4}\Pi_{\eps_4}(\bfr_3-\bfr_2)]\rangle_{\phi_2\phi'_2}.\no
\ee
After performing the averaging with respect to $\phi$ and $\phi'$ one obtains
\be
\Delta S&=&\frac{\pi^2 i}{4}\int_{\bfr,\bfp,\eps_i} \Gamma^{il}_{2,d}(\bfp,\eps_f)\Gamma^{jk}_{2,d}(\bfp,-\eps_f) \\
&&\times\tr[(\gamma_j\Lambda_{\eps_f} \Pi_{\eps_f}\gamma_i)\underline{Q_{\eps_1\eps_2;\alpha \beta}}]\no\\
&&\times \tr[(\gamma_l\Lambda_{\eps_f} \Pi_{\eps_f}\gamma_k)\underline{Q_{\eps_3\eps_4;\beta \alpha}}]\delta_{\eps_1-\eps_2,\eps_4-\eps_3}.\no\label{eq:DeltaSfinalb}
\ee
The origin of the additional factor $2$ compared to formula \eqref{eq:DeltaSintermediate} is the spin degree of freedom. Further evaluation gives
\be
\Delta S&=&i\pi^2\int_{\bfp,\eps_f} \;\sigma_f [\mathcal{D}^R_{\eps_f}(\bfp) \Gamma_{2;d}^R(\bfp,\eps_f)]^2\label{eq:DeltaSGamma22}\\
&&\int_{\bfr,\eps_i}\tr[\gamma_1\underline{Q_{\alpha\beta;\eps_1\eps_2}}(\bfr)]\tr[\gamma_2\underline{Q_{\beta \alpha;\eps_3\eps_4}}(\bfr)]\delta_{\eps_1-\eps_2,\eps_4-\eps_3}.\no
\ee
Here, the structure of the $\Gamma_2$-type interaction term is reproduced.

The result for the corrections to $\Gamma_1$ and $\Gamma_2$ from the first pair of diagrams can easily be found by comparing the obtained results to $S_\Gamma$,
\be
(\Delta \Gamma_1)_1&=&\frac{i}{\nu}\int_{\bfp,\eps_f}\sigma_f[\mathcal{D}^R_{\eps_f}(\bfp)\Gamma_{2;d}^R(\bfp,\eps_f)]^2\quad\label{eq:DeltaGamma1}\\
(\Delta \Gamma_2)_1&=&\frac{2i}{\nu}\int_{\bfp,\eps_f}\sigma_f[\mathcal{D}^R_{\eps_f}(\bfp)\Gamma_{2;d}^R(\bfp,\eps_f)]^2.\
\ee

Now let us clarify the cancellation within each pair when a contraction is not of the $\phi_2$-type. If one or both of the H-S fields $\phi_2$ and $\phi'_2$ are replaced by $\phi_0$ ($\phi_0'$) or $\phi_1$ ($\phi_1'$), then the overall spin structure of both types of terms corresponding to diagrams (a) and (b) coincide as well as their spin factors. However, the relative sign apparent from formulas \eqref{eq:DeltaSGamma1text} and \eqref{eq:DeltaSGamma22} remains, and thus those terms cancel. Therefore, the only contribution that remains is the one shown above with two amplitudes $\Gamma_2$.

The remaining pairs of diagrams can be treated in a similar way. The diagrams are displayed in Fig.~\ref{fig:Summary_Gamma}. Each pair is formed by the two diagrams labeled as (a) and (b).

The results are
\be
(\Delta \Gamma_1)_2&=&-\frac{2i}{\nu}\Gamma_2 \int_{\eps_f} \sigma_f \Gamma_{2,d}(\bfp,\eps_f)\;\mathcal{D}^2_{\eps_f}(\bfp)\qquad\\
(\Delta \Gamma_1)_3&=&\frac{2}{\nu}\Gamma_2\int |\eps_f| \Gamma_{2,d}^2(\bfp,\eps_f)  \mathcal{D}^3_{\eps_f}(\bfp)\\
(\Delta \Gamma_1)_4&=&-\frac{2}{\nu}\Gamma_2^2 \int |\eps_f| \Gamma_{2,d}(\bfp,\eps_f)\mathcal{D}^3_{\eps_f}(\bfp)\\
(\Delta \Gamma_1)_5&=&-\frac{i}{\nu}\Gamma_2^2 \int |\eps_f|\eps_f \Gamma^2_{2,d}(\bfp,\eps_f)\mathcal{D}^4_{\eps_f}(\bfp).
\ee
The previously explained relation holds for all five pairs of diagrams:
\be
(\Delta \Gamma_2)_i=2(\Delta \Gamma_1)_i,\quad i=1-5.\label{eq:pairsonefive}
\ee
Note that the amplitudes which appear in connection with the external legs of the diagrams are not dressed.
As encountered already for the first pair of diagrams, a cancellation takes place if at least one of the amplitudes $\Gamma_2$ is replaced by $\Gamma_0$ or $\Gamma_1$.

\subsection{Logarithmic integrals}
\label{subsec:logarithmic integrals}
Here, we present a list of logarithmic integrals that appear as a result of the RG transformations. As shown above, the  NL$\sigma$M preserves its original form during the course of this procedure. This implies that the obtained corrections can be rewritten in the form of RG equations for the $\it{flowing}$ (i.e., scale-dependent) parameters of the model. As we have already mentioned, the only small parameter needed for the RG analysis is $\rho$, which has the meaning of the sheet resistance determined at a given scale and measured in dimensional units.

\subsubsection{$\rho=\frac{1}{(2\pi)^2\nu D}$}
We concentrate on the long-range Coulomb interaction. In this limit, the effective interaction in the singlet channel is controlled by the inverse of the polarization operator. Even despite the screening, the resulting correction to $D$
differs substantially from the case of the short-range interaction due the frequency dependence of the polarization operator, as given by Eq.~\eqref{eq:chibarfinaln}, which cannot be ignored. One has to start with Eq.~\eqref{eq:DeltaDprelim}, and to write inside the integral
\be
i\int_{\bfp,\eps_f}\sigma_f D\bfp^2 \mathcal{D}^3_{\eps_f}(\bfp)\left[\Gamma^R_{d}(\bfp,\eps_f)-2\Gamma_{2,d}^R(\bfp,\eps_f)\right]
\ee
all dressed amplitudes in the explicit form:
\be
\mathcal{D}^3_{\eps_f}(\bfp)\left[\Gamma^R_{d}(\bfp,\eps_f)-2\Gamma_{2,d}^R(\bfp,\eps_f)\right]=\no\\
\Gamma_{0}\tilde{\mathcal{D}}_1\mathcal{D}_1\mathcal{D}_2+
\Gamma_{1}\mathcal{D}\mathcal{D}_1\mathcal{D}_2-2\Gamma_{2}\mathcal{D}^2\mathcal{D}_2.\label{eq:threecorrect}
\ee
For brevity, we omitted the arguments $\bfp,\eps_f$ in the second line. At a given $\eps_f$, the integral over ${\bf p}$ is convergent for each of the three terms, both in the limits of large and small momenta. One can, therefore, safely perform the integration over ${\bf p}$; the result is real and inversely proportional to $\eps_f$. The last fact is clear, if one takes a look at the dimension of the integrands. Next, owing to $\sigma_f$, the remaining integral over $\eps_f$ is twice the integral over the positive frequencies only. To present the integral in a form suitable for the RG-treatment, it remains to integrate within the energy shell $\lambda \Lambda _{\tau}<\eps_f<\Lambda _{\tau}$, where $\lambda<1$ and $\Lambda _{\tau}$ is the current scale in the RG-procedure. The upper cutoff of the scaling process is
$\Lambda _{\tau}\sim 1/\tau$; the lower one is discussed below. We will present all corrections as proportional to
\be
\int_{\lambda \Lambda _{\tau }}^{\Lambda _{\tau }}\frac{d\eps _{f}}{\eps _{f}}=\ln \lambda^{-1}.
\ee

Performing the integrations described above, one gets the result
\be
\frac{\Delta \rho }{\rho ^{2}}=\Big[\frac{1}{1+F_{0}^{\rho }}
f_{1}(z,z_{1})+\Gamma _{1}f_{2}(z,z_{1},z_{2})\no\\
-2\Gamma _{2}f_{2}(z,z,z_{2})\Big]\ln{\lambda ^{-1}}\label{eq:threeterms},
\ee
where
\be
f_1(a,b)&=&\frac{1}{a-b}\ln\frac{a}{b},\\
f_2(a,b,c)&=&\frac{2b}{b-c}f_1(a,b)-\frac{2c}{b-c}f_1(a,c),
\ee
together with the definition $f_1(a,a)=1/a$. The terms in Eq.~\eqref{eq:threeterms} arise from the $\Gamma_0$, $\Gamma_1$ and $\Gamma_2$-contributions as given in the second line of the expression \eqref{eq:threecorrect}.
Obviously, for a short-range interaction, the $\Gamma_0$-term should be excluded.

\subsubsection{$z$}
The most natural way to get the RG-equation for $z$ is to rewrite Eq.~\eqref{eq:Deltaz} as follows
\be
\Delta z=\frac{1}{2\pi\nu}\int_{\bfp}\sigma_f\mathcal{D}_{\eps_f}(\bfp)\left.\mathcal{V}_{\eps_f}^R(\bfp)
\right|_{\lambda\Lambda_{\tau}}^{\Lambda_{\tau}}.
\ee
The integral in ${\bf p}$ becomes convergent once the upper and lower limits are considered together. Then, the straightforward integration yields
\be
\Delta z=\rho \left[ -\frac{1}{2(1+F_{0)}^{\rho }}-\Gamma _{1}+2\Gamma _{2}\right] \ln
\lambda ^{-1}\label{eq:Deltazinterim}.
\ee
In the case of a short-range interaction, the first term should be abandoned.

Note that another way to perform the RG-procedure is to introduce a momentum cutoff besides the one in frequency.

\subsubsection{$\Gamma_1$ and $\Gamma_2$}

After uncovering the dressed amplitudes, and performing the necessary integrations, one gets:
\be
(\Delta \Gamma_1)_0&=&\Gamma_2\;\rho\ln\lambda^{-1}\label{eq:DeltaGammasresult}
\ee
and
\be
(\Delta \Gamma_1)_1&=&-\frac{\Gamma_2^2}{z_2}\;\rho\ln\lambda^{-1}\label{eq:fiveDeltaGamma1}\\
(\Delta \Gamma_1)_2&=&2 \Gamma_2^2\;f_1(z_2,z)\;\rho\ln\lambda^{-1}\no\\
(\Delta \Gamma_1)_3&=&-\frac{\Gamma_2^3}{z_2}\;f_2(z,z,z_2)\;\rho\ln\lambda^{-1}\no\\
(\Delta \Gamma_1)_4&=&\frac{\Gamma_2^3}{z}\;f_2(z_2,z_2,z)\;\rho\ln\lambda^{-1}\no\\
(\Delta \Gamma_1)_5&=&\frac{\Gamma_2^4}{(z-z_2)^2}\left(\frac{1}{z}+\frac{1}{z_2}-2f_1(z,z_2)\right)\;\rho\ln\lambda^{-1}.\no
\ee
Remarkably,
the sum of the five terms $i=1-5$ reduces to a very simple combination 
\be
\sum_{1}^{5}(\Delta \Gamma_1)_{i}=\frac{\Gamma _{2}^{2}}{z}\;\rho\ln\lambda^{-1}.
\ee
For $\Gamma_2$ we have 
\be
(\Delta \Gamma_2)_0&=&\left[\frac{1}{2(1+F_0^\rho)}+\Gamma_1\right]\;\rho \ln\lambda^{-1},
\ee
and $(\Delta \Gamma_2)_i=2(\Delta \Gamma_1)_i$ for $i=1-5$.

The corrections $(\Delta \Gamma_1)$ and $(\Delta \Gamma_2)$ can be summarized as follows:
\be
(\Delta \Gamma_1)&=&\left[\Gamma_2+\frac{\Gamma_2^2}{z}\right]\rho\ln\lambda^{-1},\label{eq:Gamma1final}\\
(\Delta \Gamma_2)&=&\left[ \frac{1}{2(1+F_{0}^{\rho })}+\Gamma _{1}+2\frac{\Gamma_2^2}{z}\right]\;\rho \ln\lambda^{-1}\label{eq:Gamma2notfinal}.
\ee

\subsubsection{$z_1=z-2\Gamma_1+\Gamma_2$}

It follows from the above results that $z_1$, which determines the dynamics in the $\rho$-channel (e.g., in the polarization operator), remains unchanged during the RG transformations. Indeed, by comparing Eq.~\eqref{eq:Deltazinterim} with Eqs.~\eqref{eq:Gamma1final} and \eqref{eq:Gamma2notfinal},
one immediately observes that
\be
\Delta z_1=0.
\ee
The initial values stated in Eq.~\eqref{eq:GammaF} in Sec.~\ref{sec:derivation}, therefore allow to determine the value of this unrenormalized combination:
\be
z_1=z-2\Gamma_1+\Gamma_2=\frac{1}{1+F_0^{\rho}}\label{eq:ident}.
\ee
This Ward identity\cite{Finkelstein83,Baranov99,DiCastro04,Finkelstein10} is important for finding the correct form of $\bar{\chi}^R_{nn}(\bfq,\omega)$, and also for establishing the universal form of the RG equations in the case of the screened long range Coulomb interaction. Indeed, in view of Eq.~\eqref{eq:gammrhodefinition}, where the interaction amplitude in the $\rho$-channel for small momenta has been defined, $\tilde{\Gamma}_\rho(\bfq\rightarrow 0)=\frac{1}{1+F_0^{\rho}}+2\Gamma_1-\Gamma_2$, one can read the obtained relation \eqref{eq:ident} as
\be
\tilde{\Gamma}_\rho(\bfq\rightarrow 0)=z.
\ee
Thus, the renormalized interaction amplitude and the parameter describing the renormalization of the frequency term in the case of the screened long-range interaction coincide, and do not depend on the nonuniversal Fermi liquid amplitudes. This is the reason why the RG equations in this case acquire a universal form.

\subsubsection{Final form of the RG equations}

We will write now the RG equations for the case of the screened Coulomb interaction. To make the equations universal, we exclude the combination $\frac{1}{2(1+F_{0}^{\rho })}+\Gamma _{1}$ using identity \eqref{eq:ident} discussed above. As a result, on can rewrite Eq.~\eqref{eq:threeterms} in the form
\be
\frac{\Delta \rho }{\rho ^{2}}=\left [ 1-3\left(\frac {z+\Gamma_2}{\Gamma_2}\ln \frac{z+\Gamma _2}{z}-1\right)\right ]\ln{\lambda ^{-1}},\label{eq:rhofinal}
\ee
where the two terms in the square brackets represent contributions of the $\rho$ (singlet) and $\sigma$ (triplet) channels, respectively. Note that the factor $3$ is typical for the triplet channel, and that these two contributions have opposite signs. With the help of Eq.~\eqref{eq:ident}, the equation describing the renormalization of $\Gamma_2$ acquires the following form
\be
(\Delta \Gamma_2)&=&\left[\frac{z}{2}+\frac{\Gamma_2}{2}+2\frac{\Gamma_2^2}{z}\right]\;\rho \ln\lambda^{-1}\label{eq:Gamma2final}.
\ee
Finally, the equation for $\Delta z$ simplifies, and takes a form in which the contribution of the two channels becomes immediately recognizable
\be
\Delta z&=&\frac{1}{2}\left[-z+3\Gamma_2\right]\;\rho \ln\lambda^{-1}.\label{eq:zfinal}
\ee
The corresponding RG equations can be obtained by taking derivatives with respect to $\ln {\lambda^{-1}} $, with all the coefficients understood as flowing parameters. These three (\emph{instead of four}) equations constitute a complete set of RG-equations describing the disordered electron liquid in the presence of the long-range Coulomb interaction. The long range character of the Coulomb interaction, i.e., the infinite amplitude in the limit $\bfq\rightarrow 0$, leads to a universal form of the RG-equations. Moreover, one may introduce a new variable $w_2=\Gamma_2/z$ which allows to decouple the equations for $\rho$ and the interaction in the $\sigma$-channel (represented now by $w_2$) from the equation for $z$:
\be
\frac{1}{\rho}\frac{d\log{\rho}}{d\ln\lambda^{-1}}&=&4-3\frac{1+w_2}{w_2}\ln(1+w_2)\\
\frac{1}{\rho}\frac{dw_2}{d\ln\lambda^{-1}}&=&\frac{(1+w_2)^2}{2},
\ee
and 
\be
\frac{1}{\rho}\frac{d z}{d\ln\lambda^{-1}}&=&\frac{z(3w_2-1)}{2}.
\ee
Although these equations were derived in the one-loop (first order in $\rho$) approximation, the observed decoupling of the equation for $z$ from the equations describing the other two RG charges (as well as the possibility of presenting the equations in terms of the ratio $\Gamma_2/z$) reflects the general structure of the NL$\sigma$M.\cite{Finkelstein84,Finkelstein90,Finkelstein10} This fact is important for the analysis of the Metal-Insulator transition. \cite{Punnoose05} The fixed point existing in the phase plane $\rho-w_2$ determines the equation for $z$ which, in turn, controls the critical behavior (as a function of temperature) at the metal-insulator transition.

\subsubsection{Lowest cutoff}

Finally, let us comment on the lowest cutoff for the RG-procedure. In the replica NL$\sigma$M the lower cutoff appears from the discreteness of the Matsubara frequencies, which are used to describe electron interactions at finite temperatures. In the Keldysh technique it happens differently. The matrix $\underline{\hat{Q}}=\hat{u}\circ \hat{U}\circ\hat{\sigma}_3\circ \hat{\overline{U}}\circ \hat{u}$, which is the main object of study in theory of interacting electrons, contains a superposition of two kinds of rotations. Matrices $U,\overline{U}$ describe fluctuations that correspond to diffusons, while matrices $u$ establish the connection of the diffusion modes with temperature. The latter matrices limit rotations of $U$ at energies smaller than $T$, and this is the way how the low-energy cutoff enters the RG-scheme. Technically, the cutoff enters due to the smoothening of the function $\sigma_f$ at $\eps_f\sim T$. The whole RG-procedure can be reformulated as a process of gradual sharpening of $\sigma_f$, starting from $1/\tau$ and up to $T$.

\section{Correlation functions and conductivity}
\label{sec:together}
We now combine the analysis presented in Secs.~\ref{sec:correlation functions} and \ref{sec:renormalization}; the RG-equations derived above will be connected with the observable quantities, such as the correlation functions and electric conductivity.

As it will be shown below, there is an important difference between the static part of the density-density correlation function $\bar{\chi}_{nn}^{st,R}$, and the static part of the spin-density spin-density correlation function $\chi_{s^is^i}^{st,R}$. Namely, $\bar{\chi}_{nn}^{st,R}=-2\nu\gamma_\bullet^\sigma=-2\nu/(1+F_0^\rho)$ remains unrenormalized, whereas $\chi_{s^is^i}^{st,R}$ becomes scale-dependent. The reason for the particular behavior of $\overline{\chi}_{nn}^{st,R}$ lies in the well known Ward identity: $\overline{\chi}_{nn}^{st,R}=-\partial n/\partial\mu$. It has been argued\cite{Finkelstein90} that the cancellation of corrections to $\partial n/\partial \mu$ is related to the fact that it is the much smaller quantity $1/\tau$ and not $\mu$ that determines the ultraviolet cut-off for the logarithmic singularities originating from the diffusive regime. As a consequence, the dependence of the density $n$ on the chemical potential $\mu$ cannot be modified by the discussed logarithmic corrections and, therefore, $\partial n/\partial \mu$ remains unchanged. (We shall demonstrate below that, technically, it is due to the cancellation of the logarithmic corrections.)
No protection of this type exists for the spin susceptibility that is determined by the static part of the spin-density spin-density correlation function and, indeed, the spin-susceptibility is renormalized. Finally, we use the density-density correlation function to obtain the Einstein relation for interacting electrons, and to relate the electric conductivity to the scaling parameter $\rho$.
\subsection{Corrections to $\gamma_\bullet^{\rho/\sigma}$ and the spin susceptibility}
\label{subsec:Corrections to gamma}
\begin{figure}
\includegraphics[width=8cm]{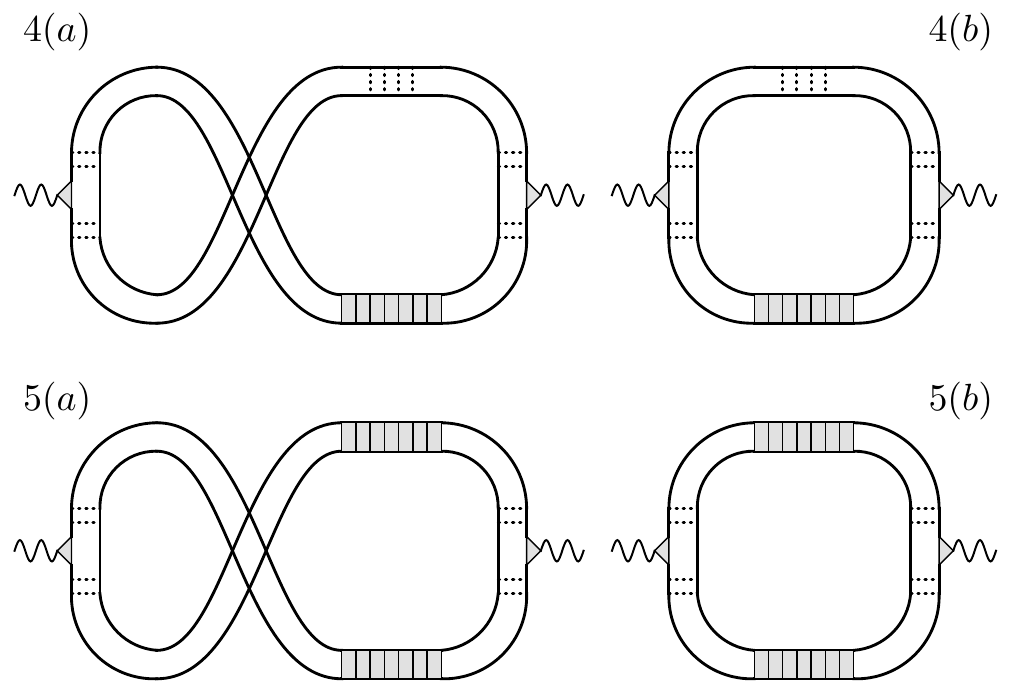}
\caption{These diagrams give rise to the corrections to $\gamma_\bullet^{\rho/\sigma}$. They are organized into two pairs, in close analogy to the corresponding diagrams in Fig.~\ref{fig:Summary_Gamma} with the same labels.}
\label{fig:static}
\end{figure}

The static parts of the correlation functions are determined by $\gamma_\bullet^{\rho/\sigma}$; compare the discussion in Sec.~\ref{sec:correlation functions}, in particular Eq.~\eqref{eq:static parts}.
We now show how these quantities are modified by the RG-corrections. One needs to find
\be
\Delta S_{\varphi\varphi}=-\frac{1}{2}\left\langle\!\left\langle S_{\varphi,2}^2 S_{int,1}\right\rangle\!\right\rangle-\frac{i}{4}\left\langle\!\left\langle S_{\varphi,2}^2 S_{int,1}^2\right\rangle\!\right\rangle.\label{eq:DeltaSphi2}
\ee
The corresponding diagrams are closely related to those presented in Fig.~\ref{fig:Summary_Gamma}, in particular to contributions $4$ and $5$ for the renormalization of the interaction amplitudes. In a similar way, when calculating the corrections to $\gamma_\bullet^{\rho/\sigma}$, one also deals with pairs of diagrams, see Fig.~\ref{fig:static}.

We present some details for the first term in Eq.~\eqref{eq:DeltaSphi2}. As mentioned, the correction consists of two parts,
\be
-\frac{1}{2}\left\langle\!\left\langle S_{\varphi,2}^2 S_{int,1}\right\rangle\!\right\rangle=A+B,
\ee
corresponding to the diagrams labeled as 4(a) and 4(b) in Fig.~\ref{fig:static}, respectively. $A$ and $B$ take the form
\be
&&A
=-\frac{i}{8}(\pi\nu)^4\sum_{n=0}^2\\
&&\left(\langle\!\langle \Tr[\underline{\vartheta}\sigma_3
\bcontraction{}{P}{P]\Tr[\underline{\phi_n}\sigma_3 }{P}
\bcontraction{PP]\Tr[\underline{\phi_n}\sigma_3 P]\tr[\underline{\phi_n}\sigma_3 }{P}{]\tr[\underline{\vartheta}\sigma_3 }{P}
\contraction{P}{P}{]\Tr[\underline{\phi_n}\sigma_3 P]\tr[\underline{\phi_n}\sigma_3 P]\tr[\underline{\vartheta}\sigma_3 P}{P}
PP]\Tr[\underline{\phi_n}\sigma_3 P]\tr[\underline{\phi_n}\sigma_3 P]\tr[\underline{\vartheta}\sigma_3 PP]\rangle\!\rangle_{\phi}\no\right.\\
&&\left.+\langle\!\langle\Tr[\underline{\vartheta}\sigma_3
\contraction{}{P}{P]\Tr[\underline{\phi_n}\sigma_3 P]\tr[\underline{\phi_n}\sigma_3 P]\tr[\underline{\vartheta}\sigma_3}{P}
\bcontraction{P}{P}{]\Tr[\underline{\phi_n}\sigma_3 }{P}
\bcontraction{PP]\Tr[\underline{\phi_n}\sigma_3 P]\tr[\underline{\phi_n}\sigma_3 }{P}{]\tr[\underline{\vartheta}\sigma_3 P}{P}
PP]\Tr[\underline{\phi_n}\sigma_3 P]\tr[\underline{\phi_n}\sigma_3 P]\tr[\underline{\vartheta}\sigma_3 PP]\rangle\!\rangle_{\phi}\right)\no\\
&=&-2\int_{x} \vec{\vartheta}^T_{\alpha\beta}(x)\gamma^2\vec{\vartheta}_{\beta\alpha}(x)\int_{\bfp,\eps_f}|\eps_f|\Gamma^R_d(\bfp,\eps_f)\mathcal{D}^3_{\eps_f}(\bfp)\no\\
&&+2\int_{x} \vec{\vartheta}^T_{\alpha\alpha}(x)\gamma^2\vec{\vartheta}_{\beta\beta}(x)\int_{\bfp,\eps_f}|\eps_f|\Gamma^R_{2,d}(\bfp,\eps_f)\mathcal{D}^3_{\eps_f}(\bfp),\no
\ee
and
\be
&&B=-\frac{i}{4}(\pi\nu)^4\sum_{n=0}^2\\
&&\langle\!\langle \Tr[\underline{\vartheta}\sigma_3
\contraction{}{P}{P]\Tr[\underline{\phi_n}\sigma_3 P]\tr[\underline{\phi_n}\sigma_3 P]\tr[\underline{\vartheta}\sigma_3 P}{P}
\bcontraction{P}{P}{]\Tr[\underline{\phi_n}\sigma_3 }{P}
\bcontraction{PP]\Tr[\underline{\phi_n}\sigma_3 P]\tr[\underline{\phi_n}\sigma_3 }{P}{]\tr[\underline{\vartheta}\sigma_3 }{P}
PP]\Tr[\underline{\phi_n}\sigma_3 P]\tr[\underline{\phi_n}\sigma_3 P]\tr[\underline{\vartheta}\sigma_3 PP]\rangle\!\rangle_{\phi}\no\\
&=&2\int_{x} \vec{\vartheta}^T_{\alpha\beta}(x)\gamma^2\vec{\vartheta}_{\beta\alpha}(x)\no\\
&&\times \int_{\bfp,\eps_f}|\eps_f|\left[\Gamma^R_{d}(\bfp,\eps_f)-2\Gamma^R_{2,d}(\bfp,\eps_f)\right]\mathcal{D}^3_{\eps_f}(\bfp).\no
\ee
In these expressions, we abbreviated $\vartheta=\gamma_\triangleleft^\rho \varphi+\gamma_{\triangleleft}^\sigma\bfvarphi\bfsigma$. Summing contributions $A$ and $B$, one gets
\be
&&-\frac{1}{2}\left\langle\!\left\langle S_{\varphi,2}^2 S_{int,1}\right\rangle\!\right\rangle\\
&=&-8(\gamma_\triangleleft^\sigma)^2\int_x \vec{\bfvarphi}^T(x)\gamma_2\vec{\bfvarphi}(x)\int_{\bfp,\eps_f}|\eps_f|\Gamma^R_{2,d}(\bfp,\eps_f)\mathcal{D}^3_{\eps_f}.\no
\ee

Two remarks are in order here. First, we see that the amplitude $\Gamma$ disappears from the final result due to a cancellation between $A$ and $B$. Second, a logarithmic correction exists only for the triplet component, the singlet part remains untouched. These two observations carry over to the calculation of the other contribution to $\Delta S_{\varphi\varphi}$, which is also organized into a pair of diagrams; see diagrams 5(a) and 5(b) in Fig.~\ref{fig:static}. The total result can conveniently be written in the form
\be
\Delta S_{\varphi\varphi}=\frac{4\nu (\gamma_\triangleleft^\sigma)^2}{\Gamma_2^2}\sum_{i=4,5}(\Delta \Gamma_1)_i\int_x\vec{\bfvarphi}^T(x)\gamma_2\vec{\bfvarphi}(x).
\ee
Comparing with $S_{\varphi\varphi}$, and using the relations for $(\Delta \Gamma)_i$ stated in Eq.~\eqref{eq:DeltaGammasresult}, one finds
\be
&&\Delta \gamma^\rho_\bullet=0,\quad\Delta \gamma^\sigma_\bullet=\frac{2\Gamma_2}{zz_2}(\gamma_\triangleleft^\sigma)^2\;g\ln\lambda^{-1}.\label{eq:Deltagamma}
\ee

The correction to $\gamma^\sigma_\bullet$ depends on the vertices $\gamma_\triangleleft^\sigma$. In Sec.~\ref{subsec:vertex corrections}, we will show that the RG-equations generalize the Fermi-liquid relations for $\gamma_\triangleleft^\sigma$ and $\gamma^\sigma_\bullet$ as follows
\be
\gamma_\triangleleft^\sigma=\gamma^\sigma_\bullet=z+\Gamma_2.
\ee
As a result, we observe that the renormalization of the electron-electron interaction in the triplet channel leads to the scale-dependent
spin susceptibility\cite{Finkelstein84,Castellani84rapid}
\be
\chi^{\sigma}=(z+\Gamma_2)\chi^{\sigma}_{free} ,
\ee
where $\chi^{\sigma}_{free}=1/2(g_L\mu_B)^2\nu$ is the unrenormalized spin susceptibility of the free electron gas.

\subsection{Vertex corrections}
\label{subsec:vertex corrections}
As we have seen in Eq.~\eqref{eq:Deltagamma}, the knowledge of the triangular vertices $\gamma_\triangleleft^{\sigma}$ is crucial for finding the static vertex $\gamma_\bullet^{\sigma}$. In addition, $\gamma_\triangleleft^{\rho/\sigma}$ also determines the dynamical correlation functions, see Eqs.~\eqref{eq:chidyn1} and \eqref{eq:chidyn2}. We will discuss the renormalization of the vertices in this section.

First of all, it is important to stress that $\gamma_\triangleleft^{\rho/\sigma}$ has been chosen as the common charge for \emph{two} vertices: the one associated with the quantum source and the one associated with the classical one. It is crucial for the overall structure of the theory that both of them are renormalized in the same way. As will be seen below, it is indeed the case.

In order to find the vertex corrections, one needs to find corrections to the term $S_{\varphi Q}$ defined in Eq.~\eqref{eq:SourcephiQ}:
\be
\Delta S_{\varphi Q}&=&i\left\langle\!\left\langle S_{\varphi,2}S_{int,1}\right\rangle\!\right\rangle-\frac{1}{2}\left\langle\!\left\langle S_{\varphi,2}S^2_{int,1}\right\rangle\!\right\rangle\\
&&-\left\langle\!\left\langle S_{\varphi,2}S_{int,1}S_{int,2}\right\rangle\!\right\rangle-\frac{i}{4}\left\langle\!\left\langle S_{\varphi,2}S^2_{int,1}S_{int,2}\right\rangle\!\right\rangle.\no
\ee
Due to the structural similarity between $S_{\varphi,2}$ and $S_{int,2}$, the calculation is very similar to the one performed for the renormalization of the interaction amplitudes, compare the corrections $(\Delta S_\Gamma)_{2-5}$ in Eqs.~\eqref{eq:DeltaSGamma2}-\eqref{eq:DeltaSGamma5}.

\begin{figure}[tb]
\includegraphics[width=8cm]{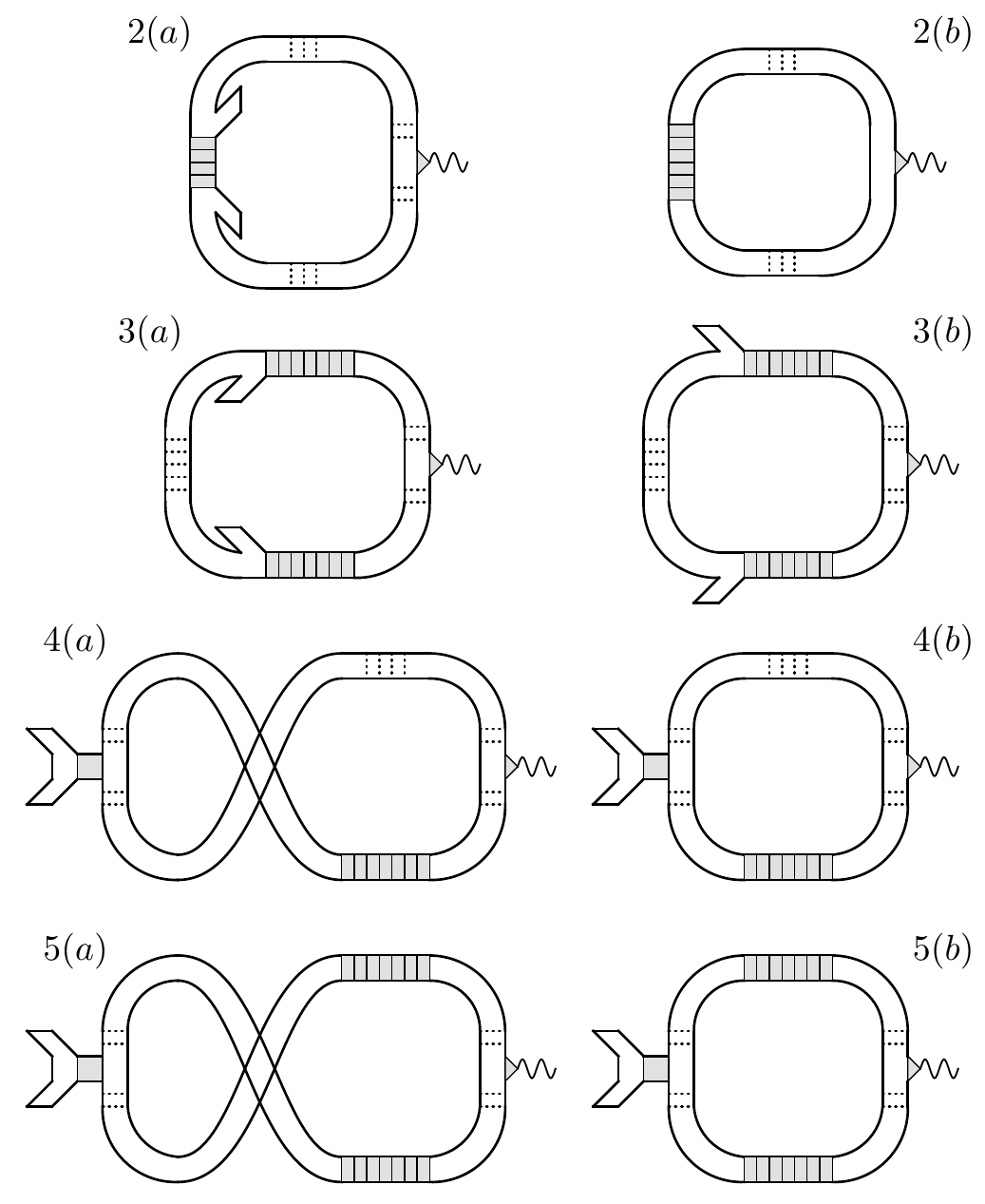}
\caption{The four pairs of diagrams relevant for the vertex corrections.}
\label{fig:Summary_Vertex}
\end{figure}

Again, the diagrams come in pairs, see Fig.~\ref{fig:Summary_Vertex}, which is structured in analogy to Fig.~\ref{fig:Summary_Gamma}. Here, we merely state the result, which can be expressed in terms of the corrections to the interaction amplitudes stated in Eq.~\eqref{eq:DeltaGammasresult}:
\be
&&\Delta S_{\varphi Q}\\
&=&2\pi\nu\Tr[\gamma_\triangleleft^\sigma\hat{\bfvarphi}\bfsigma \underline{Q}]\left[\frac{1}{2}\sum_{i=2}^3\frac{(\Delta \Gamma_1)_i}{\Gamma_2}+\sum_{i=4}^5\frac{(\Delta \Gamma_1)_i}{\Gamma_2}\right].\quad\no
\ee
It turns out that the final result is very simple
\be
\Delta \gamma_\triangleleft^\rho=0,\quad \Delta \gamma_{\triangleleft}^\sigma=\frac{2\Gamma_2}{z}\gamma_\triangleleft^\sigma \;g\ln\lambda^{-1}.
\ee

It is instructive to compare $\Delta \gamma_{\triangleleft}^\sigma$ with the correction to $z_2$,
\be
\Delta z_2=\Delta z+\Delta \Gamma_2=\frac{2\Gamma_2}{z}z_2\;g\ln\lambda^{-1}.
\ee
Since initially $z_2=\gamma_\triangleleft^\sigma=1/(1+F_0^\sigma)$, as it follows from Eqs.~\eqref{eq:initial} and \eqref{eq:GammaF}, we may conclude that the relation
\be
\gamma_\triangleleft^\sigma=z_2
\ee
holds also for the renormalized quantities.
With this information at hand, one may return to the calculation of $\Delta \gamma_\bullet^\sigma$, and finds
\be
\Delta \gamma_\bullet^\sigma=\frac{2\Gamma_2}{z} z_2\;g\ln\lambda^{-1}=\Delta z_2.
\ee
Since initially $\gamma_\bullet^\sigma=1+\Gamma_2$, one obtains that $\gamma_\triangleleft^\sigma=\gamma_\bullet^\sigma=z_2$ as it was already stated in Eq.~\eqref{eq:gammatrisigma}. Besides, the above calculations confirm that $\Delta\gamma_\triangleleft^\rho=\Delta\gamma_\bullet^\rho=0$.

Importantly, these results imply that the relations \eqref{eq:twoconditions} are indeed fulfilled. These relations make sure that the conservation laws hold at any stage of the renormalization procedure. Let us note that for the triplet channel not only the ratio $(\gamma_\triangle^\sigma)^2/\gamma_\bullet^\sigma$ equals $z_2$ but, besides, each of the quantities $\gamma_\triangle^\sigma$ and $\gamma_\bullet^\sigma$ separately. For the singlet channel, the statement $\Delta\gamma_\triangleleft^\rho=\Delta\gamma_\bullet^\rho=0$ should be supplemented with the observation that $\Delta z_1=0$. This is sufficient for the relation $z_1=(\gamma_\triangleleft^\rho)^2/\gamma_\bullet^\rho$ to hold unchanged.

\subsection{Electric conductivity}
\label{subsec:conductivity}
Combination of the continuity equation and the Kubo formula allows to extract the electric conductivity from the retarded density-density correlation function as follows:
\be
\sigma=-e^2\lim_{\omega\rightarrow 0} \lim_{\bfq\rightarrow 0} \left[\frac{\omega}{\bfq^2}\mbox{Im}\bar{\chi}^R_{nn}(\bfq,\omega)
\right]\label{eq:Kubo}.
\ee
Formula \eqref{eq:chibarfinaln} can be conveniently written as
\be
\bar{\chi}_{nn}^R(\bfq,\omega)=-\frac{\partial n}{\partial \mu}\frac{D_{FL}\bfq^2}{D_{FL}\bfq^2-i\omega},
\ee
where
\be
\quad D_{FL}=D(1+F_0^\rho).
\ee
As a result, Eq.~\eqref{eq:Kubo} leads to the Einstein relation
\be
\sigma=e^2\frac{\partial n}{\partial \mu} D_{FL}=2\nu e^2 D.
\ee

One can see that the Fermi liquid correction $1+F_0^\rho$ cancels between $D_{FL}$ and $\partial n/\partial\mu=2\nu/(1+F_0^\rho)$, so that the renormalized diffusion coefficient $D$ in the NL$\sigma$M
yields directly the electric conductivity with minimal dimensional coefficients.\cite{Finkelstein83}

\section{Conclusion}
\label{sec:conclusion}

We, thus, re-derived using the
Keldysh technique the main results of the RG theory of the disordered electron liquid.\cite{Finkelstein90,DiCastro04,Finkelstein10} Besides the set of the RG equations, the discussed items include: (i) the derivation of the Einstein relation which allows to connect the electric conductivity to the scale-dependent diffusion coefficient $D$ in the NL$\sigma$M, (ii) the expression for the renormalized spin susceptibility, (iii) a number of relations between the vertices and the interaction parameters, which in essence are the Ward identities. For understanding the overall structure of the Keldysh NL$\sigma$M, it was crucial to observe that the two vertices, the one associated with the quantum source and the one associated with the classical one, are both renormalized in the same way.

The validity of the theory has been confirmed experimentally by measuring resistance along with in-plane magnetoresistance in Si-MOSFETs at various temperatures and densities.\cite{Anissimova07,Knyazev08,Punnoose10}

We concentrated here mainly on the peculiarities induced by the matrix structure of the NL$\sigma$M in the Keldysh technique. We conclude, that apart from differences related to working with Keldysh matrices instead of replicas, the RG-procedure in both schemes are rather similar.
In subsequent papers we apply the developed technique for the calculation of the heat density-heat density correlation function, which allows us to analyze heat transport at low temperatures.

\section*{Acknowledgments}
G.~S. would like to thank K.~Takahashi and K.~B.~ Efetov for work on related subjects. The authors gratefully acknowledge the support by the Alexander von Humboldt Foundation. G.~S. also acknowledges financial support by the
Albert Einstein Minerva Center for Theoretical Physics at the Weizmann Institute of Science. A. F. is supported by the
National Science Foundation grant NSF-DMR-1006752.

\end{document}